\documentclass[smallcondensed]{svjour3}     

\usepackage[hidelinks]{hyperref}
\usepackage[utf8]{inputenc}
\usepackage{xcolor}
\usepackage{amsmath}
\usepackage{amsfonts}
\usepackage{amssymb}
\usepackage{mathrsfs}
\usepackage{fontenc}
\usepackage{subcaption}
\usepackage{epsfig}
\usepackage{epstopdf}
\usepackage{array}
\usepackage{bm}
\usepackage{tikz-cd}
\usepackage{graphicx}
\usepackage{adjustbox}
\usepackage{comment}
\usepackage{csquotes}
\usepackage{soul}

\newcommand{\cf}{\textit{cf.} }
\renewcommand{\qed}{$\hfill\blacksquare$}

\newcommand{\half}{\frac{1}{2}}

\newcommand{\B}[1]{\boldsymbol{#1}}
\newcommand{\bb}[1]{\mathbb{#1}}
\newcommand{\cl}[1]{\mathcal{#1}}

\newcommand{\tr}{\mathrm{tr}}

\newcommand{\TwoTwoMat}[4]{
	\begin{pmatrix}
		#1 & #2 \\
		#3 & #4
\end{pmatrix}}

\newcommand{\TwoVecArr}[2]{
	\begin{matrix}
		#1\\
		{\color{gray}#2} 
\end{matrix}}

\newcommand{\TwoVecText}[2]{
	\begin{matrix}
		#1\\
		{\color{gray}\mathrm{#2}} 
\end{matrix}}

\newcommand{\map}[3]{{#1}:{#2}\rightarrow {#3}}
\newcommand{\fullmap}[5]{
	\begin{split}
		#1 : {#2} &\rightarrow {#3}\\
		{#4} &\mapsto {#5}
	\end{split}
}

\newcommand{\sym}{\mathrm{sym}}
\newcommand{\asym}{\mathrm{skew}}

\newcommand{\spFn}[1]{C^\infty(#1)}
\newcommand{\spVec}[1]{\Gamma(T#1)}
\newcommand{\spFrm}[2]{\Omega^{#2}(#1)}
\newcommand{\spFrmB}[1]{\spFrm{\cl{B}}{#1}}
\newcommand{\spFrmS}[1]{\spFrm{\cl{S}}{#1}}
\newcommand{\spVecPhi}{\Gamma(\varphi^*T\cl{S})}
\newcommand{\spVecS}{\Gamma(T\cl{S})}
\newcommand{\spVecB}{\Gamma(T\cl{B})}

\newcommand{\spvecFrmB}[1]{\spFrm{\cl{B};T\cl{B}}{#1}}
\newcommand{\spcovFrmB}[1]{\spFrm{\cl{B};T^*\cl{B}}{#1}}
\newcommand{\spvecFrmS}[1]{\spFrm{\cl{S};T\cl{S}}{#1}}
\newcommand{\spcovFrmS}[1]{\spFrm{\cl{S};T^*\cl{S}}{#1}}
\newcommand{\spvecFrmPhi}[1]{\Omega_\varphi^{#1}(\cl{B};T\cl{S})}
\newcommand{\spcovFrmPhi}[1]{\Omega_\varphi^{#1}(\cl{B};T^*\cl{S})}

\newcommand{\spvecFrmBbnd}[1]{\spFrm{\partial\cl{B};T\cl{B}}{#1}}
\newcommand{\spcovFrmBbnd}[1]{\spFrm{\partial\cl{B};T^*\cl{B}}{#1}}
\newcommand{\spvecFrmSbnd}[1]{\spFrm{\partial\cl{S};T\cl{S}}{#1}}
\newcommand{\spcovFrmSbnd}[1]{\spFrm{\partial\cl{S};T^*\cl{S}}{#1}}
\newcommand{\spvecFrmPhibnd}[1]{\Omega_\varphi^{#1}(\partial\cl{B};T\cl{S})}
\newcommand{\spcovFrmPhibnd}[1]{\Omega_\varphi^{#1}(\partial\cl{B};T^*\cl{S})}

\newcommand{\spFrmM}[2]{\spFrm{M;#2}{#1}}
\newcommand{\spFrmMf}[2]{\Omega_f^{#1}(M;#2)}

\newcommand{\intB}{\int_{\cl{B}}}
\newcommand{\intS}{\int_{\cl{S}}}

\newcommand{\spC}{\mathscr{C}}
\newcommand{\spMetB}{\cl{M}(\cl{B})}
\newcommand{\spIsomA}{\mathrm{Isom}(\mathscr{A})}
\newcommand{\spSymTensB}{\Gamma(ST^0_2\cl{B})}
\newcommand{\spSymTensS}{\Gamma(ST^0_2\cl{S})}
\newcommand{\bndB}{\partial\cl{B}}
\newcommand{\cfgSp}{\mathscr{C}}
\newcommand{\cfgT}{\varphi_t}
\newcommand{\cfgInvT}{\varphi{\scriptstyle^{-1}_t}}
\newcommand{\cfgP}{\varphi_t^*}
\newcommand{\cfgPFwd}{\varphi_{t,*}}
\newcommand{\cfgPB}{\varphi_{t,b}^*}
\newcommand{\cfgPref}{\varphi_{0}^*}
\newcommand{\Finvr}{F{\scriptstyle^{-1}}}

\newcommand{\mFM}{\tilde{\mu}}
\newcommand{\mFC}{\hat{\mu}}
\newcommand{\mFS}{\mu}

\newcommand{\mDS}{\rho}
\newcommand{\mDC}{\hat{\rho}}
\newcommand{\mDM}{\tilde{\rho}}

\newcommand{\vFS}{\omega_{\gS}}
\newcommand{\vFC}{\hat{\omega}_{\gC}}
\newcommand{\vFCt}{\hat{\omega}_{\gC_t}}
\newcommand{\vFM}{\tilde{\omega}_{\gMref}}

\newcommand{\vS}{v}
\newcommand{\vM}{\tilde{v}}
\newcommand{\vC}{\hat{v}}
\newcommand{\parx}[1]{\frac{\partial}{\partial x^{#1}}}
\newcommand{\parX}[1]{\frac{\partial}{\partial X^{#1}}}
\newcommand{\dx}[1]{\mathrm{d}x^{#1}}
\newcommand{\dX}[1]{\mathrm{d}X^{#1}}

\newcommand{\vfS}{v^\flat}
\newcommand{\vfM}{\tilde{v}^\flat}
\newcommand{\vfC}{\hat{v}^\flat}

\newcommand{\gS}{g}
\newcommand{\gM}{\tilde{g}}
\newcommand{\gC}{\hat{g}}
\newcommand{\gMref}{G}
\newcommand{\nabS}{\nabla}
\newcommand{\nabM}{\tilde{\nabla}}
\newcommand{\nabC}{\hat{\nabla}}
\newcommand{\Lie}[2]{\cl{L}_{#1}{#2}}
\newcommand{\divrS}{\mathrm{div}}
\newcommand{\divrC}{\widehat{\mathrm{div}}}
\newcommand{\divrM}{\widetilde{\mathrm{div}}}

\newcommand{\epS}{\varepsilon}
\newcommand{\epC}{\hat{\varepsilon}}
\newcommand{\epfS}{\varepsilon^\flat}
\newcommand{\epfC}{\hat{\varepsilon}^\flat}

\newcommand{\extd}{\mathrm{d}}
\newcommand{\cfgPleg}[1]{\varphi_{\mathrm{#1}}^*}
\newcommand{\cfgFleg}[1]{\varphi_{\mathrm{#1},*}}

\newcommand{\wedgedot}{\ \dot{\wedge}\ }
\newcommand{\extcdS}{\extd_{\nabS}}
\newcommand{\extcdC}{\hat{\extd}_{\nabC}}
\newcommand{\extcdM}{\tilde{\extd}_{\nabM}}

\newcommand{\duPair}[3]{\left\langle  #1 | #2 \right\rangle_{#3}}

\newcommand{\inPair}[3]{\left\langle  #1 , #2 \right\rangle_{#3}}

\newcommand{\hodgeS}{\star^\flat}
\newcommand{\invhodgeS}{\star^\sharp}
\newcommand{\hodgeC}{\hat{\star}^\flat}
\newcommand{\invhodgeC}{\hat{\star}^\sharp}
\newcommand{\hodgeM}{\tilde{\star}^\flat}
\newcommand{\invhodgeM}{\tilde{\star}^\sharp}

\newcommand{\flatV}{\flat_\mathrm{v}}

\newcommand{\Ekin}{\mathscr{E}_\mathrm{kin}}
\newcommand{\Eint}{\mathscr{E}_\mathrm{int}}
\newcommand{\Pst}{\mathscr{P}_\mathrm{st}}
\newcommand{\Sbound}{{\partial\cl{S}}}
\newcommand{\Bbound}{{\partial\cl{B}}}
\newcommand{\stS}{\cl{T}}
\newcommand{\stC}{\widehat{\cl{T}}}
\newcommand{\stM}{\widetilde{\cl{T}}}
\newcommand{\ptr}{i_{\mathrm{f}}^*}

\newcommand{\momM}{\widetilde{\cl{M}}}
\newcommand{\momC}{\widehat{\cl{M}}}
\newcommand{\momS}{\cl{M}}

\newcommand{\stTauS}{\tau}
\newcommand{\stTauC}{\hat{\tau}}
\newcommand{\stTauM}{\tilde{\tau}}

\newcommand{\stSigS}{\sigma}
\newcommand{\stSigC}{\hat{\sigma}}
\newcommand{\stSigM}{\tilde{\sigma}}

\newcommand{\EintC}{\cl{\hat{E}}}
\newcommand{\EintM}{\cl{\tilde{E}}}
\newcommand{\EintS}{\cl{E}}
\newcommand{\parD}[2]{\frac{\partial #1}{\partial #2}}
\newcommand{\gradEintC}{{\left(\parD{\EintC}{\gC}\right)}^\flat}
\newcommand{\gradEintM}{{\left(\parD{\EintM}{F}\right)}}
\newcommand{\gradEintS}{{\left(\parD{\EintS}{\gS}\right)}^\flat}

\newcommand{\andrea}[1]{{\color{black}#1}}
\newcommand{\stefano}[1]{{\color{black}#1}}
\newcommand{\fede}[1]{{\color{black}#1}}

\newcommand{\erwin}[1]{{\color{black}#1}}

\graphicspath{{Figures/}}

\title{Intrinsic nonlinear elasticity:\\ An exterior calculus formulation}
\author{Ramy Rashad, Andrea Brugnoli, Federico Califano, Erwin Luesink, Stefano Stramigioli}

\institute{R. Rashad, F. Califano, S. Stramigioli \at
	Robotics and Mechatronics Department, University of Twente, The Netherlands.\\
	\email{\{r.a.m.rashadhashem, f.califano, s.stramigioli\}@utwente.nl}           
	\and
	A. Brugnoli \at
	Institute of Mathematics, Technische Universit\"at Berlin, Germany.\\
	\email{brugnoli@math.tu-berlin.de}
	\and
	E. Luesink \at
	Mathematics of Operations Research, University of Twente, The Netherlands.\\
	\email{e.luesink@utwente.nl}
}

\date{\today}
\authorrunning{Rashad et al.}
\titlerunning{Intrinsic nonlinear elasticity}

\begin{document}

\maketitle

\begin{abstract}
	In this paper we formulate the theory of nonlinear elasticity in a geometrically intrinsic manner using exterior calculus and bundle-valued differential forms. 
	We represent kinematics variables, such as velocity and rate-of-strain, as intensive vector-valued forms while kinetics variables, such as stress and momentum, as extensive covector-valued pseudo-forms. 
	We treat the spatial, material and convective representations of the motion and show how to geometrically convert from one representation to the other.
	Furthermore, we show the equivalence of our exterior calculus formulation to standard formulations in the literature based on tensor calculus.
	In addition, we highlight two types of structures underlying the theory. First, the principle bundle structure relating the space of embeddings to the space of Riemannian metrics on the body, and how the latter represents an intrinsic space of deformations. Second, the de Rham complex structure relating the spaces of bundle-valued forms to each other.
	
\end{abstract}


\newpage

\section{Introduction}
Identifying the underlying structure of partial differential equations is a fundamental topic in modern treatments of continuum mechanics and field theories in general. Not only does every discovery of a new structure provide a better mathematical understanding of the theory, but such hidden structures are fundamental for analysis, discretization, model order-reduction, and controller design. Throughout the years, many efforts were made to search for the geometric, topological and energetic structures underlying the governing equations of continuum mechanics and we aim in this paper to contribute to this search.

\subsubsection*{Geometric structure}
The first endeavor in this journey began around 1965 by the work of C. Truesdell and W. Noll \cite{Truesdell1966TheMechanics} on one side and V. Arnold \cite{arnold1965topologie} on the other side, \erwin{where the focus of the latter is on fluid mechanics.} 
The common factor in both works was differential geometry which introduced new insights to fluid mechanics and elasticity in addition to simplifying many complications that are inherent in classical coordinate-based formulations.
The starting point in this geometric formulation of elasticity is to represent the configuration of an elastic body as an embedding $\map{\varphi}{\cl{B}}{\mathscr{A}}$ of the body manifold $\cl{B}$ into the ambient space $\mathscr{A}$.

From a conceptual point of view, an elastic body during a deformation process is characterized by a few physical variables (e.g. velocity, momentum, strain, and stress) and constitutive equations that relate these variables to each other. \erwin{One of the challenges in nonlinear elasticity is to understand the motion and deformation separately. In literature there is an abundance of mathematical representations addressing this issue, but usually feature the same physical variables}.
A recurrent theme in the literature is to unify these different representations and show how they are related to each other using tools of differential geometry. 

One reason for this multiplicity of representations is that one can represent each physical variable with respect to an observer attached to $\cl{B}$ (known as the \textit{convective} representation), an observer attached to $\mathscr{A}$ (known as the \textit{spatial} representation), or using two-point tensor fields on both $\cl{B}$ and $\mathscr{A}$ (known as the \textit{material} representation).
Even though all three representations are equivalent, each has its own advantages since some parts of the theory are more intuitive or have simpler expressions in one representation compared to the others. Provided that one can juggle between the three representations in a clear way that respects their geometric nature, there should be no problem in principle. In this respect, the differential geometric concepts of pullback and pushforward have proven to be essential for this smooth transition between the convective, material and spatial representations.

One of the important principles in geometric mechanics is that of \textit{intrinsicality} emphasized by Noll \cite{noll1974new}.
In his work, it was highlighted that the matter space $\cl{B}$ should be conceptually and technically distinguished from any of its configurations in the ambient space $\mathscr{A}$. With this separation, one can identify which concepts are \textit{intrinsic} to the elastic body and which are dependent on some arbitrary \textit{reference configuration}.
An important feature of this formulation is that the body manifold $\cl{B}$ does not have an intrinsic metric and is merely a continuous assembly of particles equipped only with a mass measure. \erwin{In other words, the body manifold is a space that merely contains information about matter, but not of scale, angles or distances.} On the other hand, a (constant) metric on $\cl{B}$ depends on the choice of reference configuration and thus is a non-intrinsic property.
Equipping the body manifold with a Riemannian structure is in fact another source of multiplicity of mathematical representations in the literature. One clear example of its consequences is in representing strain and stress.

Intuitively speaking, strain is the difference between any two states of deformation (i.e. a relative deformation) and not necessarily that one of them is an unloaded (stress-free) reference configuration.
In the literature one can find a very large number of tensor fields that are used to describe the state of deformation. The most common ones are the right Cauchy-Green and Piola tensor fields, used in convective representations, and the left Cauchy-Green and Almansi tensor fields, used in spatial representations.
Using the Riemannian metrics on $\cl{B}$ and $\mathscr{A}$ one can then define more tensorial-variants of these tensor fields by raising and lowering their indices.
Each of these deformation tensor fields gives rise to a different definition of strain and consequently a different stress variable.
The stress representations can be even doubled by distinguishing between mass-dependent and mass-independent versions (e.g. the Kirchoff and Cauchy stress tensor fields in the spatial representation).

Using tools from differential geometry, one can see that all the aforementioned representations of deformation states are equivalent to only one intrinsic quantity! Namely, the pullback of the Riemannian metric of $\mathscr{A}$ onto $\cl{B}$ by the embedding $\varphi$. 
This time-dependent metric on $\cl{B}$ is an intrinsic quantity that allows one to define strain without referring to an undeformed reference configuration.
Based on this geometric insight, it was further discovered by P. Rougee \cite{Rougee2006AnStrain} that the space of Riemannian metrics on $\cl{B}$, denoted by $\spMetB$, played a fundamental role in the intrinsic formulation of finite-strain theory. In particular, it was shown that a point on the infinite-dimensional Riemannian manifold $\spMetB$ represents a state of deformation while the rate of strain and stress are elements of the tangent and cotangent spaces, respectively, at a point in $\spMetB$.

The construction of the Riemannian structure of $\spMetB$ has led to many findings and is still an active area of research. The most profound one being that one cannot simply define the strain to be the subtraction of two states of deformations (e.g. as in \cite[Sec. 1.3]{Marsden1994MathematicalElasticity}). Instead, one should take the curvature of $\spMetB$ into account which led to the introduction of the logarithmic strain measure \cite{Fiala2011GeometricalMechanics}.
Another important finding is that the numerous objective stress rates used in hypo-elasticity are equivalent to covariant differentiation on $\spMetB$ \cite{Kolev2021ObjectiveMetrics} and not all of them are derivable from a Lie derivative as claimed in \cite[Sec. 1.6]{Marsden1994MathematicalElasticity}.

\subsubsection*{Topological structure}
An important feature of the geometric approach to continuum mechanics is the separation between metric-dependent and topological metric-free operations. Identifying the underlying topological structure of the governing equations is fundamental for both analysis and discretization as well as it has the advantage of being applicable to both classical and relativistic theories \cite{Segev2013NotesFields}.

Physical variables in continuum mechanics are naturally associated to integral quantities on either $\cl{B}$ or its configuration $\varphi(\cl{B})$ in the ambient space. Mass, kinetic energy, strain energy, and stress power are examples of such quantities.
These variables are in fact densities that should be integrated over $\cl{B}$ or $\varphi(\cl{B})$ in order to yield a real number.
This integration process is metric-independent and the theory of integration over manifolds implies that the natural mathematical objects to represent these densities are differential forms \cite{Frankel2019ThePhysics}. Similar to a function that can be naturally evaluated at a point, a differential $k$-form can be naturally evaluated on $k$-dimensional space.

In contrast to traditional formulations of continuum mechanics using vector and tensor calculus, exterior calculus based on differential forms highlights this difference between topology and geometry. Furthermore, it provides an elegant machinery for differential and integral calculus that not only unifies numerous operations and identities of tensor calculus, but also generalizes them to arbitrary dimensions and coordinates.

It was shown by the work of Frankel \cite{Frankel2019ThePhysics} and Kanso et.al \cite{Kanso2007OnMechanics} that one needs to use bundle-valued differential forms for representing solid and fluid mechanics using exterior calculus.
In particular, their work highlighted that tensor fields used to represent the physical variables have in fact \textit{two legs} that should be distinguished from each other; a \enquote{form} leg and a \enquote{bundle-value} leg.
The use of bundle-valued forms clarified more the difference between the spatial and material representations and showed that one can go back and forth by pulling-back or pushing-forward the form leg only leaving the bundle-valued leg untouched.


An important application of studying the topological structure of continuum mechanics is structure-preserving discretization which aims to develop numerical schemes that represent the underlying smooth structures at the discrete level. The celebrated de Rham complex is a typical example of such topological structure which is fundamental for the development of Finite Element Exterior Calculus \cite{arnold2018finite} and Discrete Exterior Calculus \cite{hirani2003discrete}.
The underlying complex structure of linear and nonlinear elasticity has been thoroughly studied in \cite{Angoshtari2015DifferentialMechanics,Angoshtari2016HilbertElasticity,yavari2020applications} and its application for developing numerical schemes is an active area of research \cite{Yavari2008OnElasticity,FaghihShojaei2018Compatible-strainElasticity,FaghihShojaei2019Compatible-strainElasticity}.

\subsubsection*{Objectives and main result of this paper}

\begin{figure}
	\centering
	\includegraphics[width=\columnwidth]{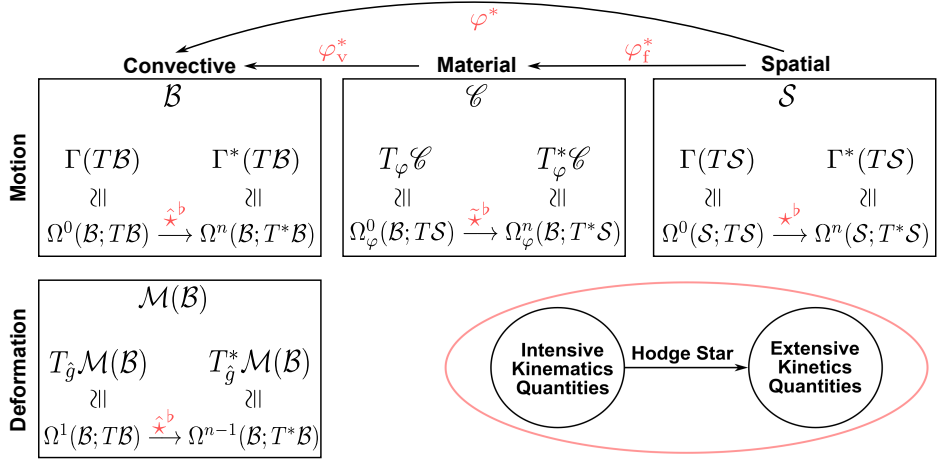}
	\caption{Overview of our intrinsic geometric formulation of nonlinear elasticity}
	\label{fig:main_result}
\end{figure}

In this paper we focus on the formulation of nonlinear elasticity using exterior calculus in a geometrically intrinsic manner.
Throughout the paper we aim to highlight the underlying geometric and topological structures of nonlinear elasticity while treating the spatial, material and convective representations of the theory. 
An overview of our formulation and the main result is depicted in Fig. \ref{fig:main_result}.

The major contribution of this paper lies in its holistic approach that combines 1) the intrinsicality principle of \cite{noll1974new}, 2) the geometric formulation of deformation using the space of Riemannian metrics $\spMetB$ by \cite{Rougee2006AnStrain,Fiala2011GeometricalMechanics}, and 3) the exterior calculus formulation using bundle-valued forms by \cite{Kanso2007OnMechanics}.
Compared to \cite{Rougee2006AnStrain,Fiala2011GeometricalMechanics}, the novelty of our work lies in its coordinate-free treatment using exterior calculus. In addition, we highlight the principal fiber bundle structure relating the space of Riemannian metrics $\spMetB$ to the configuration space $\spC$ of embeddings from $\cl{B}$ to $\mathscr{A}$. This hidden structure allows one to decompose the motion of the elastic body into a pure deformation and a pure rigid body motion. In addition, it justifies why the description of constitutive stress-strain relations can be most conveniently done in the convective representation.
Compared to \cite{Kanso2007OnMechanics,Angoshtari2013GeometricElasticity}, the novelty of our work is that we treat all three representations of the motion, show how they are related in exterior calculus, and emphasize their underlying de Rham complexes.

Our formulation of nonlinear elasticity is distinguished by its minimalistic nature which simplifies the theory to its essential intrinsic coordinate-free parts.
We show how the kinematics are naturally described using the tangent bundles $T\spC$ and $T\spMetB$ in addition to the space of vector fields $\spVecB$ and $\spVecS$, on $\cl{B}$ and $\cl{S}$ respectively. This will include velocities and rate-of-strain variables.
By identifying these kinematic quantities with appropriate \textit{intensive vector-valued forms}, the momentum and stress variables will be naturally represented as \textit{extensive covector-valued pseudo-forms}, by topological duality. Furthermore, using Riesz representation theorem, we will construct appropriate Hodge-star operators that will relate the different variables to each other.
Not only does our intrinsic formulation reflect the geometric nature of the physical variables, but also the resulting expressions of the dynamics are compact, in line with physical intuition, and one has a clear recipe for changing between the different representations.
Finally, in order to target a wider audience than researchers proficient in geometric mechanics, we present the paper in a pedagogical style using several visualizations of the theory and include coordinate-based expressions of the abstract geometric objects.

The outline of the paper is as follows: In Sec. 2, we present an overview of the \textbf{motion kinematics} of an elastic body in an intrinsic coordinate-free manner along with a separate subsection for the coordinate-based expressions.
In Sec. 3, we discuss the \textbf{deformation kinematics} highlighting the role of the space of Riemannian metrics for describing deformation of an elastic body.
In Sec. 4, we discuss the \textbf{mass structure} associated to the body which relates the kinematics variables to the kinetics ones and highlight the intrinsicality of using mass top-forms instead of mass density functions.
In Sec. 5, bundle-valued forms and their \textbf{exterior calculus} machinery will be introduced and shown how they apply to nonlinear elasticity. 
In Sec. 6, we present the dynamical \textbf{equations of motion} formulated using exterior calculus and then we show their equivalence to \textbf{standard formulations} in the literature in Sec. 7.
In Sec. 8, we discuss the principal bundle structure relating the configuration space to the space of Riemannian metrics in addition to the underlying de Rham complex structure of bundle-valued forms.
Finally, we conclude the paper in Sec. 9.


\section{Intrinsic motion kinematics}\label{sec:kinematics}
In this section we recall the geometric formulation of the kinematic variables and operations that describe motion of an elastic body in a \stefano{completely} coordinate-free manner. In such intrinsic treatment we do not identify the abstract body with a reference configuration in the ambient space.
We first describe the three representations of the motion and the various relations to go from one representation to the other in a coordinate-free manner.
The corresponding coordinate-based expression will be presented in a separate section.
It is assumed that the reader is familiar with the geometric formulation of elasticity and differential geometry, especially the topics of differential forms and fiber bundles. Due to its relevance to work, we provide in the appendix a summary of fiber bundles while further background and details can be found in \cite{Frankel2019ThePhysics,Marsden1994MathematicalElasticity,Truesdell1966TheMechanics}.

\subsection{Configuration and velocity}
\begin{figure}
	\centering
	\begin{subfigure}[t]{0.58\textwidth}
		\includegraphics[width=\columnwidth]{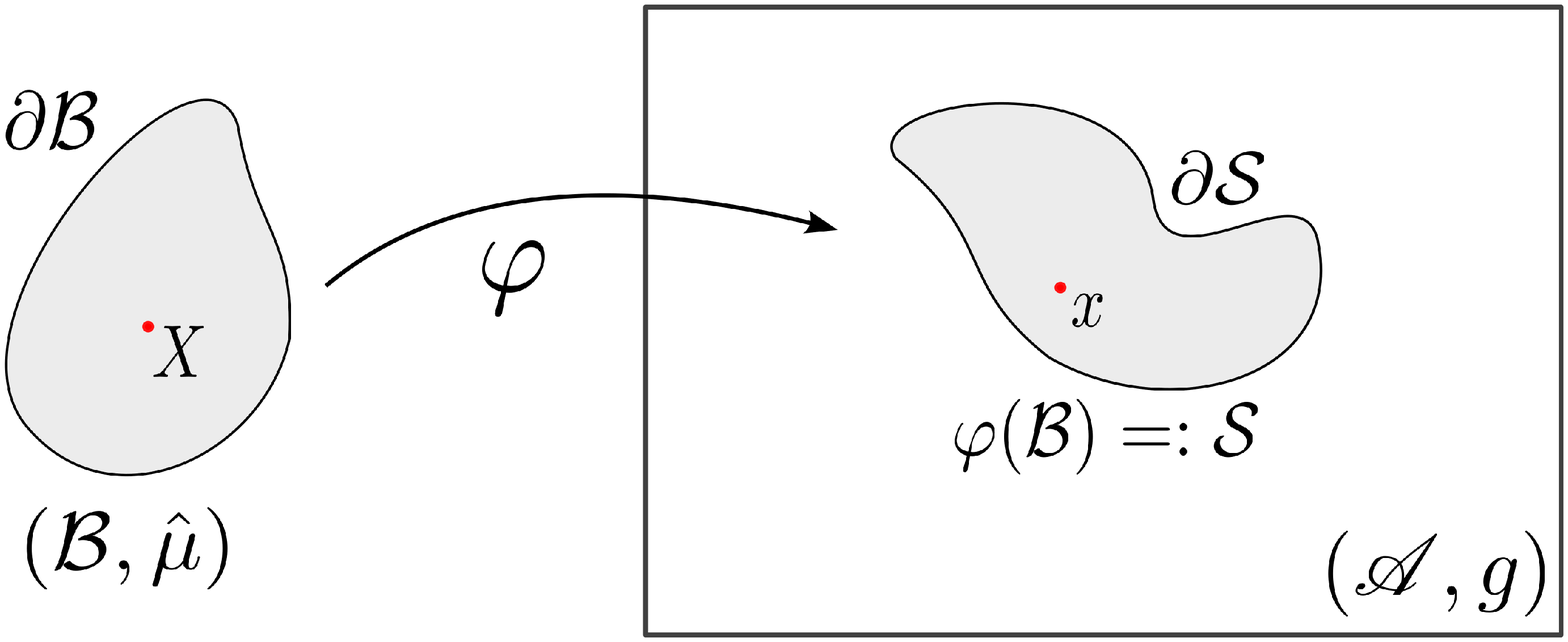}
		\caption{configuration}
		\label{fig:cfg}
	\end{subfigure}
	$\qquad$
	\begin{subfigure}[t]{0.35\textwidth}
		\includegraphics[width=\columnwidth]{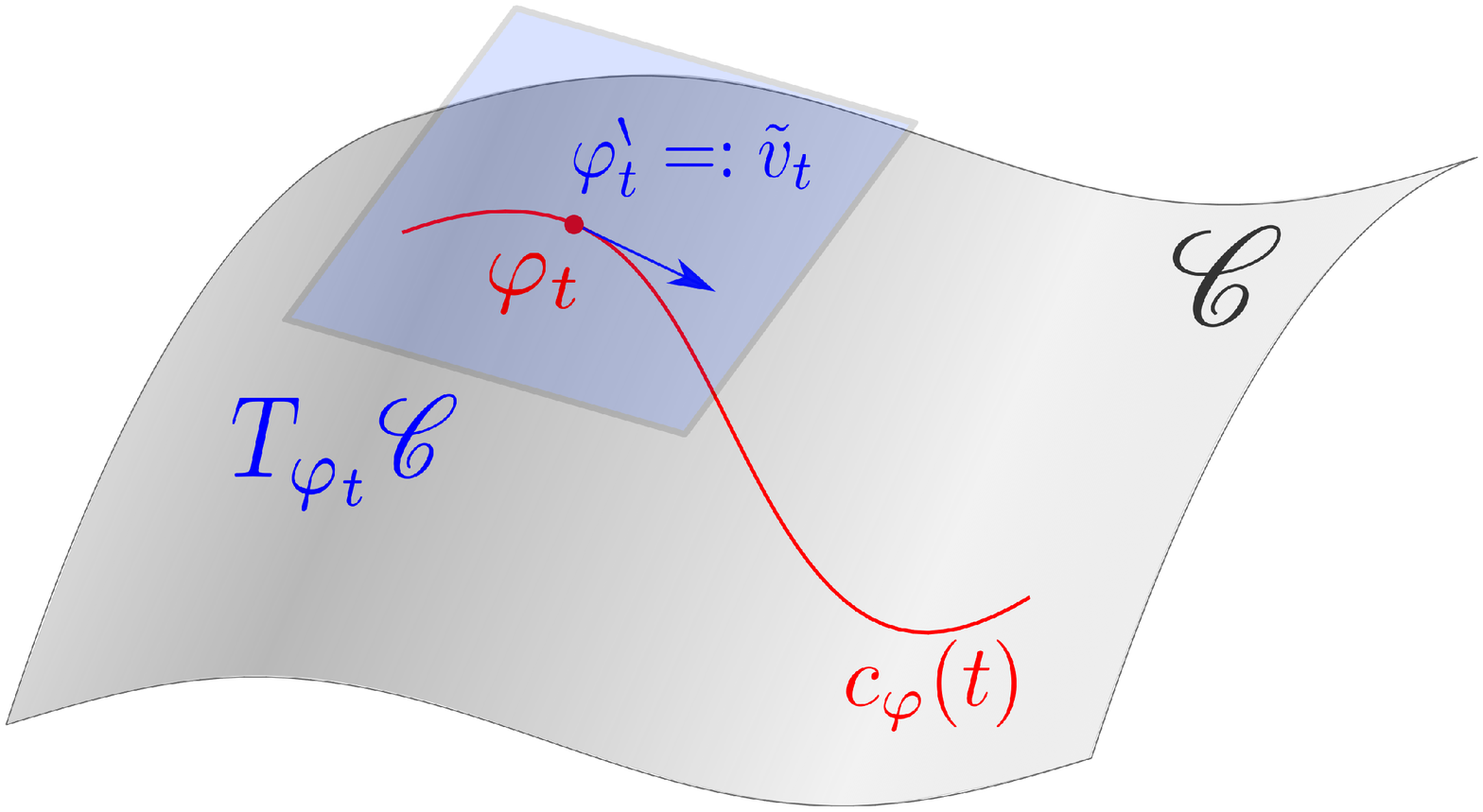}
		\caption{configuration space}
		\label{fig:cfg_space}
	\end{subfigure}
	\caption{Illustration of the embedding $\map{\varphi}{\cl{B}}{\mathscr{A}}$ of the elastic body $\cl{B}$ in the ambient space $\mathscr{A}$ and its motion as a curve in the configuration space $\mathscr{C}$.}
\end{figure}

The geometric setting for an elastic body undergoing a deformation is as follows. The material points of the body comprise mathematically a three-dimensional compact and orientable smooth manifold $\cl{B}$ with boundary $\bndB$.
This body manifold is equipped with a mass-form $\mFC \in \spFrmB{3}$ representing \stefano{the material property of mass in the body,} and we denote by $X\in\cl{B}$ a material particle.
The ambient space in which this body deforms is represented by a three-dimensional smooth oriented manifold $(\mathscr{A},g)$ with $g$ denoting its Riemannian metric. 
Therefore following the work of Noll \cite{noll1974new}, \stefano{we completely split the body with its \textit{material properties} from the embodying space with its \textit{geometric properties}}. The structures of $\cl{B}$ and $\mathscr{A}$ express these constant physical properties associated to each entity.
In this work, we will focus on the case $\mathrm{dim}(\cl{B}) =3$. At the end of this paper we will comment on how to treat other cases. 

The configuration of the elastic body is represented by a smooth orientation-preserving embedding
$\map{\varphi}{\cl{B}}{\mathscr{A}}$, which represents a placement of the body in the ambient space.
With reference to Fig. \ref{fig:cfg}, we will denote the image of the whole body by $\cl{S}:= \varphi(\cl{B}) \subset \mathscr{A}$ and we will denote by $x\in \cl{S}$ the spatial points of the body. Since $\dim(\cl{B}) = 3$, also $\dim(\cl{S})=3$.
The configuration space is thus the set $\cfgSp:= \text{Emb}^\infty(\cl{B},\mathscr{A})$ of smooth embeddings of $\cl{B}$ in $\mathscr{A}$ which can be equipped with the structure of a \fede{an infinite dimensional} differential manifold \cite{Abraham1988ManifoldsApplications}.
A motion of the elastic body is represented by a smooth curve $\map{c_\varphi}{\bb{R}}{\cfgSp}$, as illustrated in Fig. \ref{fig:cfg_space}.
Using the fact that an embedding is a diffeomorphism onto its image, $c_\varphi(t)=:\cfgT$ represents a \fede{one-parameter family} of diffeomorphisms $\map{\cfgT}{\cl{B}}{\cl{S}}$.

\begin{figure}
	\centering
	\begin{subfigure}{0.6\textwidth}
		\includegraphics[width=\columnwidth]{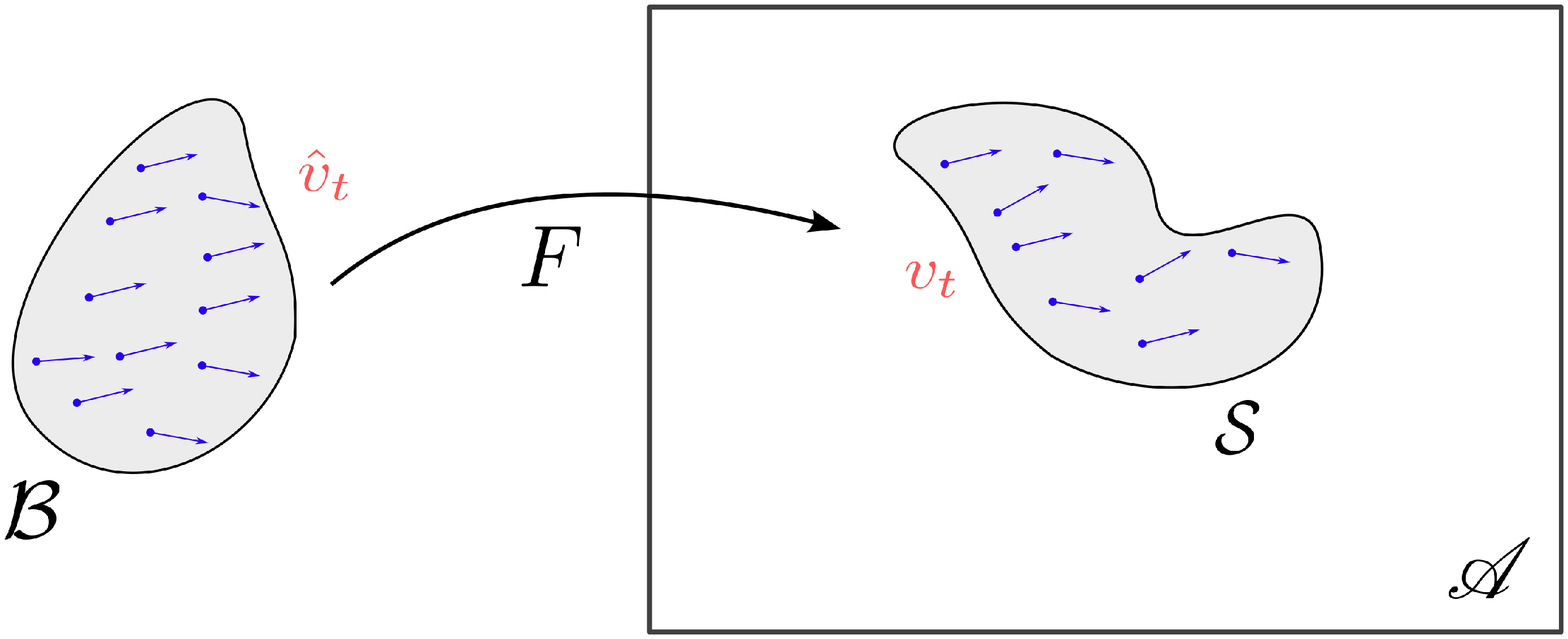}
		\caption{spatial and convective velocity fields}
	\end{subfigure}
	$\qquad$
	\begin{subfigure}{0.3\textwidth}
		\includegraphics[width=\columnwidth]{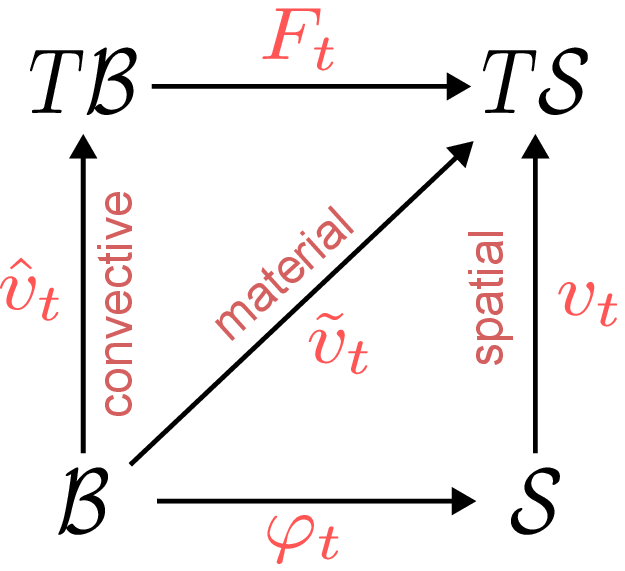}
		\caption{commutative diagram}
	\end{subfigure}
	\caption{Illustration of the body's velocity field in its spatial representation $\vS_t\in \spVecS$ and its convective representation $\vC_t\in \spVecB$ and their corresponding commutative diagram.}
	\label{fig:velocities}
\end{figure}

The tangent vector to the curve $c_\varphi$ at a given configuration $\cfgT$ is denoted by $\vM_t \in T_{\cfgT}\cfgSp$ which defines a map $\map{\vM_t}{\cl{B}}{T\cl{S}}$ such that
\begin{equation}\label{eq:def_mat_velocity}
	\vM_t : X\in\cl{B} \mapsto \left.\frac{d}{ds}\right |_{s=t} \varphi_s(X) \in T_{\cfgT(X)}\cl{S}.
\end{equation}
Thus, the tangent space $T_{\cfgT}\cfgSp$ is canonically identified with the (infinite-dimensional) vector space $\spVecPhi$, the space of vector fields over the map $\varphi$ (i.e. sections of the induced bundle $\varphi^*T\cl{S}$ as discussed in Appendix \ref{appendix:bundles}).
We refer to $\vM_t \in \spVecPhi$ as the material (Lagrangian) velocity field which describes the infinitesimal motion of the body.
This motion can be also described by the \enquote{true} vector fields $\vS_t\in \spVecS$ or $\vC_t\in \spVecB$ (\cf Fig. \ref{fig:velocities}) defined by
\begin{equation}\label{eq:vel_relations}
	\vS_t := \vM_t \circ \cfgInvT, \qquad\qquad \vC_t := T\cfgInvT \circ \vM_t = T\cfgInvT \circ \vS_t \circ \cfgT,
\end{equation}
with $\map{T\cfgInvT}{T\cl{S}}{T\cl{B}}$ denoting the tangent map of $\cfgInvT$. 
While $\vS_t$ is referred to as the spatial velocity field, $\vC_t$ is referred to as the convective velocity field.
Using the notation of pullbacks, the material, spatial and convective representations of the body's velocity are related by
$$\vC_t = \cfgP(\vS_t), \qquad \qquad \vM_t = \cfgPB(\vS_t),$$
where by $\cfgPB$ we mean pullback of the base point of $\vS_t$ considered as a map $\map{\vS_t}{\cl{S}}{T\cl{S}}$. 

The material velocity field $\vM_t \in \spVecPhi$ is an example of a two-point tensor field over the map $\map{\cfgT}{\cl{B}}{\cl{S}}$ (\cf Appendix \ref{appendix:bundles}). Another important example of a two-point tensor field is the tangent map of $\cfgT$ which is usually denoted by $F_t\in \Gamma(T^*\cl{B} \otimes \varphi^*T\cl{S})$ and called the deformation gradient. Thus, $F_t := T\cfgT$.
Note that both $\vM_t$ and $F_t$ are regarded as functions of $X\in \cl{B}$ and not $x\in \cl{S}$. Thus, at every $X\in\cl{B}$, we have that $\map{F_t(X)}{T_X\cl{B}\times T^*_{\cfgT(X)}\cl{S}}{\bb{R}}$ defines a ${\small\TwoTwoMat{0}{1}{1}{0}}$ two-point tensor while $\vM_t(X)$ defines a ${\small\TwoTwoMat{0}{1}{0}{0}}$ two-point tensor, both over the map $\cfgT$.

\subsection{Riemannian structure on $\cl{B}$}\label{sec:metric_structure_B}

While $\cl{B}$ represents the abstract (metric-free) assembly of material particles, its embedding in the ambient space $\mathscr{A}$ is what enables observation and measurement of physical properties and deformation using the metric (inner product) structure of $\mathscr{A}$ which allows quantifying lengths and angles.
The metric inherited by the associated configuration $\cl{S} = \cfgT(\cl{B})$ from the ambient space $\mathscr{A}$ and its corresponding Levi-Civita connection are denoted, respectively, by:
$$\map{\gS}{\spVecS\times \spVecS}{\spFn{\cl{S}}}, \qquad \map{\nabS}{\spVecS\times \Gamma(T_s^r {\cal S})}{\Gamma(T_s^r {\cal S})}.$$
We will refer to $\gS$ as the spatial metric and to $\nabS$ as the spatial connection.

Every configuration $\cfgT\in \mathscr{C}$ induces a Riemannian metric structure on $\cl{B}$ characterized by 
$$\map{\gC}{\spVecB\times \spVecB}{\spFn{\cl{B}}}, \qquad \map{\nabC}{\spVecB\times \Gamma(T_q^p \cl{B}) }{\Gamma(T_q^p \cl{B})},$$
where $\gC:= \cfgP(\gS)$ denotes the convective metric defined such that
\begin{equation}
	\gC(\hat{u}_1,\hat{u}_2) = \gS(\cfgPFwd \hat{u}_1,\cfgPFwd \hat{u}_2) \circ \cfgT, \qquad \qquad \forall \hat{u}_1,\hat{u}_2 \in \spVecB,
\end{equation}
while $\nabC$ is the associated Levi-Civita connection of $\gC$.
For the case of vector fields (i.e. $p=1,q=0$), $\nabC$ is given by:
\begin{equation}\label{eq:cov_der_def_conv}
	\nabC_{\hat{u}_1} \hat{u}_2 = \cfgP(\nabS_{(\cfgPFwd \hat{u}_1)} \cfgPFwd \hat{u}_2), \qquad \qquad \forall \hat{u}_1,\hat{u}_2 \in \spVecB.
\end{equation}
The extension of the definition (\ref{eq:cov_der_def_conv}) to more general tensor bundles is done in the usual manner using the Leibniz rule \cite[Sec. 6.3]{Schutz1980GeometricalPhysics}.
We denote by $\spMetB$ the set of all Riemannian metrics $\gC$ on $\cl{B}$ which plays an important role in finite-strain theory, as will be shown later.

\begin{remark}[\textbf{Constant metric on $\cl{B}$}]\label{remark:metric_free_B}

	Note that it is quite insightful technically to differentiate between the abstract body manifold with its intrinsic structure and its observations in the ambient space \cite{noll1974new}.
	In order to do so, one should refrain from identifying $\cl{B}$ with some reference configuration $\varphi_0(\cl{B})$ for a given choice of embedding $\map{\varphi_0}{\cl{B}}{\mathscr{A}}$.
	
	In many geometric treatments of nonlinear elasticity, one finds that the body manifold is equipped with a constant Riemannian structure, denoted by $G$ in \cite{Marsden1994MathematicalElasticity,Simo1984StressElasticity.,Simo1988ThePlates,Yavari2006OnElasticity}.
	This metric is in fact inherited from $\mathscr{A}$ which can be seen from
	$$G\equiv \gC_0:= \cfgPref(\gS),$$
	and thus is a non-intrinsic quantity that depends on the arbitrary choice of the reference configuration $\varphi_0$.

	This constant metric $G$ usually makes appearance in the material representation only and its existence in fact adds unnecessary ambiguity to the theory.
	For example, $G$ is sometimes used to create tensorial variants (i.e. pull indices up or down) of variables represented in the convective description, making it non-intrinsic.
	What usually causes more ambiguity is that usually it is assumed that $\varphi_0$ is some sort of identity map (between different spaces) and thus $\cl{B}$ is identically $\varphi_0(\cl{B})$ and consequently $G$ is the same as $\gS$, which makes no sense!
	
	As we will show throughout this paper, one can formulate the governing equations of nonlinear elasticity without requiring this extra structure on $\cl{B}$. 
	Using extensive variables in contrast to the more common intensive variables, we will show later that even the material representation can be described in an intrinsic manner.\\
\end{remark}

While $(\gS,\nabS)$ are used for spatial tensor fields and $(\gC,\nabC)$ are used for convective tensor fields, the analogous objects used for two-point tensors that appear in the material representation are 
$$\map{\gM}{\spVecPhi\times \spVecPhi}{\spFn{\cl{B}}}, \qquad \map{\nabM}{\spVecB\times \Gamma(T_q^p \cl{B}\otimes\varphi^* T_s^r \cl{S} )}{\Gamma(T_q^p \cl{B}\otimes\varphi^* T_s^r \cl{S})}.$$
The material metric $\gM$ is induced on $\cl{B}$ by a configuration $\cfgT\in \mathscr{C}$ and is defined by $\gM:= \cfgPB(\gS)$ such that
\begin{equation}
	\gM(\tilde{u}_1,\tilde{u}_2) = \gS(\tilde{u}_1\circ \cfgInvT,\tilde{u}_2\circ \cfgInvT) \circ \cfgT, \qquad \qquad \forall \tilde{u}_1,\tilde{u}_2 \in \spVecPhi.
\end{equation}
Furthermore, every $\cfgT$ induces on $\cl{B}$ the connection $\nabM$ which allows covariant differentiation of two point-tensors along true vector fields on $\cl{B}$.
For the case of a vector field over the map $\cfgT$ (i.e. $p=q=s=0, r=1$), $\nabM$ is constructed only using the spatial connection $\nabS$ by:
\begin{equation}\label{eq:cov_der_def_mat}
	\nabM_{\hat{u}} \tilde{w} := \cfgPB(\nabS_{(\cfgPFwd \hat{u})} (\tilde{w} \circ \cfgInvT)), \qquad \forall \hat{u}\in \spVecB, \tilde{w}\in \spVecPhi.
\end{equation}
On the other hand, the extension of the definition (\ref{eq:cov_der_def_mat}) to generic tensor bundles requires the convective connection $\nabC$ \cite{Grubic2014TheManifold}.
For instance, for the case $p=r=1, q=s=0$ we have that $\nabM_{\hat{u}} \tilde{P} \in \Gamma(T\cl{B}\otimes\varphi^* T\cl{S})$ is defined by
{
	\newcommand{\ahat}{{\hat{\alpha}}}
	\newcommand{\uhat}{{\hat{u}}}
	\newcommand{\btld}{{\tilde{\beta}}}
	
\begin{equation}\label{eq:cov_der_def_mat_tensor}
	\nabM_\uhat \tilde{P}(\ahat,\btld) := \uhat(\tilde{P}(\ahat,\btld)) - \tilde{P}(\nabC_\uhat \ahat, \btld) - \tilde{P}(\ahat,\nabM_\uhat \btld),
\end{equation}
for any $\uhat\in \spVecB, \ahat\in \spVec{^*\cl{B}}, \btld \in \Gamma(\varphi^*T^*\cl{S}).$
}
In elasticity, (\ref{eq:cov_der_def_mat}) is used for covariant differentiation of the material velocity field, while (\ref{eq:cov_der_def_mat_tensor}) is used to define the divergence of the first Piola-Kirchhoff stress tensor field.

\begin{remark}[\textbf{The material metric and connection}] \label{remark:material_metric}
	
	i) While $\gC_t \in \spMetB$ is a \enquote{true} time-dependent Riemannian metric on $\cl{B}$ with $\nabC$ being its associated Levi-Civita connection, neither $\gM$ is a Riemannian metric on $\cl{B}$ nor is $\nabM$ the Levi-Civita connection of $\gM$.
	
	ii) In the extension of $\nabM$ for high order tensor fields in (\ref{eq:cov_der_def_mat_tensor}), the convective connection $\nabC$ on $\cl{B}$ is necessary. In principle, one could either use the time-varying connection
	(\ref{eq:cov_der_def_conv}) associated to the current metric $\gC_t := \cfgP(\gS)$ \cite{Grubic2014TheManifold} or use the constant connection associated to the reference metric $G := \cfgPref(\gS)$ \cite{Yavari2006OnElasticity,Marsden1994MathematicalElasticity} (defined similar to (\ref{eq:cov_der_def_conv}) using $\varphi_0$ instead).
	While the former option allows a fully intrinsic description, it suffers from the mixing of the convective and material representation.
	Thus, the standard choice in the literature is to extend $\nabM$ using the reference metric $G$.
	
	iii) An important benefit of our formulation based on bundle-valued forms is that we will extend the definition of (\ref{eq:cov_der_def_mat}) to high order tensor fields in a different way compared to (\ref{eq:cov_der_def_mat_tensor}) which will not require the metric structure of $\cl{B}$.
	This point will be discussed further in Remarks \ref{remark:extd_coord} and \ref{remark:intrinsic_material}.\\
\end{remark}

Each of the metrics above induces the standard index-lowering ($\flat$ map) and index-raising ($\sharp$ map) actions by associating to each vector field a unique covector field, i.e. a section of the cotangent bundle, which we refer to as a one-form.
By linearity, these actions extend also to arbitrary tensor-fields.
The appearance of tensorial variants of physical variables occurs frequently in geometric mechanics in general. 
For instance, the one-forms\footnote{In the same manner $\vM$ is not a true vector field, $\vfM$ is not a true one-form.} associated with the spatial, convective and material velocity fields are defined respectively by
\begin{align*}
	\vfS(u) :=& \gS(\vS,u) \in \spFn{\cl{S}}, && \forall u \in \spVecS\\
	\vfC(\hat{u}) :=& \gC(\vC,\hat{u}) \in  \spFn{\cl{B}}, && \forall\hat{u} \in \spVecB\\
	\vfM(\tilde{u}) :=& \gM(\vM,\tilde{u}) \in \spFn{\cl{B}}, && \forall\tilde{u} \in \spVecPhi.
\end{align*}
With an abuse of notation, we shall denote the associated index lowering $(\flat)$ and index raising $(\sharp)$ maps to $\gS$, $\gC$ and $\gM$ by the same symbols as it will be clear from the context.
However, when we want to explicitly mention which metric is used we will use the notation $\vfS= \gS\cdot \vS, \  \vfC= \gC\cdot \vC, \ \vfM= \gM\cdot \vM$ and conversely $\vS= \gS^{-1}\cdot \vfS,\ \vC= \gC^{-1}\cdot \vfC, \ \vM= \gM^{-1}\cdot \vfM.$

\subsection{Connection-based operations}\label{sec:velgrad_accel}

The connection plays an important role in continuum mechanics and is used for defining a number of key physical quantities and operations. In particular, 1) the covariant differential, 2) the divergence operator , and 3) the material derivative, which will be introduced next.

\subsubsection{Covariant differential and divergence of tensor fields}
Let $P\in\spVec{^r_s\cl{S}},\hat{P}\in\spVec{^p_q\cl{B}},$ and $\tilde{P}\in\Gamma(T_q^p \cl{B}\otimes\varphi^* T_s^r \cl{S})$ be arbitrary tensor fields.
One important observation is that their covariant derivatives $\nabS_{u} P, \nabC_{\hat{u}} \hat{P}$ and $\nabM_{\hat{u}} \tilde{P}$ along any $u \in \spVecS$ and $\hat{u}\in \spVecB$ depend only on the values of $u$ and $\hat{u}$ 
\stefano{in the point where the operation is evaluated as a section and not in any point close by}
(which is in contrast to the Lie derivative operation for example).
Thus, the connections $\nabS,\nabC,\nabM$ can be interpreted as differential operators 
\begin{align}
	&\map{\nabS}{\spVec{^r_s\cl{S}}}{\spVec{_{s+1}^r\cl{S}}}\nonumber\\
	&\map{\nabC}{\spVec{^p_q\cl{B}}}{\spVec{_{q+1}^p\cl{B}}}\nonumber\\
	&\map{\nabM}{\Gamma(T_q^p \cl{B}\otimes\varphi^* T^r_s \cl{S})}{\Gamma(T^p_{q+1} \cl{B}\otimes\varphi^* T^r_s \cl{S})} .\label{eq:connection_operator_1}
\end{align}
In Sec. \ref{sec:ext_calc}, we will show how these connections will be extended to define differential operators for bundle-valued forms.

An important physical quantity that uses the construction above is the velocity gradient which is a 2-rank tensor field defined as the covariant differential applied to the velocity field. The spatial, convective and material representations of the velocity gradient are denoted by
$$\nabS\vS \in \spVec{^*\cl{S}\otimes T\cl{S}},\qquad  \nabC\vC \in \spVec{^*\cl{B}\otimes T\cl{B}}, \qquad \nabM\vM \in \Gamma(T^*\cl{B} \otimes \varphi^*T\cl{S})$$
Note that while $\nabS\vS$ and $\nabC\vC$ are (1,1) tensor-fields over $\cl{S}$ and $\cl{B}$, respectively, $\nabM\vM$ is a ${\small\TwoTwoMat{0}{1}{1}{0}}$ two-point tensor-field over $\varphi$.

The connections $\nabS$ and $\nabC$ are 
\stefano{by definition Levi-Civita connections}
compatible with the metrics $\gS$ and $\gC$ respectively such that $\nabS \gS = 0$ and $\nabC \gC = 0$ at all points. 
Similarly, the compatibility of $\nabM$ and $\gM$ is straightforward to check, as we will show later in Sec. \ref{sec:coord_kinematics}.
An important consequence of this compatibility is that the index raising and lowering operations commute with covariant differentiation \cite[Pg. 80]{Marsden1994MathematicalElasticity}.
For example, the covariant form of the velocity gradients above are equivalent to the covariant differential of their corresponding one-form velocity fields.
The spatial, convective and material covariant velocity gradients are given, respectively, by
\begin{equation}\label{eq:cov_vel_grad_relation}
	\begin{split}
		(\nabS\vS)^\flat &= \nabS\vfS \in \spVec{_2^0\cl{S}},\\
		(\nabC\vC)^\flat &= \nabC\vfC \in \spVec{_2^0\cl{B}},\\
		(\nabM\vM)^\flat &= \nabM\vfM \in \Gamma(T^*\cl{B}\otimes\varphi^*T^*\cl{S}).
	\end{split}
\end{equation}

These covariant velocity gradients will play an important role in subsequent developments.
One important property is that one can decompose $\nabS\vfS$ and $\nabC\vfC$ into symmetric and skew-symmetric parts
as shown in the following proposition.
\begin{proposition}\label{prop:covDiff_decomp}
	The covariant differential of $\vfS\in \spVecS$ and $\vfC\in \spVecB$ can be expressed as
	\begin{align}
		\nabS \vfS=& \sym(\nabS \vfS) + \asym(\nabS \vfS) = \half \Lie{\vS}{\gS} + \half \extd \vfS,\label{eq:vel_grad_identity_S_}\\
		\nabC \vfC=& \sym(\nabC \vfC) + \asym(\nabC \vfC) = \half \Lie{\vC}{\gC} + \half \extd \vfC,\label{eq:vel_grad_identity_C_}
	\end{align}
	with $\Lie{\vS}{\gS} \in \spSymTensS$ and $\Lie{\vC}{\gC}\in \spSymTensB$ being symmetric $(0,2)$ tensor fields over $\cl{S}$ and $\cl{B}$, respectively.
	Furthermore, the 2-forms $\extd \vfS \in \spFrmS{2} \subset \spVec{^0_2\cl{S}}$ and $\extd \vfC \in \spFrmB{2} \subset \spVec{^0_2\cl{B}}$ are considered as generic (0,2) tensor field in the equations above.
	
	Using Cartan's homotopy (magic) formula
	\begin{equation}\label{eq:Cartan_Lie_deriv}
		\Lie{u}{} = \extd \circ \iota_u +\iota_u \circ\extd,
	\end{equation}
	a corollary of the above identities is that 
	\begin{align}
		\nabS_{\vS} \vfS =& \Lie{{\vS}}{\vfS} - \half \extd\iota_{\vS}\vfS \label{eq:vel_grad_identity_S}\\
		\nabC_{\vC} \vfC =& \Lie{{\vC}}{\vfC} - \half \extd\iota_{\vC}\vfC.\label{eq:vel_grad_identity_C}
	\end{align}
\end{proposition}
\begin{proof}
	See Appendix \ref{propProof:covDiff_decomp}.
\end{proof}


\begin{remark}
	Note that identity (\ref{eq:vel_grad_identity_S_}) appears in \cite{Gilbert2023AMechanics} with a minus on the term $\half\extd \vfS$ instead of a plus. The reason in this discrepancy is due to the opposite convention used in defining $\nabS \vfS$. While we consider the $\nabS$ to be the first leg and $\vfS$ to be the second leg (\cf Table \ref{table:motion_kinematics}), the authors in \cite{Gilbert2023AMechanics} consider $\nabS$ to be the second leg and $\vfS$ to be the first leg.
	Furthermore, identity (\ref{eq:vel_grad_identity_S_}) appears also in \cite{Kanso2007OnMechanics} without the $\half$ factor which is clearly incorrect.
	
\end{remark}

The divergence of any spatial tensor field , for $r\geq 1, s\geq 0$, is constructed by contracting the last contravariant and covariant indices of $\nabS P$.
Similarly, the divergence of any convective tensor field $\hat{P}\in\spVec{^p_q\cl{B}}$,  for $p\geq 1, q\geq 0$, will be constructed from $\nabC \hat{P}$ while the divergence of any material tensor field $\tilde{P}\in\Gamma(T^p_q \cl{B}\otimes\varphi^* T^r_s \cl{S})$, for $p\geq 1, q,r,s\geq 0$, will be constructed from $\nabM\tilde{P}$.
We will denote the divergence of $P,\hat{P}$ and $\tilde{P}$ respectively by
$$\divrS(P)\in\spVec{_s^{r-1}\cl{S}}, \qquad \divrC(\hat{P})\in\spVec{_q^{p-1}\cl{B}}, \qquad \divrM(\tilde{P})\in\Gamma(T^{p-1}_q \cl{B}\otimes\varphi^* T^r_s \cl{S}).$$
Examples of such tensor fields that will appear in this paper are the divergence of the spatial and convective velocities $\divrS(\vS)\in \spFn{\cl{S}}$ and $\divrC(\vC)\in \spFn{\cl{B}}$, in addition to the divergence of the stress tensors.

\subsubsection{Material time derivative}
Another important quantity in continuum mechanics that is also defined using the connection 
is the material time derivative, denoted by $D_t$, which describes the rate of change of a certain physical quantity of a material element as it undergoes a motion along the curve $c_\varphi$. Thus, $D_t$ is used for describing the rate of change of two point tensor fields in the material representation.
One can geometrically define such derivative by pulling back the spatial connection along a curve similar to the standard formulation of the geodesic equation on a Riemannian manifold \cite{Kolev2021ObjectiveMetrics,bullo2019geometric}.
Two cases are of interest in our work, the material time derivative of the material velocity $\vM_t$ and the deformation gradient $F_t$.
For the reader's convenience, we include in Appendix \ref{append:mat_time_deriv} the construction for the case of a generic vector field over a curve.

Let $I\subset\bb{R}$ be a time interval.
For any fixed point $X\in \cl{B}$, the configuration map $\map{\cfgT}{\cl{B}}{\cl{S}}$ defines a curve $\map{\varphi_X}{I}{\cl{S}}$ in $\cl{S}$. Similarly, one can consider the material velocity to be a map $\map{\vM_X}{I}{T\cl{S}}$ such that $\vM_X(t) = \vS_t(\varphi_X(t)) \in T_{\varphi_X(t)}\cl{S}$. Thus, we have that $\vM_X \in \Gamma(\varphi_X^* T\cl{S})$ to be a vector field over the map $\varphi_X$.
Further, let $\map{\varphi_X'}{I}{T\cl{S}}$ denote the tangent curve of $\varphi_X$.
Then, the material time derivative $D_t \vM_X \in \Gamma(\varphi_X^* T\cl{S})$ is defined as (\cf (\ref{eq:mat_deriv_def}) in Appendix \ref{append:mat_time_deriv}):
$$D_t \vM_X(t) := (\nabS_{\varphi_X'(t)} v)(\varphi_X(t)) \in T_{\varphi_X(t)}\cl{S}.$$
By extension to all points in $\cl{B}$, one can define $D_t \vM_t\in \Gamma(\varphi_t^* T\cl{S})$.

A key quantity that is defined using the material derivative is the acceleration vector field associated with the motion $\cfgT$, denoted in the material representation by
$\tilde{a}_t := D_t \vM_t \in \Gamma(\varphi_t^* T\cl{S}).$
By defining the spatial and convective representations of the acceleration by $a_t := \tilde{a}_t \circ \cfgInvT \in \spVecS$ and $\hat{a}_t := \cfgP(a_t)\in \spVecB$, one has that\cite{Simo1988ThePlates}
$$a_t = \partial_t \vS_t + \nabS_{\vS_t} \vS_t, \qquad \qquad
\hat{a}_t = \partial_t \vC_t + \nabC_{\vC_t} \vC_t.$$

The extension of the material time derivative to the deformation gradient $F_t \in \Gamma(T^*\cl{B} \otimes \varphi_t^*T\cl{S})$ and higher order material tensor fields is more involved. The reader is referred to \cite{Fiala2020ObjectiveRevised} and \cite[Ch. 2.4, Box 4.2]{Marsden1994MathematicalElasticity}.
A key identity that will be used later is that the material time derivative of the deformation gradient $D_t F_t \in \Gamma(T^*\cl{B} \otimes \varphi_t^*T\cl{S})$ is equal to the material velocity gradient\cite{Fiala2020ObjectiveRevised}
\begin{equation}\label{eq:dt_F}
	D_t F = \nabM \vM \in \Gamma(T^*\cl{B} \otimes \varphi^*T\cl{S}).
\end{equation}

\subsection{Coordinate-based expressions}\label{sec:coord_kinematics}
While the motivation of this work is to formulate nonlinear elasticity using purely geometric coordinate-free constructions as much as possible, it is sometimes instructive to understand certain identities and perform certain calculations using coordinate-based expressions.
Furthermore, the coordinate-based expressions are essential for computational purposes. Nevertheless, caution should be taken as one might be misguided by a purely coordinate-based construction.
We believe both treatments are complementary and the maximum benefit is achieved by switching between them correctly.

The coordinate-based description of the motion is achieved by introducing coordinate functions $\map{\B{X}^I}{\cl{U}\subset\cl{B}}{\bb{R}}$ and $\map{\B{x}^i}{\cl{V}\subset\cl{S}}{\bb{R}}$ , for $i,I\in\{1,\cdots,3\}$, that assign to each physical point $X\in \cl{B}$ and $x\in\cl{S}$ the coordinates $(X^1,X^2,X^3)\in \bb{R}^3$ and $(x^1,x^2,x^3)\in \bb{R}^3$, respectively.
These coordinate systems induce the basis $\left\{ \parx{i} \right\}$ and $\left\{ \parX{I} \right\}$ for the tangent spaces $T_x\cl{S}$ and $T_X\cl{B}$, respectively, and the dual basis 
$\{\dx{i}\}$ and $\{\dX{I}\}$ for the cotangent spaces $T^*_x\cl{S}$ and $T^*_X\cl{B}$, respectively. In what follows we shall use Einstein's summation convention over repeated indices.

One in general needs not to use such coordinate-induced bases and one could refer to arbitrary bases.
In our work we shall opt for this generality.
In particular, the generic tensor fields $P\in\spVec{^1_1\cl{S}}$, $\hat{P}\in\spVec{^1_1\cl{B}}$, and $\tilde{P}\in\Gamma(T_1^1 \cl{B}\otimes\varphi^* T_1^1 \cl{S})$ are expressed locally at the points $x\in\cl{S}$ and $X\in \cl{B}$ as
\begin{align*}
	P|_x &= P^i_j(x)\ e_i|_x \otimes e^j|_x\\
	\hat{P}|_X &= \hat{P}^I_J(X)\ E_I|_X \otimes E^J|_X\\
	\tilde{P}|_X &=\tilde{P}^{I i}_{J j}(X)\ E_I|_X \otimes E^J|_X \otimes e_i|_{\varphi(X)} \otimes e^j|_{\varphi(X)},
\end{align*}
where $\{e_i|_x\}$ and $\{E_I|_X \}$ denote arbitrary bases for $T_x\cl{S}$ and $T_X\cl{B}$, respectively, while $\{e^i|_x\}$ and $\{E^I|_X \}$ denote their corresponding dual bases such that their pairing is the Kronecker delta symbol:
$e^j|_x(e_i|_x) = \delta_i^j$ and $E^J|_X(E_I|_X) = \delta_I^J .$
We denote by $P^i_j\in \spFn{\cl{S}}$ and $\hat{P}^I_J,\tilde{P}^{I i}_{J j} \in \spFn{\cl{B}}$ the component functions of the tensor fields in the arbitrary basis.

It is important to note the partial $\varphi$-dependence (thus time dependence) nature of the basis for material tensor fields in contrast to spatial and convective ones. This is a fundamental property and it implies that one needs to be cautious when defining time derivatives of material quantities (\cf Sec. \ref{sec:velgrad_accel}) and transforming between representations in coordinates.

A summary of the local expressions of the motion kinematics quantities introduced so far can be found in Table \ref{table:motion_kinematics}. For notational simplicity, we will omit the time and base point dependency when writing local expressions, unless needed.

\begin{table}
	\centering
	\small
	{\renewcommand{\arraystretch}{1.5}
	\begin{tabular}{p{2.2cm}|l|l|l}
		\hline
		&	\textbf{Convective} & \textbf{Material} & \textbf{Spatial}\\\hline
		Metric&	 $\gC = \gC_{IJ} E^I\otimes E^J$ & $\gM = \gM_{ij} e^i|_\varphi\otimes e^j|_\varphi$ & $\gS = \gS_{ij} e^i\otimes e^j$\\\hline
		Velocity & $\vC = \vC^I E_I$ & $\vM = \vM^i e_i|_\varphi$& $\vS = \vS^i e_i$\\\hline
		Velocity  & $\vfC = \vC_I E^I $ & $\vfM = \vM_i e^i|_\varphi$ & $\vfS = \vS_i e^i$\\
		one-form&$\vC_I:= \gC_{IJ}\vC^J$& $\vM_i:= \gM_{ij}\vM^j$ & $\vS_i:= \gS_{ij} \vS^j$\\\hline
		Velocity   & $\nabC\vC = \nabC_I\vC^J E^I\otimes E_J$ & $\nabM\vM = \nabM_I\vM^i E^I\otimes e_i|_\varphi$& $\nabS\vS = \nabS_i\vS^j e^i\otimes e_j$\\
		gradient& $\nabC_I\vC^J:= \frac{\partial\vC^J}{\partial X^I} + \hat{\Gamma}^J_{IK}\vC^K$ & $\nabM_I\vM^i:= \frac{\partial\vM^i}{\partial X^I} + F^j_I (\Gamma^i_{jk} \circ \varphi) \vM^k$ & $\nabS_i\vS^j:=\frac{\partial \vS^j}{\partial x^i} + \Gamma^j_{ik}\vS^k$ \\\hline
		Covariant   & $\nabC\vfC = \nabC_I\vC_J E^I\otimes E^J$ & $\nabM\vfM = \nabM_I\vM_i E^I\otimes e^i|_\varphi$& $\nabS\vfS = \nabS_i\vS_j e^i\otimes e^j$\\
		velocity gradient & $\nabC_I\vC_J:= \frac{\partial\vC_J}{\partial X^I} - \hat{\Gamma}^K_{IJ}\vC_K$ & $\nabM_I\vM_i:= \frac{\partial\vM_i}{\partial X^I} - F^j_I (\Gamma^k_{ij} \circ \varphi) \vM_k$ & $\nabS_i\vS_j:=\frac{\partial \vS_j}{\partial x^i} - \Gamma^k_{ij}\vS_k$ \\\hline
		Acceleration  & $\hat{a} = \hat{a}^J E_J$ & $\tilde{a} = \tilde{a}^j e_j|_\varphi$& $a = a^j e_j$ \\
		& $\hat{a}^J := \partial_t \vC^J + \vC^I\nabC_I\vC^J$ & $\tilde{a}^j := \partial_t \vM^j + (\Gamma^j_{ik} \circ \varphi) \vM^i\vM^k$ & $a^j := \partial_t \vS^j + \vS^i\nabS_i\vS^j$ \\
		\hline
		Velocity \mbox{divergence}  & $\divrC(\vC) = \nabC_I \vC^I$ && $\divrS(\vS) = \nabS_i \vS^i$ \\
		\hline
	\end{tabular}
	}
	\caption{Local coordinate-based expressions of motion kinematics quantities}
	\label{table:motion_kinematics}
\end{table}

The tangent map $\map{F := T\cfgT}{T\cl{B}}{T\cl{S}}$, which is commonly referred to as the deformation gradient and denoted by $F$, and its inverse $\Finvr$ play a key role in coordinate expressions of the pullback and pushforward operations.
In a local chart, $F$ and $\Finvr$ are given by the Jacobian matrix of partial derivatives of the components of $\cfgT$ and $\cfgInvT$, respectively, in  that chart:
\begin{align}
	F|_X &= F^i_I(X) \left.\parx{i}\right|_{\cfgT(X)} \otimes \dX{I}|_X,\\
	\Finvr|_x &= (\Finvr)^I_i(x) \left.\parX{I}\right|_{\cfgInvT(x)} \otimes \dx{i}|_x,
\end{align}
where $F^i_I(X) := \frac{\partial \cfgT^i}{\partial X^I}(X)$ and $ (\Finvr)^I_i(x) := \frac{\partial (\cfgInvT)^I}{\partial x^i}(x)$ whereas 
$$\map{\cfgT^i:= \B{x}^i\circ \cfgT}{\cl{B}}{\bb{R}}, \qquad \qquad \map{(\cfgInvT)^I:= \B{X}^I \circ \cfgInvT}{\cl{S}}{\bb{R}}.$$
The time derivative of $\cfgT^i$ is equal to the components of the material velocity field in the chart induced basis, i.e. $\vM^i(X) := \frac{\partial \cfgT^i}{\partial t}(X)$.
In a generic basis, the components of $F$, $\Finvr$ and $\vM$ are related to the ones defined above using the usual tensor transformation rules.


Using $F$ and $\Finvr$, we can now relate the convective, material and spatial representations as follows:
In local coordinates, the components of the three metrics are related by
$$\gC_{IJ} = F^i_I F^j_J \gM_{ij}, \qquad \qquad \gM_{ij} = \gS_{ij}\circ \cfgT.$$
The components of the velocities $\vS,\vC,$ and $\vM$ are related by
$$\vS^i = \vM^i \circ \cfgInvT, \qquad \qquad \vC^I = ((\Finvr)^I_i \circ \cfgT) \vM ^i = ((\Finvr)^I_i \circ \cfgT) (\vS^i \circ \cfgT).$$
While the components of the velocity one-forms  $\vfS,\vfC$ and $\vfM$ are related by
$$\vS_i = \vM_i \circ \cfgInvT, \qquad \qquad \vC_I = F^i_I \vM_i = F^i_I (\vS_i \circ \cfgT).$$
Furthermore, it is straightforward to assess in local components that 
\begin{equation}\label{eq:equality_metrics_B}
	\gM(\vM,\vM) = \gM_{ij}\vM^i\vM^j = \gM_{ij}F^i_I F^j_J \vC^I\vC^J = \gC_{IJ}\vC^I\vC^J = \gC(\vC,\vC).
\end{equation}


An essential ingredient for the local expressions (in a coordinate chart) of operations based on $\nabS$ and $\nabC$ are the Christoffel symbols $\Gamma^i_{jk}$ and $\hat{\Gamma}^I_{JK}$ associated with the spatial and convective metrics $\gS$ and $\gC$, respectively.
See for example, the application of $\nabS$ and $\nabC$ on $\vS,\vfS$ and $\vC,\vfC$  in Table \ref{table:motion_kinematics}, respectively.
On the other hand, caution is required when dealing with the material connection which in general involves the deformation gradient $F$ in addition to $\nabS$ and $\nabC$, as shown in (\ref{eq:cov_der_def_mat}-\ref{eq:cov_der_def_mat_tensor}).
For example, the local expressions of $\nabM \vM$ and $\nabM \vfM$ can be found in Table \ref{table:motion_kinematics}, wheres the rank-three material tensor field $\nabM \tilde{P} \in \Gamma(T^1_1\cl{B}\otimes\varphi^* T\cl{S})$, introduced before in (\ref{eq:cov_der_def_mat_tensor}), has local components
\begin{equation}\label{eq:coord_exp_cov_M_P}
	\nabM_J \tilde{P}^{Ij} =\frac{\partial \tilde{P}^{Ij}}{\partial X^J} + \tilde{P}^Kj  \hat{\Gamma}^I_{JK} + \tilde{P}^Ik (\Gamma^j_{ik} \circ \varphi) F^i_J.
\end{equation}
As mentioned before in Remark \ref{remark:material_metric}, it is very common to use time-independent Christoffel symbols $\hat{\Gamma}^I_{JK}$ derived from the reference metric $G:=\gC_0$ when treating material variables. In this way one can avoid mixing the convective and material representations.

The connections $\nabS$ and $\nabC$ are naturally compatible with the metrics $\gS$ and $\gC$ respectively such that at any point one has that $\nabS_k \gS_{ij} = 0$ and $\nabC_K \gC_{IJ} = 0$. Similarly, the compatibility of $\nabM$ and $\gM$ is straightforward to check since $\nabM_K \gM_{ij} = F^k_K (\nabS_k g_{ij} \circ \varphi) = 0.$
An important consequence of this compatibility is that the index raising and lowering operations commute with covariant differentiation \cite[Pg. 80]{Marsden1994MathematicalElasticity}. Therefore, we have that
$$\nabS_j\vS_i = \nabS_j(\gS_{ik}\vS^k) = \vS^k\nabS_j(\gS_{ik}) + \gS_{ik}\nabS_j\vS^k = \gS_{ik}\nabS_j\vS^k,$$
and similarly for $\nabC_J\vC_I$ and $\nabM_J\vM_i$.

\section{Intrinsic deformation kinematics}\label{sec:deformation}

%
%
%
%

Now we turn attention to the kinematics of deformation and its geometric formulation. We highlight in this section the important role of the space of Riemannian metrics $\spMetB$ and how it intrinsically represents the space of deformations of the body. This allows us to define the geometric representations of strain and rate-of-strain.
The principle bundle structure relating the configuration space $\spC$ to $\spMetB$ will be discussed later in Sec. \ref{sec:structures}.

\subsection{Space of deformations}
Analogously to a classical spring in $\bb{R}^3$, the strain of an elastic body is roughly speaking the difference between any two states of deformation.
In light of Remark \ref{remark:metric_free_B}, the measurement of distances, and thus geometric deformation, is achieved using the metric $\gS$ inherited by $\cl{S}$ from the ambient space $\mathscr{A}$.
At every point $x\in\cl{S}$, the value of $\gS$ at $x$ determines an inner product of any two vectors attached to that point and thus establishes a geometry in its vicinity.
The only intrinsic way to do the same directly on $\cl{B}$ is by the pullback of $\gS$ by $\varphi$ which gives rise to a $\varphi$-dependent mechanism for measuring lengths and angles of material segments \cite{Rougee2006AnStrain}.
Therefore, the Riemannian metric $\gC:= \varphi^*(\gS) \in \spMetB$ serves as an intrinsic state of deformation while the space of Riemannian metrics $\spMetB$ is the corresponding space of deformations.

The state space $\spMetB$ has been extensively studied in the literature due to its importance in the geometric formulation of elasticity. The interested reader is referred to \cite{Rougee2006AnStrain,Fiala2011GeometricalMechanics,Fiala2016GeometryAnalysis,Kolev2021AnConstraints,Kolev2021ObjectiveMetrics}.
This space has been shown to have an infinite dimensional manifold structure and is an open convex set in the infinite dimensional vector space $\spSymTensB$ of symmetric $(0,2)$ tensor fields over $\cl{B}$.
Furthermore, it has been shown that $\spMetB$ is itself a Riemannian manifold with constant negative curvature~\cite{Fiala2011GeometricalMechanics}.

\begin{figure}
	\centering
	\includegraphics[width=0.9 \textwidth]{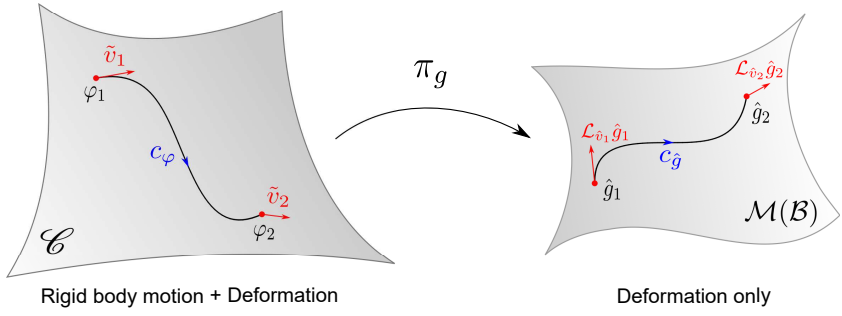}
	\caption{Illustration of the motion of the elastic body as both a curve on $\mathscr{C}$ and a curve on $\spMetB$ using the map $\pi_g$.}
	\label{fig:curve_spMetB}
\end{figure}

Consider the map 
\begin{equation*}
	\fullmap{\pi_g}{\mathscr{C}}{\spMetB}{\varphi_t}{\cfgT^*(\gS) =: \gC_t,}
\end{equation*}
that associates to any configuration $\cfgT$ a Riemannian metric on $\cl{B}$.
With reference to Fig. \ref{fig:curve_spMetB}, a curve $c_{\varphi}:t\mapsto \cfgT$ in the configuration space $\mathscr{C}$ (which represents a motion of the elastic body) induces the curve $c_{\gC}:t\mapsto \gC_t = \pi_g(\cfgT)$ in the space of metrics $\spMetB$.
The tangent vector to the curve $c_{\gC}$ at any point $\gC_t\in \spMetB $ can be calculated using the tangent map of $\pi_g$ or equivalently using properties of the Lie derivative as
\begin{equation}\label{eq:rate_of_strain_Lie}
	T_{\gC_t}\spMetB \ni\partial_t \gC_t
	= \partial_t(\cfgP\gS) = \cfgP(\Lie{\vS_t}{\gS}) = \Lie{(\cfgP\vS_t)}{\cfgP\gS} = \Lie{\vC_t}{\gC_t}.
\end{equation}
Thus, the tangent space $T_{\gC_t}\spMetB$ is canonically identified with the vector space $\spSymTensB$. Furthermore, one can show that the tangent bundle $T\spMetB$ is in fact trivial \cite{Kolev2021ObjectiveMetrics}, i.e. it is equivalent to the product space $T\spMetB = \spMetB\times \spSymTensB$.
This is in contrast to the tangent bundle $T\mathscr{C}$ of the configuration space which is not trivial \cite{Simo1988ThePlates}.
In Sec. \ref{sec:structures}, we will show how the map $\pi_g$ factors out rigid body motions from $c_{\varphi}$, such that the curve $c_{\gC}$ represents only deformation of the body which leads to the principle bundle structure relating $\spC$ to $\spMetB$.

\begin{remark}\label{remark:tangent_MB}
	Note that at any point $\gC$ in the space of Riemannian metrics $\spMetB$, one can arbitrarily change the tensor type of the tangent vector $\partial_t \gC$. Thus, in principle one can also identify $T_{\gC}\spMetB$ with the vector spaces $\Gamma(ST_0^2\cl{B})$ or $\Gamma(T_1^1\cl{B})$. We shall later use this arbitrariness such that we identify $T_{\gC}\spMetB$ with vector-valued forms and consequently $T^*_{\gC}\spMetB$ with covector-valued forms.
\end{remark}

\subsection{Logarithmic strain measure}

With the above construction, the strain can be now defined as the relative deformation between any states $\gC_1,\gC_2\in \spMetB$. However, the space of deformations $\spMetB$ does not have a vector space structure. For instance, the positive-definiteness property of a metric in $\spMetB$ is not closed under subtraction \cite{Fiala2011GeometricalMechanics}.
Therefore, one cannot simply define an arbitrary finite strain as the subtraction of $\gC_1$ and $\gC_2$, while for example the classical Green St. Venant or Euler-Almansi strain tensor fields are only valid for (infinitesimally) small strains \cite{Fiala2011GeometricalMechanics}.

The correct geometric definition of strain is the \enquote{shortest motion} between $\gC_1$ and $\gC_2$ on $\spMetB$, i.e. the geodesic connecting these two points.
It has been shown in \cite{Rougee2006AnStrain,Fiala2016GeometryAnalysis,Kolev2021ObjectiveMetrics} that this construction leads to the logarithmic strain measure defined by:
$$\hat{\delta}(\gC_1,\gC_2) := \half \mathrm{Log}_{\gC_1}(\gC_2),$$
where $\mathrm{Log}_{\gC_1}$ denotes the inverse of the Riemannian exponential map $\map{\mathrm{Exp}_{\gC_1}}{T_{\gC_1}\spMetB}{\spMetB}$ corresponding to the geodesic flow starting at the point $\gC_1\in\spMetB$.

Now if we turn attention back to the elastic body's motion described by the curves $\map{c_\varphi}{t}{\cfgT}$ and $\map{c_{\gC}}{t}{\gC_t}$, one can define at any $t$ the convective strain tensor field to be
$$\hat{\delta}_t := \half \mathrm{Log}_{\gC_0}(\gC_t),$$
as the relative deformation between the current state $\gC_t$ and a reference state $\gC_0$.
However, it is important to note that on the manifold $\spMetB$ there is no privileged deformation state $\gC_0$ that would allow us to define the strain $\hat{\delta}_t$ in an intrinsic way.
One common choice is to select $\gC_0$ to be the initial value of the convective metric at $t=0$.

In hyper-elasticity, one usually proceeds by defining the strain energy functional using $\hat{\delta}_t$ and then define stress as some \enquote{gradient} of this functional with respect to $\hat{\delta}_t$ (i.e. the convective counterpart of the Doyle-Erickson formula).
However, an energy functional defined using $\hat{\delta}_t$ or $\gC_t$ would only differ by a constant offset that corresponds to the strain energy of the reference state $\gC_0$. The stress on the other hand, is identical in both cases.
Thus, when defining constitutive relations of the stress, as we shall show later, it suffices to use the state of deformation $\gC_t$.

\subsection{Rate-of-strain}

We finally conclude by presenting the rate-of-strain (1,1) tensor fields which in the convective and spatial representations are denote, respectively, by
$$\epC_t\in \spVec{^1_1\cl{B}},\qquad\qquad\epS_t\in \spVec{^1_1\cl{S}}.$$
Their corresponding 2-covariant variants are given by
\begin{equation*}
	\epfC_t := \gC_t \cdot \epC_t = \half \Lie{\vC_t}{\gC_t} \in \spSymTensB,\qquad\qquad
	\epfS_t := \gS \cdot \epS_t = \half \Lie{\vS_t}{\gS} \in \spSymTensS,
\end{equation*}
which are symmetric 2-rank tensor fields.
From (\ref{eq:vel_grad_identity_S_}-\ref{eq:vel_grad_identity_C_}), one can see that the rate-of-strain variables $\epfC_t$ and $\epfS_t$ are also equal to the symmetric component of the covariant velocity gradients $\nabC\vfC_t$ and $\nabS\vfS_t$, respectively.
Furthermore, in line of Remark \ref{remark:tangent_MB}, we can consider both $\epfC_t$ and $\epC_t$ to be technically tangent vectors at the point $\gC_t$ in $\spMetB$.

An interesting distinction between $\epfC_t$ and $\epfS_t$ is that only the convective variable is a time-derivative of some deformation state-variable:
\begin{equation}\label{eq:gC_strain_rate}
	\partial_t \gC_t  = 2 \epfC_t = 2\,\sym(\nabC\vfC_t),
\end{equation}
which is not the case for its spatial counterpart.
In the material representation, things get more interesting since symmetry of the velocity gradient $\nabM\vfM_t$ cannot be defined in the first place due to its geometric nature of being a two-legged tensor.
Consequently, one does not have a material rate-of-strain tensor field.

In summary, we conclude that only in the convective representation, one has a proper geometric state of deformation, which along with its rate-of-change, encodes the necessary information for an intrinsic description of deformation.
We shall come back to this point later in Sec. \ref{subsec:const_eqns} when we discuss the constitutive equations that relate stress to the deformation kinematics.



\section{Mass and volume properties} \label{sec:mass_volume}

Now we turn attention to the mass structure that is associated to the abstract manifold $\cl{B}$. This structure provides the link between kinematics and kinetics quantities of an elastic body and is fundamental to the subsequent intrinsic formulation of nonlinear elasticity we present in this paper. Therefore, we shall discuss it in detail in this section.

\subsection{Mass, volume and mass density}

Following \cite{Kolev2021AnConstraints}, we define the associated mass measure to $\cl{B}$ by a top form $\mFC\in\spFrmB{3}$ that we refer to as the body (convective) mass form.
This top-form assigns to any $\cl{U}\subseteq \cl{B}$ a non-negative scalar $\mathrm{m}(\cl{U})$ that quantifies the physical mass of $\cl{U}$ and is defined by
$$\mathrm{m}(\cl{U}) := \int_\cl{U} \mFC \in \bb{R}^+.$$
For every embedding $\cfgT$, there is an induced time-dependent mass form on $\cl{S}$ defined by
\begin{equation*}
	\mFS_t := \cfgPFwd (\mFC) \in \spFrmS{3},
\end{equation*}
such that, using the change of variables theorem, we have that\\
\begin{equation}\label{eq:mass_body}
\int_{\cfgT(\cl{U})} \mFS_t = \int_\cl{U} \mFC = \mathrm{m}(\cl{U}).
\end{equation}
We refer to $\mFS_t$ as the spatial mass form.

It is important to note that the aforementioned two mass-forms are all that one requires to formulate the governing equations of nonlinear elasticity in an intrinsic way as we shall show later.
The mass of $\cl{B}$ is a fundamental physical property that is independent of its configuration in the ambient space $\mathscr{A}$. Thus, it is assigned to $\cl{B}$ \textit{a priori} as an extra structure and it doesn't inherit it from $\mathscr{A}$. On the other hand, to represent the mass forms $\mFC$ and $\mFS_t$ as scalar-densities, one then needs to introduce a volume measure which will allow us to define these scalar mass densities as the ratio of mass to volume. Such volume measure is not intrinsic to $\cl{B}$ and thus must be inherited from $\mathscr{A}$ via an embedding.

Given its inherited metric structure, any configuration $\cl{S}$ of the body has a volume form $\vFS \in \spFrmS{3}$ that is induced by the Riemannian metric $g$.
Since all top forms on a manifold are proportional to each other, the time-dependent spatial mass density function $\mDS_t\in\spFn{\cl{S}}$ is implicitly defined such that
\begin{equation}
	\mFS_t = \mDS_t \vFS.
\end{equation}
Similarly, the induced metric $\gC_t = \cfgP(\gS)$ induces on $\cl{B}$ a volume form $\vFCt = \cfgP(\vFS) \in \spFrmB{3}$ that allows one to define the time-dependent convective mass density function $\mDC_t \in \spFn{\cl{B}}$ such that
\begin{equation}\label{eq:mass_form_conv}
	\mFC = \mDC_t \vFCt.
\end{equation}
The spatial and convective mass densities are related by
$\mDC_t = \mDS_t \circ \cfgT,$
which follows from comparing (\ref{eq:mass_form_conv}) to
\begin{equation}\label{eq:mass_density_relations}
	\mFC = \cfgP(\mFS) = \cfgP(\mDS_t \vFS) = (\mDS_t \circ \cfgT)\cfgP(\vFS) = (\mDS_t \circ \cfgT) \vFCt.
\end{equation}

Now the interesting question is does one have another intrinsic representation of the above mass top-forms and mass density functions that can be used for the material representation of the motion? The answer is no !
If one needs to define integral quantities (e.g. kinetic and strain energies) using material variables, then the mass top form $\mFC$ suffices since it allows integration on $\cl{B}$. For notational convenience, we shall denote the body mass form $\mFC$ by $\mFM$ when used for the material representation, i.e. $\mFM \equiv \mFC$.
On the other hand, since $\cl{B}$ does not have an intrinsic volume form, a material mass density function cannot be defined !
What can be done in principle is to also use $\mDC_t$ for the material representation. However, as mentioned before in Remark \ref{remark:material_metric}, such mixing of material and convective representations is usually avoided.
What is common in the literature is that one uses a reference configuration $\varphi_0(\cl{B})$ that induces on $\cl{B}$ a reference metric $\gMref:= \cfgPref(\gS)$ which in turn induces the volume form $\vFM\in \spFrmB{3}$.
Using this extra structure, one can define a (time-independent) material mass density $\mDM \in \spFn{\cl{B}}$ such that
\begin{equation}\label{eq:mass_form_mat}
	\mFM = \mDM \vFM.
\end{equation}
Using the Jacobian of $\cfgT$, denoted by $J_{\cfgT}\in \spFn{\cl{B}}$ and defined such that
\begin{equation}\label{eq:Jacobian_def}
	\vFCt = \cfgP(\vFS) = J_{\cfgT} \vFM, \qquad \implies \qquad J_{\cfgT} = \det(F_t) \frac{(\det (\gS) \circ \cfgT)}{\det(\gMref)},
\end{equation}
one has the standard relations between the material mass density and the other representations:
$\mDM = J_{\cfgT} \mDC_t = J_{\cfgT}(\mDS_t \circ \cfgT),$
which follows from substituting (\ref{eq:Jacobian_def}) in (\ref{eq:mass_density_relations}) and
comparing to (\ref{eq:mass_form_mat}). 
Table provides a summary of the mass and volume quantities introduced in this section.

\begin{table}
	\centering
	\begin{tabular}{p{3cm}|l|l|l}
		\hline
		&	\textbf{Convective} & \textbf{Material} & \textbf{Spatial}\\\hline
		Mass form [M] & $\mFC$ &  $\mFM$ &  $\mFS_t$ \\
		Volume form [L$^3$] & $\vFCt$ &  $\vFM$ &  $\vFS$\\
		Mass density [M L$^{-3}$] & $\mDC_t$ &  $\mDM$ &  $\mDS_t$ \\\hline
	\end{tabular}
	\caption{The different representations of the mass and volume quantities. Time dependency is denoted by a $t$-subscript and the physical units of mass and length are denoted by [M] and [L], respectively.}
	\label{table:mass_vol_quantities}
\end{table}

\begin{remark}[\textbf{Orientation and pseudo-forms}]\label{remark:orientation}
	
	In order for the description of the mass of the body in (\ref{eq:mass_body}) to be physically acceptable, one requires that the integration of $\mFC$ and $\mFS_t$ over their respective domains to be invariant with respect to a change of orientation (e.g. using a right-hand rule instead of a left-hand rule). This imposes that $\mFC$ and $\mFS_t$ to change sign when the orientation is reversed such that the integral always leads to a positive value of mass.
	
	This leads to two classes of differential forms: those that change sign with a reverse of orientation, and those that do not.
	We refer to the former as \textit{pseudo-forms} and the latter as \textit{true-forms} following \cite{Frankel2019ThePhysics}. Other terminology in the literature include \textit{outer-oriented} and \textit{inner-oriented} forms \cite{Gerritsma2014Structure-preservingModels,palha2014physics} and \textit{twisted} and \textit{straight} forms \cite{bauer2016new}.
	The mass forms $\mFC$ and $\mFS_t$ are then imposed to be \textit{pseudo-forms}. The same also holds for the volume forms introduced above.
	
	This distinction of the orientation-nature of physical quantities is a topic that is usually neglected. However, for structure-preserving discretization, the exploitation of this distinction is currently an active area of research. The interested reader is referred to our recent work \cite{Brugnoli2022DualCalculus,Brugnoli2023FiniteSystems}. \\
\end{remark}

\subsection{Conservation of mass and volume}

The conservation of mass in the three representations of motion are summarized in the following result.

\begin{proposition}\label{prop:cons_mass}
	Let the curve $c_\varphi(t) \in \mathscr{C}$ denote a motion of the elastic body. Conservation of mass requires that along $c_\varphi(t)$ we have that
	\begin{align*}
		&\partial_t \mFM = 0, \qquad \qquad &&\partial_t \mDM = 0 &&&\mathrm{(material)}\\
		&\partial_t \mFC = 0, \qquad \qquad &&\partial_t \mDC_t + \mDC_t \divrC(\vC_t)  = 0 &&&\mathrm{(convective)}\\
		&\partial_t \mFS_t + \Lie{{\vS_t}}{\mFS_t} = 0, \qquad \qquad &&\partial_t \mDS_t + \Lie{{\vS_t}}{\mDS_t} + \mDS_t \divrS(\vS_t) = 0 &&&\mathrm{(spatial)}
	\end{align*}
	where $\divrC(\vC_t)\in \spFn{\cl{B}}$ denotes the divergence of the convective velocity field and $\divrS(\vS_t)\in \spFn{\cl{S}}$ denotes the divergence of the spatial velocity field. 
\end{proposition}
\begin{proof}
	See \cite{Abraham1988ManifoldsApplications,Marsden1994MathematicalElasticity}.
	\qed
\end{proof}

The conservation of mass expression in terms of $\mFS_t$ states that the spatial mass form $\mFS_t$ is an advected quantity of the motion while its corresponding mass density $\mDS_t$ is not. Furthermore, one can see that the evolution equation of the convective density $\mDC_t$ depends on $\vC_t$ which explains why it is not appealing to use $\mDC_t$ in the material representation, as mentioned earlier.

In incompressible elasticity, one has the additional constraint that along $c_\varphi(t)$ the convective volume form is constant and equal to its value\footnote{ which is usually chosen as the reference configuration} at $t=0$. 
Consequently, 
$$0 = \partial_t\vFCt = \partial_t (J_{\cfgT} \vFM) =  \partial_t\cfgP(\vFS) = \cfgP (\Lie{{\vS_t}}{\vFS}) = \Lie{{\vC_t}}{\vFCt}.$$
Since $\Lie{{\vC_t}}{\vFCt}= \divrC(\vC_t) \vFCt$ and $\Lie{{\vS_t}}{\vFS} = \divrS(\vS_t) \vFS$, the incompressibility condition in the convective, material and spatial representations, respectively, is expressed as
$$\divrC(\vC_t)  = 0, \qquad \partial_t J_{\cfgT} = 0, \qquad \divrS(\vS_t) =0.$$
As a consequence, from Prop. \ref{prop:cons_mass} one sees that in incompressible elasticity both mass forms, volume forms and mass densities of the convective and material representation become constant. Furthermore, they coincide with each other if the reference configuration is chosen as the initial configuration.
On the other hand, in the spatial representation, the mass form, volume-form and mass density become advected quantities of the motion.

\subsection{Extensive and intensive physical quantities}
From a thermodynamical perspective, the properties of any physical system can be classified into two classes: \textit{extensive} and \textit{intensive} properties. Intensive properties are those quantities that do not depend on the amount of material in the system or its size. Examples of intensive quantities for an elastic body include velocity, velocity gradient, acceleration and mass density.
In contrast, the value of extensive properties depends on the size of the system they describe, such as the mass and volume of the elastic body.
The ratio between two extensive properties generally results in an intensive value (e.g., mass density is the ratio of mass and volume).

The distinction between intensive and extensive physical quantities is fundamental in our geometric formulation of nonlinear elasticity.
The kinematic quantities introduced in Sec. \ref{sec:kinematics} and \ref{sec:deformation} are all of intensive nature.
The mass properties of the body relates these kinematics quantities to the kinetics ones, such as momentum and stress which will be introduced later.
However, here one can choose whether to represent the mass structure of the body using the extensive mass top-forms or the intensive mass densities. In the first case, the resulting momentum and stress representations are extensive, while in the second, they are intensive. The common Cauchy, first and second Piola-Kirchhoff stress tensors are all examples of intensive stress representations.

Based on the intrinsicality and technical advantage of mass top-forms compared to mass densities, we shall opt in our work to represent kinetics in terms of extensive quantities. We will demonstrate how this choice will yield a completely intrinsic formulation of the governing equations with many technical advantages compared to the more common descriptions. Next, we discuss the exterior calculus tools needed for this formulation.

\section{Exterior calculus formulation}\label{sec:ext_calc}

There have been several attempts in the literature to formulate nonlinear elasticity and continuum mechanics in general in a geometrically consistent way.
The approach we opt for in this work is to use exterior calculus for representing the governing equations of nonlinear elasticity by formulating the corresponding physical variables as differential forms \cite{Kanso2007OnMechanics,Gilbert2023AMechanics}.
Compared to other approaches that rely on tensor fields \cite{Yavari2006OnElasticity}, tensor field densities \cite{Grubic2014TheManifold}, or tensor distributions \cite{Kolev2021AnConstraints}, the use of differential forms highlights the geometric nature of the physical variables associated to their intrinsic integral quantities over the elastic body's domain $\cl{B}$ and its boundary $\bndB$.
Furthermore, this natural geometric structure has proven fundamental for deriving efficient and stable discretization schemes of nonlinear elasticity by preserving this structure at the discrete level. The interested reader is referred to \cite{Yavari2008OnElasticity} for numerical schemes based on discrete exterior calculus and to \cite{FaghihShojaei2018Compatible-strainElasticity,FaghihShojaei2019Compatible-strainElasticity} for schemes based on finite-element exterior calculus.

To formulate nonlinear elasticity using exterior calculus bundle-valued differential forms are required. They are a generalization of the more common scalar-valued differential forms.
While scalar-valued forms are only applicable for theories that use anti-symmetric tensor-fields (such as electromagnetism), bundle-valued forms will allow us to incorporate symmetric tensor fields, used for strain and stress variables, as well as two-point tensor fields, used for the material representation of the variables.
First, we start with a generic introduction to bundle-valued differential forms and then we show how this applies to the nonlinear elasticity problem.
We refer the reader to Appendix \ref{appendix:bundles} for an introduction to fibre bundles and notations used.

\subsection{Bundle-valued differential forms}
Let $M$ be a smooth manifold of dimension $n$ and $\bb{E}\rightarrow M$ be a smooth vector bundle over $M$. 
Recall that a scalar-valued differential $k$-form on $M$ is an element of $\spFrm{M}{k} := \Gamma(\Lambda^k T^*M)$, for $k\in \{0,\cdots,n\}$ (\cf Appendix \ref{appendix:bundles}).
A $\bb{E}$-valued differential $k$-form on $M$ is a multilinear map that associates to each $p\in M$ an element of 
$\Lambda^k T_p^*M \otimes \bb{E}_p$, 
i.e. a $k$-form with values in $\bb{E}_p$.
We will denote the space of $\bb{E}$-valued differential $k$-forms by 
$$\spFrmM{k}{\bb{E}}:= \Gamma(\Lambda^k T^*M \otimes_{\textrm{id}} \bb{E}).$$
For the case $k=0$, an $\bb{E}$-valued 0-form is simply a section of the bundle $\bb{E}$, i.e. $\spFrmM{0}{\bb{E}} = \Gamma(\bb{E})$. For the cases $\bb{E}=TM$ and  $\bb{E}=T^*M$, we shall refer to $\spFrmM{k}{TM}$ and $\spFrmM{k}{T^*M}$ as \textit{vector-valued forms} and \textit{covector-valued forms}, respectively.

Let $N$ be another smooth manifold and $\bb{F}\rightarrow N$ be a smooth vector bundle over $N$. 
An $\bb{F}$-valued differential $k$-form over the map $\map{f}{M}{N}$ is a multilinear map that associates to each $p\in M$ an element of $\Lambda^k T_p^*M \otimes \bb{F}_{f(p)}$. We will denote the space of $\bb{F}$-valued differential $k$-forms over $f$ by 
$$\spFrmMf{k}{\bb{F}}:= \Gamma(\Lambda^k T^*M \otimes_{f}\bb{F}).$$
For the case $k=0$, $\spFrmMf{0}{\bb{F}} = \Gamma(f^*\bb{F})$.
For the case $\bb{F}=TN$, elements of $\spFrmMf{k}{TN}$ are ${\small\TwoTwoMat{0}{1}{k}{0}}$ two-point tensor fields over $f$ that we shall refer to as \textit{vector-valued forms over $f$}. Similarly, elements of $\spFrmMf{k}{T^*N}$ are ${\small\TwoTwoMat{0}{0}{k}{1}}$ two-point tensor fields over $f$ that we shall refer to as \textit{covector-valued forms over $f$}. 

A generic vector-valued $k$-form $\zeta \in \spFrmM{k}{TM}$ and a generic covector-valued $k$-form $\cl{X}\in\spFrmM{k}{T^*M}$ are expressed locally as
$$\zeta = \zeta^i \otimes e_i, \qquad \qquad \cl{X} = \cl{X}_i \otimes e^i,$$
where each $\zeta^i,\cl{X}_i \in \spFrm{M}{k}$, for $i\in \{1,\cdots,n\}$, is an ordinary $k$-form on $M$, 
while $e_i|_p$ denotes an arbitrary basis for ${T_pM}$ and $e^i|_p$ denotes its corresponding dual basis. \stefano{This shows also that the combination is correspondent to a tensor product giving rise to elements of the dimension $n$ times the dimension for the definition of each $\zeta^i$ and $\cl{X}^j$ which is also $n$}.
We shall notationally distinguish between vector-valued and covector-valued forms in our work by denoting the latter using upper-case symbols.
A trivial vector-(or covector-) valued $k$-form is one which is equivalent to the tensor product of a vector-field (or covector-field) and an ordinary $k$-form.
For example, we say that $\zeta \in \spFrmM{k}{TM}$ is trivial if it is composed of a  vector-field $u \in \Gamma(TM)$ and a $k$-form $\alpha \in \spFrm{M}{k}$ such that 
$$\zeta = \alpha\otimes u = \left( \frac{1}{k!}\alpha_{i_1\cdots i_k} e^{i_1}\wedge \cdots \wedge e^{i_k}\right) \otimes u^j e_j.$$

The usual operations such as raising and lowering indices with the $\sharp$ and $\flat$ maps as well as the pullback operators can be applied to either leg of a bundle-valued form.
Instead of using a numerical subscript to indicate which leg the operation is applied to (as in \cite{Kanso2007OnMechanics}), we shall use an f\textit{-subscript} to indicate the form-leg and a v\textit{-subscript} to indicate the value-leg.
For example, for $\zeta = \alpha\otimes u \in \spFrmM{k}{TM}$, we have that $\flatV(\zeta) = \alpha\otimes u^\flat \in \spFrmM{k}{T^*M}$ and $i^*_\mathrm{f}(\zeta) = i^*(\alpha)\otimes u\in \spFrm{\partial M;TM}{k}$, where $\map{i}{\partial M}{M}$ denotes the inclusion map.

\subsubsection*{Wedge-dot and duality product}

The \textit{wedge-dot} product 
$$\map{\wedgedot}{\spFrmM{k}{TM}\times\spFrmM{l}{T^*M}}{\spFrm{M}{k+l}}$$ 
between a vector-valued and a covector-valued form is, by definition, a standard duality product of the covector and vector parts from the value legs and a standard wedge product between the form legs. For example, if one considers the trivial forms $\zeta = \alpha\otimes u \in \spFrmM{k}{TM}$ and $\cl{X} = \beta \otimes \gamma \in \spFrmM{l}{T^*M}$ then 
$\zeta\wedgedot \cl{X} := \gamma(u) \alpha\wedge\beta \in \spFrm{M}{k+l}.$
It is useful to note that the construction of $\wedgedot$ does not use any metric structure and is thus a topological operator.
An important distinction between the $\wedgedot$ product and the standard wedge product of scalar-valued forms is that the pullback map does not distribute over $\wedgedot$ \cite{califano2022energetic}.
However, the pullback does distribute over the form-legs as usual. For example, if we consider the inclusion map $\map{i}{\partial M}{M}$, then we have that
\begin{equation}\label{eq:wedgedot_dist_property}
	i^*(\zeta\wedgedot \cl{X}) = \ptr(\zeta)\wedgedot\ptr(\cl{X}) \in \spFrm{\partial M}{k+l}.
\end{equation}

The duality pairing between a vector-valued $k$-form $\zeta\in \spFrmM{k}{TM}$ and a covector-valued $n-k$-form $\cl{X} \in \spFrmM{n-k}{T^*M}$ is then defined as
\begin{equation}\label{eq:duality_pairing_def}
	\duPair{\cl{X}}{\zeta}{M} := \int_M \zeta\wedgedot\cl{X} \in \bb{R}.
\end{equation}

In the same line of Remark \ref{remark:orientation}, if the duality pairing (\ref{eq:duality_pairing_def}) represents a physical integral quantity that should always be positive, then the $n$-form $\zeta\wedgedot\cl{X}$ should be a pseudo-form. This requires that either $\zeta$ is a vector-valued pseudo-form and $\cl{X}$ is a covector-valued true-form or that $\zeta$ is a vector-valued true-form and $\cl{X}$ is a covector-valued pseudo-form. As we shall discuss in the coming section, we will always have the second case in our work where this duality pairing will represent physical power between a kinematics quantity, represented as a vector-valued true-form, and a kinetics quantity, represented as a covector-valued pseudo-form.\\

\subsubsection*{Exterior covariant derivative}

Let $(\boldsymbol{g},\nabla)$ be the associated Riemannian metric and Levi-Civita connection to $M$, respectively.
Similar to the property (\ref{eq:connection_operator_1}), the connection can be interpreted as a differential operator on the space of vector-valued 0-forms, i.e. $\map{\nabla}{\spFrmM{0}{TM}}{\spFrmM{1}{TM}}$.
Differentiation of generic vector-valued forms is achieved using the exterior covariant derivative operator $\map{\extcdS^k}{\spFrmM{k}{TM}}{\spFrmM{k+1}{TM}}$ which extends the action of $\nabla$.

The exterior covariant derivative $\extcdS^k$ of any $\alpha \in \spFrmM{k}{TM}$ is defined by \andrea{\cite[Ch. 3]{Angoshtari2013GeometricElasticity} \cite[Def. 5.1.]{Quang2014TheAlgebra}}
\begin{equation}\label{eq:ext_cov_der_def}
	\begin{split}
		(\extcdS^k\alpha)(u_0,\cdots,u_k) =& \sum_{i=0}^{k} (-1)^i \nabS_{u_i}(\alpha(u_0, \cdots, \underline{u_i}, \cdots, u_k)) \\
		+& \sum_{0\leq i<j<k} (-1)^{i+j} \alpha(\Lie{u_i}{u_j}, u_0, \cdots, \underline{u_i}, \cdots, \underline{u_j}, \cdots, u_k),
	\end{split}
\end{equation}
for all vector fields $u_j\in \spVec{M}$, for $j \in \{0,\cdots,k\}$, where an underlined argument indicates its omission.
For a generic bundle-valued 0-form, $\extcdS^0$ is simply the covariant derivative, whereas for the case of scalar-valued k-forms $\extcdS^0$ degenerates to the exterior derivative. Hence the name, exterior covariant derivative.

For illustration, the expressions of $\extcdS^0,\extcdS^1$ and $\extcdS^2$ applied respectively to any $\alpha\in \spFrmM{0}{TM},\beta\in \spFrmM{1}{TM},$ and $\gamma\in \spFrmM{2}{TM}$ are given by
\begin{align*}
	(\extcdS^0\alpha)(u_0)&= \nabS_{u_0}\alpha &&\in \spVec{M},\\
	(\extcdS^1\beta)(u_0,u_1)&= \nabS_{u_0}(\beta(u_1)) - \nabS_{u_1}(\beta(u_0)) - \beta(\Lie{u_0}{u_1}) &&\in \spVec{M},\\
	(\extcdS^2\gamma)(u_0,u_1,u_2)&= \nabS_{u_0}(\gamma(u_1,u_2)) - \nabS_{u_1}(\gamma(u_0,u_2)) + \nabS_{u_2}(\gamma(u_0,u_1)) &&\\
	&- \gamma(\Lie{u_0}{u_1},u_2) + \gamma(\Lie{u_0}{u_2},u_1) - \gamma(\Lie{u_1}{u_2},u_0) &&\in \spVec{M}.
\end{align*}
%

One key property of the exterior covariant derivative is that it satisfies the Leibniz rule, i.e. 
\begin{equation}\label{eq:Leibniz_ext_cov}
	\extcdS(\zeta\wedgedot\cl{X}) = \extcdS\zeta\wedgedot\cl{X} + (-1)^k \zeta\wedgedot\extcdS\cl{X}, \qquad \forall \zeta\in \spFrmM{k}{TM}, \cl{X}\in\spFrmM{l}{T^*M},
\end{equation}
where the first $\extcdS$ on the left-hand-side degenerates to an exterior derivative on $(k+l)$ forms.
By combining the Leibniz rule (\ref{eq:Leibniz_ext_cov}) with Stokes theorem and (\ref{eq:wedgedot_dist_property}), the \textit{integration by parts} formula using exterior calculus is expressed as
\begin{equation}\label{eq:integ_by_parts}
	\intS \extcdS\zeta\wedgedot\cl{X} + \zeta\wedgedot\extcdS\cl{X} = \int_{\Sbound} i^*(\zeta\wedgedot\cl{X}) = \int_{\Sbound} \ptr(\zeta)\wedgedot\ptr(\cl{X}).
\end{equation}

\begin{remark}[\textbf{Local expression of exterior covariant derivative}]\label{remark:extd_coord}
	
	In local coordinates, the vector valued forms $\extcdS^0\alpha,\extcdS^1\beta,\extcdS^2\gamma$ above are expressed as
	\begin{align}
		\extcdS^0\alpha &= \alpha^i_{;a} e^a \otimes e_i  &&\in \spFrmM{1}{TM} \nonumber\\
		\extcdS^1\beta &= 2\beta^i_{[a;b]} \left(\frac{e^a\wedge e^b}{2}\right) \otimes e_i  &&\in \spFrmM{2}{TM}\nonumber\\
		\extcdS^2\gamma &= 3\gamma^i_{[ab;c]} \left(\frac{e^a\wedge e^b\wedge e^c}{3!}\right) \otimes e_i  &&\in \spFrmM{3}{TM},\label{eq:ext_cov_stress_express}
	\end{align}
	where the square brackets indicate anti-symmetrization and the semi-colon indicates covariant differentiation such that $\alpha^i_{;a} := \nabS_a\alpha^i, \beta^i_{a;b} := \nabS_b\beta^i_{a},$ and $\gamma^i_{ab;c} := \nabS_c\gamma^i_{ab}$.
	Thus, one can see that the exterior covariant derivative is constructed via an anti-symmetrization process of the covariant derivative on the form leg.
	Furthermore, the covariant differentiation is applied to vector fields and not to higher order tensor fields. For instance, $\gamma^i_{ab;c} = \partial_c(\gamma^i_{ab}) + \Gamma_{kc}^i \gamma^k_{ab}$. Thus, the component functions $\gamma^i_{ab}$ are differentiated as a collection of vector fields, indexed by $a$ and $b$, and not as a (1,2) tensor field.
	This is a fundamental difference between exterior covariant differentiation and covariant differentiation of high-order tensor fields (e.g. in (\ref{eq:cov_der_def_mat_tensor})) which will have a significant impact on the intrinsicality of the material equations of motion. This point will be further discussed in Remark \ref{remark:intrinsic_material}.
\end{remark}


\subsubsection*{Hodge star operator}

{\newcommand{\hodeStar}{\star^\flat}
In analogy to the standard construction for scalar-valued forms (\cf \cite[Ch.6]{arnold2018finite}), we can construct a Hodge-star operator that maps vector-valued forms to covector-valued pseudo-forms in the following manner.
Let $\B{\mu}\in\spFrm{M}{n}$ be some top-form on $M$.
The duality product (\ref{eq:duality_pairing_def}) defines a linear functional that maps $\spFrmM{n-k}{T^*M}$ to $\bb{R}$. By the Riesz representation theorem, we can introduce the Hodge-star operator
$$\map{\hodeStar}{\spFrmM{k}{TM}}{\spFrmM{n-k}{T^*M}},$$
such that 
\begin{equation}\label{eq:def_Hodge_star}
	\zeta \wedgedot \hodeStar \xi = \inPair{\zeta}{\xi}{\B{g}} \B{\mu}, \qquad \qquad \forall\zeta,\xi \in \spFrmM{k}{TM},
\end{equation}
where $\map{\inPair{\cdot}{\cdot}{\B{g}}}{\spFrmM{k}{TM}\times \spFrmM{k}{TM}}{\spFn{M}}$ denotes the point-wise inner product of vector-valued forms treated as $(1,k)$ tensor fields over $M$.
The action of $\hodeStar$ is equivalent to an index lowering operation using $\B{g}$ on the value-leg and a standard Hodge-operator with respect to $\B{\mu}$ on the form-leg.
When needed, we shall denote this dependency explicitly by $\hodeStar[\B{g},\B{\mu}]$.
For example, if $\zeta=\alpha\otimes u$ and $\xi = \bar{\alpha}\otimes\bar{u}$ are trivial vector-valued $k$-forms, then  
$\hodeStar \xi = \hodeStar (\bar{\alpha}\otimes\bar{u}) = \star\bar{\alpha}\otimes \B{g}\cdot\bar{u},$
whereas
$$\inPair{\zeta}{\xi}{\B{g}} = \B{g}_{ab} u^a \bar{u}^b \B{g}^{i_1 j_1} \cdots \B{g}^{i_k j_k} \alpha_{i_1\cdots i_k}\bar{\alpha}_{j_1\cdots j_k} \in \spFn{M},$$
with $\B{g}_{ij},\B{g}^{ij}\in \spFn{M}$ denoting components of the metric and inverse metric tensors.
Finally, we denote the inverse Hodge star of $\hodeStar$ by
$$\map{\star^\sharp}{\spFrmM{n-k}{T^*M}}{\spFrmM{k}{TM}}.$$

\begin{remark}[\textbf{Material properties in Hodge-star}]
	
	It is important to note that the above definition of the Hodge star (\ref{eq:def_Hodge_star}), the top form $\B{\mu}$ is not necessarily equal to the volume form $\omega_{\B{g}}$ induced by the metric, but of course proportional to it by some scalar function. This general definition of the Hodge star allows the incorporation of material properties as is done for example in electromagnetism \cite{bossavit1998computational}. In the coming section, we will include the mass top-forms of the elastic body, introduced in Sec. \ref{sec:mass_volume}, inside the Hodge-star operator. In this way, the Hodge-star will be used to map \textit{intensive kinematics quantities}, expressed as vector-valued forms, to \textit{extensive kinetics quantities}, expressed as covector-valued pseudo-forms.
\end{remark}

\begin{remark}[\textbf{Extension to tensor-valued differential forms}]
	The construction presented so far for vector-valued forms and covector-valued pseudo-forms can be extended to any complementary pair of $(p,q)$-tensor valued forms and $(q,p)$-tensor valued pseudo-forms. While the wedge-dot product would be unchanged, the Hodge star operator should be extended such that the value leg valence is transformed from $(p,q)$ to $(q,p)$.
\end{remark}

}

\subsection{Application to nonlinear elasticity}

\begin{table}
	\centering
	\begin{tabular}{|c|c|c|c|}
		\hline
		&	Convective & Material & Spatial\\
		& 	$\spvecFrmB{k}$		&	$\spvecFrmPhi{k}$		& $\spvecFrmS{k}$ \\\hline
		$k=0$& $\vC,\hat{a}$ & $\vM,\tilde{a}$& $\vS,a$\\
		$k=1$  & $\nabC\vC, \hat{\epsilon}$ & $\nabM\vM, F$& $\nabS\vS, \epsilon$\\
		\hline
	\end{tabular}
	\caption{Bundle-valued differential forms representation of the kinematics physical quantities}
	\label{table:bv_forms_kinematics}
\end{table}

In terms of bundle-valued forms, we can fully formulate the theory of nonlinear elasticity as follows. First, all kinematics quantities introduced in Sec. \ref{sec:kinematics} and \ref{sec:deformation} will be treated as \textit{intensive vector-valued forms} (\cf Table \ref{table:bv_forms_kinematics}). In particular, convective quantities will belong to $\spvecFrmB{k}$, spatial quantities will belong to $\spvecFrmS{k}$, while material quantities will belong to $\spvecFrmPhi{k}$. 

The velocity fields will be treated as vector-valued 0-forms with the underlying 0-form being their component functions.
Thus, we identify
$$\spVecB \cong \spvecFrmB{0}, \qquad T_\varphi \spC  = \spVecPhi \cong \spvecFrmPhi{0}, \qquad \spVecS \cong \spvecFrmS{0}.$$
The local expressions of all three velocities seen as vector-valued 0-forms is given by
$$\vC = \vC^I \otimes E_{I}, \qquad \qquad \vfM = \vM^i \otimes e_{i}|_\varphi, \qquad \qquad \vS = \vS^i \otimes e_{i}$$
which is in contrast to their expressions seen as vector fields in (\ref{table:motion_kinematics}).
The spatial and convective velocity gradients are considered as vector-valued 1-forms in $\spvecFrmS{1}$ and $\spvecFrmB{1}$, respectively. Whereas the material representation of the velocity gradient as well as the deformation gradient ($F$) are elements of $\spvecFrmPhi{1}$.
The velocity one forms and covariant velocity gradients will be treated as covector-valued (true) forms and are related to their covariant counterparts by applying the $\flat$ operation to the value-leg:
\begin{equation*}
	\begin{split}
		\vfC &= \flatV(\vC),\\
		\nabC\vfC &= \flatV(\nabC\vC),
	\end{split}
	\qquad
	\begin{split}
		\vfM &= \flatV(\vM),\\
		\nabM\vfM &= \flatV(\nabM\vM),
	\end{split}
	\qquad
	\begin{split}
		\vfS &= \flatV(\vS),\\
		\nabS\vfS &= \flatV(\nabS\vS).
	\end{split}
\end{equation*}
As for the rate of strain tensor fields $\epC$ and $\epS$, we will consider them as vector-valued one-forms and thus we identify $T_{\gC}\spMetB \cong \spvecFrmB{1}$.
In this manner, we can identify the cotangent space $T^*_{\gC}\spMetB$ by $\spcovFrmB{n-1}$, which will be the space of stresses as discussed later.

The most important technical advantage of our formulation using bundle-valued forms is that the transition from one representation to the other has a clear unified rule for all physical variables. In particular,
\begin{itemize}
	\item the transition from the spatial to material representation is performed by pulling-back the \textit{form-leg} only using $\cfgPleg{f}$.
	\item the transition from the material to convective representation is performed by pulling-back the \textit{value-leg} only using $\cfgPleg{v}$.
	\item the transition from the spatial to convective representation is performed by pulling-back \textit{both legs} using $\cfgPleg{} = \cfgPleg{v} \circ \cfgPleg{f}$.
\end{itemize}
The reverse transition is simply using the corresponding pushforward maps.
Therefore, we can rewrite the relations between the spatial, convective and material velocity fields as
\begin{equation}\label{eq:vel_relations_forms}
	\vM = \cfgPleg{f}(\vS), \qquad \qquad  \vC = \cfgPleg{v}(\vM) , \qquad \qquad  \vC = \cfgPleg{}(\vS) .
\end{equation}
Similarly, the velocity gradients are related by
\begin{equation}\label{eq:velGrad_relations_forms}
	\nabM \vM = \cfgPleg{f}(\nabS\vS), \qquad \qquad \nabC \vC = \cfgPleg{v}(\nabM\vM), \qquad \qquad \nabC \vC = \cfgPleg{}(\nabS\vS).
\end{equation}
The same relations also hold for the velocity one-forms, the covariant velocity gradients, and the accelerations.
In fact, one has (by construction) the commutative properties
\begin{equation}\label{eq:comm_prop_cov_derv}
	\nabC \circ \cfgPleg{} = \cfgPleg{}\circ \nabS, \qquad \qquad \nabM \circ \cfgPleg{f} = \cfgPleg{f}\circ \nabS, \qquad \qquad \nabC \circ \cfgPleg{v} = \cfgPleg{v}\circ \nabM.
\end{equation}

The exterior covariant derivatives used for spatial, convective and material variables are denoted, respectively, by:
\begin{align}
	&\map{\extcdS^k}{\spvecFrmS{k}}{\spvecFrmS{k+1}}\nonumber\\
	&\map{\extcdC^k}{\spvecFrmB{k}}{\spvecFrmB{k+1}}\nonumber\\
	&\map{\extcdM^k}{\spvecFrmPhi{k}}{\spvecFrmPhi{k+1}} .\label{eq:extd_cov_deriv_operators}
\end{align}
One defines $\extcdS^k$ using the spatial connection $\nabS$ in (\ref{eq:ext_cov_der_def}) by application on vector fields $u_j\in \spVecS$.
As for $\extcdC^k$ and $\extcdM^k$, they are defined using the convective and material connections $\nabC$ and $\nabM$ , respectively, by application on vector fields $\hat{u}_j\in \spVecB$.
One important property of the exterior covariant derivative is that it commutes with pullbacks \cite{Quang2014TheAlgebra} similar to the covariant derivative (\ref{eq:comm_prop_cov_derv}).
Therefore, we have that
\begin{equation}\label{eq:comm_prop_ext_cov_derv}
	\extcdC \circ \cfgPleg{} = \cfgPleg{}\circ \extcdS, \qquad \qquad \extcdM \circ \cfgPleg{f} = \cfgPleg{f}\circ \extcdS, \qquad \qquad \extcdC \circ \cfgPleg{v} = \cfgPleg{v}\circ \extcdM.
\end{equation}

Using the associated mass top-form and metric of each representation we will define three Hodge-star operators that allow us to relate intensive kinematics variables to extensive kinetics variables of the elastic body.
These include the momentum and stress variables.
The spatial, convective and material Hodge stars will be denoted respectively by:
\begin{align}
	&\map{\hodgeS}{\spvecFrmS{k}}{\spcovFrmS{n-k}},\nonumber\\
	&\map{\hodgeC}{\spvecFrmB{k}}{\spcovFrmB{n-k}},\nonumber\\
	&\map{\hodgeM}{\spvecFrmPhi{k}}{\spcovFrmPhi{n-k}}, \label{eq:hodge_stars}
\end{align}
with their metric and mass form dependencies stated by 
$\hodgeS[\gS,\mFS], \hodgeC[\gC,\mFC], \hodgeM[\gM,\mFM],$
and constructed similar to (\ref{eq:def_Hodge_star}).
A consequence of such dependency is that the spatial and convective Hodge stars will be time-dependent, which needs to be considered when differentiating in time. 
A summary of our proposed geometric formulation using bundle-valued forms is depicted in Fig. \ref{fig:bund_val_forms_nonlinear_elasticity}.

\begin{figure}
	\centering
	\begin{tikzcd}
		& {\color{red}\mathrm{Convective}}& {\color{red}\mathrm{Material}} & {\color{red}\mathrm{Spatial}} \\
		{\TwoVecText{\color{red}\mathrm{Intensive}}{\color{red}\mathrm{Kinematics}}}&\spvecFrmB{k}  \arrow[from=r,"\cfgPleg{v}"] \arrow[d, "\hodgeC"] & \spvecFrmPhi{k}  \arrow[from=r,"\cfgPleg{f}"] \arrow[d, "\hodgeM"] & \spvecFrmS{k}  \arrow[d, "\hodgeS"] \\
		{\TwoVecText{\color{red}\mathrm{Extensive}}{\color{red}\mathrm{Kinetics}}}&\spcovFrmB{n-k}   \arrow[from=r,"\cfgPleg{v}"]  & \spcovFrmPhi{n-k} \arrow[from=r,"\cfgPleg{f}"]   & \spcovFrmS{n-k}
	\end{tikzcd}
	\caption{Our proposed formulation of nonlinear elasticity using bundle-valued forms}
	\label{fig:bund_val_forms_nonlinear_elasticity}
\end{figure}

\section{Dynamical equations of motion}\label{sec:dynamics}

Now we turn attention to the governing equations of motion of nonlinear elasticity and the underlying energy balance laws using exterior calculus.
The main feature of these equations is that the momentum and stress variables will be represented as \textbf{extensive covector-valued pseudo-forms}.
In this paper, we do not present a formal derivation of these equations but instead show their equivalence to the common standard formulations in the literature.
In this manner, we avoid overloading this paper with all the technicalities involved in the derivation process.
In a future sequel of this paper, we shall present the derivation of these equations from first principles in the port-Hamiltonian framework and highlight the underlying energetic structure, similar to our previous works on fluid mechanics \cite{Rashad2021Port-HamiltonianEnergy,Rashad2021Port-HamiltonianFlow,Rashad2021ExteriorModels,califano2021geometric}.

\subsection{Overall energy balance}


We start first by a generic statement of the balance of energy, or first law of thermodynamics, which is the most fundamental balance law from which the governing equations can be derived in numerous methods, e.g. by postulating covariance \cite{Kanso2007OnMechanics}, by variational principles \cite{Gilbert2023AMechanics}, by Lagrangian reduction\cite{Gay-Balmaz2012ReducedMechanics}, or by Hamiltonian reduction \cite{Simo1988ThePlates}.

Let $\cl{U}\subseteq \cl{B}$ denote a nice open set of the body and let $\Ekin$ and $\Eint$ denote the kinetic and internal energies of that set, respectively.
Furthermore, let $\Pst$ denote the rate of work done (power) on the surface $\partial\cl{U}$ due to stress.
The first law of thermodynamics is then expressed as:
\begin{equation}\label{eq:ThD_first_law}
	\frac{\extd}{\extd t} (\Ekin + \Eint) = \Pst,
\end{equation}
which states that the rate of increase of total energy of any portion $\cl{U}$ of the body $\cl{B}$ equals the mechanical power supplied to that portion from surface traction on its boundary $\partial\cl{B}$.
For simplicity, we will exclude any body forces, which can be trivially added.
We also focus only on the mechanical aspect of the motion. Thus, for clarity of exposition, we exclude non-mechanical power exchange with other physical domains (e.g. thermo-elastic and piezo-electric effects).

Each of $\Ekin,\Eint,$ and $\Pst$ is an integral quantity that depends on certain kinematics and kinetics variables in addition to the mass properties of the elastic body.
For physical compatibility, their respective integrands are required to be psuedo-forms such that these integral quantities always have positive value under a change of orientation of $\cl{U}$.
The explicit expression of the energy balance law (\ref{eq:ThD_first_law}) depends on a number of choices:
\begin{enumerate}
\item\textit{Spatial, material or convective description}

	For both the material and convective representations the integration is performed over $\cl{U}\subseteq \cl{B}$ with respect to the mass measure defined by $\mFC=\mFM$.
	In case the spatial representation is used, the domain of integration will be $\cfgT(\cl{U}) \subseteq \cl{S}$ and the mass measure is defined by $\mFS_t$.
	
\item\textit{The pairing operation and mathematical representation of kinematics and kinetics quantities}
	
	The common choice in the literature is to use tensor fields \cite{Yavari2006OnElasticity} or  tensor field densities \cite{Grubic2014TheManifold}.
	In our work we will be representing kinematics quantities as vector valued forms while kinetics quantities as covector-valued pseudo forms. Their corresponding pairing is given by the wedge-dot product (\ref{eq:duality_pairing_def}).
	
\item\textit{Intensive or extensive description}
	
	The common choice in the literature is to separate the extensive mass structure from both kinematics and kinetics variables, and thus representing both as intensive quantities.
	What we aim for is to include the mass structure into the kinetics variables such that they are extensive quantities.
\end{enumerate}

\subsection{Extensive representation of stress}

In the exterior calculus formulation of continuum mechanics \cite{Frankel2019ThePhysics,Kanso2007OnMechanics}, one postulates the existence of the stress as a \textit{covector-valued $(n-1)$ pseudo-form}, in the same manner one postulates the existence of the traction force field in the classic Cauchy stress theorem.
The convective, material and spatial representation of this stress variable are denoted respectively by
$$\stC \in \spcovFrmB{n-1}, \qquad \qquad \stM \in \spcovFrmPhi{n-1}, \qquad \qquad \stS \in \spcovFrmS{n-1},$$
which are related to each other by
\begin{equation}\label{eq:stress_relations}
	\stM = \cfgPleg{f} (\stS), \qquad \qquad \stC = \cfgPleg{v} (\stM), \qquad \qquad \stC = \cfgPleg{} (\stS),
\end{equation}
as depicted in Fig. \ref{fig:stress_relations}.
In a local chart for $n=3$, the stresses are expressed as
$$	\stC = \underbrace{(\half\stC_{KAB}\ E^A\wedge E^B)}_{=:\stC_K}\otimes E^K,\qquad \qquad 
\stM = \underbrace{(\half\stS_{kAB}\ E^A\wedge E^B)}_{=:\stM_k}\otimes e^k|_\varphi,$$
\begin{equation}\label{eq:ext_stress_expr}
	\stS = \underbrace{(\half\stS_{kab}\ e^a\wedge e^b)}_{=:\stS_k}\otimes e^k,
\end{equation}
where each $\stC_K,\stM_k \in \spFrmB{2}$ and $\stS_k\in\spFrmS{2}$ is a two-form, while $\stC_{KAB},\stS_{kAB} \in \spFn{\cl{B}}$ and $\stS_{kab} \in \spFn{\cl{S}}$ denote their respective component functions.

\begin{figure}
	\centering
	\begin{tikzcd}
		& {\color{red}\mathrm{Convective}}& {\color{red}\mathrm{Material}} & {\color{red}\mathrm{Spatial}} \\
		{\color{red}\mathrm{Stress}}&\TwoVecArr{\spcovFrmB{n-1}}{\stC}  \arrow[from=r,"\cfgPleg{v}"] \arrow[d, "\ptr"] & \TwoVecArr{\spcovFrmPhi{n-1}}{\stM}  \arrow[from=r,"\cfgPleg{f}"] \arrow[d, "\ptr"] & \TwoVecArr{\spcovFrmS{n-1}}{\stS}  \arrow[d, "\ptr"] \\
		{\color{red}\mathrm{Trace\ of\ Stress}}&\TwoVecArr{\spcovFrmBbnd{n-1}}{\stC|_\Bbound}  \arrow[from=r,"\cfgPleg{v}"]  & \TwoVecArr{\spcovFrmPhibnd{n-1}}{\stM|_\Bbound} \arrow[from=r,"\cfgPleg{f}"]   & \TwoVecArr{\spcovFrmSbnd{n-1}}{\stS|_\Sbound}
	\end{tikzcd}
	\caption{Extensive stress representation using bundle-valued forms}
	\label{fig:stress_relations}
\end{figure}

The pairing of stress, as a covector-valued form with velocity, as a vector-valued form, results in an $(n-1)$ form that when integrated on any surface yields the rate of work done by stress on that surface.
With the stress being a pseudo-form, the sign of the $(n-1)$ form, and thus the integral, changes automatically under a change of orientation of the surface. This corresponds to the change of sign of the surface normal in the classic approach \cite{Kanso2007OnMechanics}.

In the spatial representation, this pairing would be expressed as $\vS\wedgedot\stS \in \spFrmS{n-1}$. On the boundary of the spatial configuration $\Sbound$, the surface stress power would be expressed as $\Pst = \int_\Sbound i^*(\vS\wedgedot\stS)$, where $\map{i}{\Sbound}{\cl{S}}$ denotes the spatial inclusion map.
From (\ref{eq:wedgedot_dist_property}), we could express $\Pst$ as
$$\Pst = \int_\Sbound \vS|_\Sbound \wedgedot \stS|_\Sbound, $$
where 
$$\vS|_\Sbound := \ptr(\vS) \in \spvecFrmSbnd{0}, \qquad \qquad \stS|_\Sbound:= \ptr(\stS) \in \spcovFrmSbnd{n-1},$$ 
denote the (partial) pullback of the spatial velocity and stress on the boundary under the spatial inclusion map $\map{i}{\Sbound}{\cl{S}}$.
The variables $\vS|_\Sbound$ and $\stS|_\Sbound$ represent the boundary conditions of the problem.


Similarly, in the material representation one can show, using the change of variables formula and the fact that $\Sbound = \varphi(\Bbound)$, that the surface stress power is expressed by
\begin{align*}
	\Pst = \int_{\Sbound} \vS|_\Sbound \wedgedot \stS|_\Sbound = \int_\Bbound \cfgPleg{}(\vS|_\Sbound \wedgedot \stS|_\Sbound) = \int_\Bbound \cfgPleg{f}(\vS|_\Sbound) \wedgedot \cfgPleg{f}(\stS|_\Sbound)= \int_\Bbound \vM|_\Bbound \wedgedot \stM|_\Bbound,
\end{align*}
where 
$$\vM|_\Bbound:= \ptr(\vM) = \cfgPleg{f}(\vS|_\Sbound)\in \spvecFrmPhibnd{0} \qquad \stM|_\Bbound:= \ptr(\stM) = \cfgPleg{f}(\stS|_\Sbound)\in \spcovFrmPhibnd{n-1}$$
denote the (partial) pullback of the material velocity and stress on the boundary under the body inclusion map $\map{i}{\Bbound}{\cl{B}}$, which we also denote by $i$ with an abuse of notation.

Similarly, in the convective representation one can show, using duality in addition to (\ref{eq:stress_relations}) and (\ref{eq:vel_relations_forms}), that the surface stress power is expressed by
\begin{align*}
	\Pst = \int_\Bbound \vM|_\Bbound \wedgedot \stM|_\Bbound =  \int_\Bbound \vM|_\Bbound \wedgedot \cfgFleg{v}(\stC|_\Bbound)  =  \int_\Bbound \cfgPleg{v}(\vM|_\Bbound) \wedgedot \stC|_\Bbound = \int_\Bbound \vC|_\Bbound \wedgedot \stC|_\Bbound 
\end{align*}
where 
$$\vC|_\Bbound:= \ptr(\vC) = \cfgPleg{v}(\vM|_\Bbound)\in \spvecFrmBbnd{0} \qquad \stC|_\Bbound:= \ptr(\stC) = \cfgPleg{v}(\stM|_\Bbound)\in \spcovFrmBbnd{n-1}$$
denote the (partial) pullback of the convective velocity and stress on the boundary under the body inclusion map $\map{i}{\Bbound}{\cl{B}}$.

\subsection{Extensive representation of momentum}

Instead of expressing the motion of the body using the intensive velocity variable, one can use instead the extensive momentum defined as the Hodge-star of the velocity.
The convective, material and spatial representations of this momentum variable are denoted respectively by
\begin{equation}\label{eq:vel_mom_relation}
	\momC:= \hodgeC \vC \in \spcovFrmB{n}, \qquad \momM:= \hodgeM \vM \in \spcovFrmPhi{n}, \qquad \momS:= \hodgeS \vS \in \spcovFrmS{n},
\end{equation}
which are related to each other by
\begin{equation}\label{eq:momentum_relations}
	\momM = \cfgPleg{f} (\momS), \qquad \qquad \momC = \cfgPleg{v} (\momM), \qquad \qquad \momC = \cfgPleg{} (\momS).
\end{equation}
In a local chart for $n=3$, the convective, material and spatial momentum variables are expressed as
\begin{equation}\label{eq:mom_expressions}
	\momC = \underbrace{\gC_{IJ} \vC^J \mFC}_{=:\momC_I}\otimes E^I,\qquad \momM = \underbrace{\gM_{ij} \vM^j \mFM}_{=:\momM_i}\otimes e^i|_\varphi,\qquad \momS = \underbrace{\gS_{ij} \vS^j \mFS}_{=:\momS_i}\otimes e^i,
\end{equation}
where each $\momC_I,\momM_i\in\spFrmB{3}$ and $\momS_i\in\spFrmS{3}$ is a top-form.

The pairing of momentum, as a covector-valued form with velocity, as a vector-valued form, results in an $n$-form that when integrated on any volume yields twice its kinetic energy.
Thus, the kinetic energy of the whole body is expressed in the spatial, material, and convective representations respectively as
\begin{equation}\label{eq:Ekin_mom_ext}
	\Ekin = \intS \half \vS \wedgedot \momS = \intB \half \vM \wedgedot \momM = \intB \half \vC \wedgedot \momC.
\end{equation}

\begin{remark}[\textbf{Covector-valued forms vs. tensor densities}]
	
	Note that both the convective and spatial momentum variables are trivial covector-valued forms that can be identified, respectively, with the tensor densities $\mFC\otimes\vfC$ and $\mFS\otimes\vfS$.
	On the other hand, the material momentum is not trivial. Even though one can express it equivalently as the tensor density $\mFM\otimes\vfM$, it is clearly not an element of $\spcovFrmPhi{n}$, since $\vfM$ is not a true vector field.
	Furthermore, in order for the spatial-to-material transformation in (\ref{eq:momentum_relations}) to be valid, one must consider the form-leg and value-leg of $\momS$ to be $\momS_i$ and $e^i$, as indicated in (\ref{eq:mom_expressions}), and not as $\mFS$ and $\vfS$. Similarly, the material-to-convective transformation in (\ref{eq:momentum_relations}) requires the form-leg and value-leg of $\momC$ to be $\momC_I$ and $E^I$ and not as $\mFC$ and $\vfC$.
	Thus, one should keep in mind such technical differences when using bundle-valued forms compared to tensor densities, used for example in \cite{Simo1988ThePlates,Grubic2014TheManifold}.
\end{remark}

\subsection{Equations of motion}\label{subsec:EoM}
We now present the equations of motion for the spatial, convective and material descriptions.
Each description has a local balance of momentum relating the momentum and stress variables in addition to one extra unique equation.
The spatial description has an advection equation for $\mFS$, the convective description has an advection equation for $\gC$, while the material description has a reconstruction equation for $\varphi$.
As mentioned earlier, we will not provide a formal derivation of these equations in this paper.
Instead, we delegate them to a future sequel and we settle for showing their equivalence to each standard formulations in the literature.

\begin{proposition}[\textbf{Spatial}]
	The equations of motion governing the extensive variables $(\mFS,\momS,\stS)\in \spFrmS{n}\times \spcovFrmS{n}\times \spcovFrmS{n-1}$ are given by
	\begin{align}
		\partial_t \mu =& - \extd (\iota_{\vS}\mFS),\label{eq:EoM_S_mass}\\
		\partial_t \momS =& - \extcdS(\iota_v \mu\otimes\vfS) + \extcdS \stS,\label{eq:EoM_S_mom}
	\end{align}
where $\vS = \invhodgeS \momS \in \spvecFrmS{0}$.
Furthermore, the balance of the total energy is expressed as
\begin{equation}\label{eq:EnergyBal_S}
	\frac{\extd}{\extd t}\intS \half \invhodgeS \momS \wedgedot \momS + \EintS(x,F,\gS)= \int_\Sbound \vS|_\Sbound \wedgedot \stS|_\Sbound,
\end{equation}
where $\map{\EintS}{\cl{S}\times\spvecFrmPhi{1}\times\cl{M}(\cl{S})}{\spFrmS{n}}$ is the internal energy density function in the spatial representation.
\end{proposition}

The first equation above represents the conservation of mass while the second one represents the local balance of momentum in terms of the extensive variable $\momS$. 
The form of the equations of motion in (\ref{eq:EoM_S_mass}-\ref{eq:EoM_S_mom}) is often called the \textit{conservation form}.
In such equations one can see clearly that the mass flux is identified by $\iota_v\mFS$ while the momentum flux is identified by the covector-valued $(n-1)$ form $\iota_v \mu\otimes\vfS$, which has the same geometric nature as the stress $\stS$ \cite{Gilbert2023AMechanics}.
The internal energy density function $\EintS$ and its dependencies will be discussed later in Sec. \ref{subsec:const_eqns}.

Finally, one can show that the rate of change of the kinetic energy along trajectories $(\mFS(t),\momS(t))$ of (\ref{eq:EoM_S_mass}-\ref{eq:EoM_S_mom}) satisfies
\begin{equation}\label{eq:Ekin_S_balance}
	\frac{\extd}{\extd t} \Ekin = \intS \vS\wedgedot \extcdS \stS,
\end{equation}
which states that the rate of change of kinetic energy is equal to the work done due to stress forces and shows that the momentum flux term does not contribute to the power balance \cite{Gilbert2023AMechanics}.

\begin{remark}[\textbf{Advection form of momentum balance}]
	
	Consider the following identity relating the exterior covariant derivative with the Lie derivative of a trivial covector-valued top-form \cite{Gilbert2023AMechanics}
	\begin{equation}\label{eq:Lie_deriv_identity}
		\Lie{u}{(\omega\otimes\alpha)} = \extcdS(\iota_u \omega \otimes \alpha) + \omega\otimes (\nabS u \wedgedot \alpha),
	\end{equation}
	$\forall u\in \spVec{M}, \alpha\in \spFrm{M}{1}, \omega\in \spFrm{M}{n},$
	while $\map{\iota_u}{\spFrmS{k}}{\spFrmS{k-1}}$ denotes the standard interior product of scalar-valued forms.
	
	Using the (\ref{eq:Cartan_Lie_deriv},\ref{eq:Lie_deriv_identity}) along with the identity $\nabS\vS\wedgedot\vfS = \half \extd \iota_{\vS}\vfS$, one can also express the spatial equations of motion as 
	\begin{align}
		\partial_t \mu =& - \Lie{\vS}{\mFS},\\
		\partial_t \momS =& - \Lie{\vS}{\momS} + \mFS\otimes\half\extd \iota_{\vS}\vfS + \extcdS \stS,\label{eq:EoM_S_mom_adv}
	\end{align}
	which is often referred to as the \textit{advection form} of the equations.
	It is interesting to note some resemblance between (\ref{eq:Lie_deriv_identity}) and Cartan's formula (\ref{eq:Cartan_Lie_deriv}) for scalar-valued forms.
\end{remark}


\begin{proposition}[\textbf{Convective}]
	The equations of motion governing $\gC \in \spMetB$ and the extensive variables $(\momC,\stC)\in \spcovFrmB{n}\times \spcovFrmB{n-1}$ are given by
	\begin{align}
		\partial_t \gC &= \Lie{\vC}{\gC},\label{eq:EoM_C_metric} \\
		\partial_t \momC &= \half  \mFC\otimes\extd \iota_{\vC} \vfC + \extcdC \stC,\label{eq:EoM_C_mom}
	\end{align}
	where $\vC = \invhodgeC \momC \in \spvecFrmB{0}$.
	Furthermore, the balance of the total energy is expressed as
	\begin{equation}\label{eq:EnergyBal_C}
		\frac{\extd}{\extd t}\intB \half \invhodgeC \momC \wedgedot \momC + \EintC(X,\gC)= \int_\Bbound \vC|_\Bbound \wedgedot \stC|_\Bbound,
	\end{equation}
	$\map{\EintC}{\cl{B}\times\spMetB}{\spFrmB{n}}$ is the internal energy density function in the convective representation.
\end{proposition}
Equation (\ref{eq:EoM_C_metric}) represents the advection of the convective metric (with respect to $-\vC$) while (\ref{eq:EoM_C_mom}) represents the local balance of momentum in terms of the extensive variable $\momC$.
Finally, the rate of change of the kinetic energy along trajectories $(\gC(t),\momC(t))$ of (\ref{eq:EoM_C_metric}-\ref{eq:EoM_C_mom}) satisfies
\begin{equation}\label{eq:Ekin_C_balance}
	\frac{\extd}{\extd t} \Ekin = \intB \vC\wedgedot \extcdC \stC.
\end{equation}

\begin{proposition}[\textbf{Material}]
	The equations of motion governing $\varphi \in \spC$ and the extensive variables $(\momM,\stM)\in \spcovFrmPhi{n}\times \spcovFrmPhi{n-1}$ are given by
	\begin{align}
		\partial_t \varphi &= \vM,\label{eq:EoM_M_cfg}  \\
		D_t \momM &= \extcdM \stM,\label{eq:EoM_M_mom}
	\end{align}
	where $\vM = \invhodgeM \momM \in \spvecFrmPhi{0}$.
	Furthermore, the balance of the total energy is expressed as
	\begin{equation}\label{eq:EnergyBal_M}
		\frac{\extd}{\extd t}\intB \half \invhodgeM \momM \wedgedot \momM + \EintM(X,F)= \int_\Bbound \vM|_\Bbound \wedgedot \stM|_\Bbound,
	\end{equation}
	where $\map{\EintM}{\cl{B}\times\spvecFrmPhi{1}}{\spFrmB{n}}$ is the material internal energy density function.
\end{proposition}
Equation (\ref{eq:EoM_M_cfg}) represents the reconstruction equation of the configuration $\varphi$ whereas (\ref{eq:EoM_M_mom}) represents the local balance of momentum in terms of the extensive variable $\momM$.
Note that in contrast to (\ref{eq:EoM_S_mom},\ref{eq:EoM_C_mom}), the momentum balance (\ref{eq:EoM_M_mom}) is expressed in terms of the material derivative.
Finally, the rate of change of the kinetic energy along trajectories $(\varphi(t),\momM(t))$ of (\ref{eq:EoM_M_cfg}-\ref{eq:EoM_M_mom}) satisfies
\begin{equation}\label{eq:Ekin_M_balance}
	\frac{\extd}{\extd t} \Ekin = \intB \vM\wedgedot \extcdM \stM.
\end{equation}

\subsection{Constitutive equation and internal energy}\label{subsec:const_eqns}
We conclude this section by discussing how constitutive equations for determining the stress are included in our formulation.
We do not aim for a concise treatment of this involved topic in this paper.
Instead, we aim to highlight here the form of the equations in exterior calculus, the difference between the three representations, and how only the convective representation of the constitutive equations allows a complete description, following up the discussion of Sec. \ref{sec:deformation}.
Thus, for simplicity, we only treat the case of pure hyper elasticity and neglect any memory or rate effects.
For an introduction to the subject of constitutive theory, the reader is referred to \cite[Ch.3]{Marsden1994MathematicalElasticity}.


\subsubsection{Convective}

Using the integration by parts formula (\ref{eq:integ_by_parts}) and combining (\ref{eq:EnergyBal_C}) and (\ref{eq:Ekin_C_balance}), one can see that the conservation of energy implies that the rate of change of the internal energy should satisfy 
\begin{equation}\label{eq:Eint_balance}
	\frac{\extd}{\extd t} \Eint = \intB \nabC\vC\wedgedot\stC.
\end{equation}
Thus, for the equations of motion to be well-posed, one requires a closure relation between $\stC,\nabC\vC,$ and $\Eint$.
As discussed in Sec. \ref{sec:deformation}, the convective metric $\gC$ allows an intrinsic description of the deformation's state. Thus, one can define the internal strain energy as a functional of $\gC$:
\begin{equation}
	\Eint[\gC] := \intB \EintC(X,\gC),
\end{equation}
where $\map{\EintC}{\cl{B}\times\spMetB}{\spFrmB{n}}$ is the internal energy density function, while dependence of body points allows modeling non-homogeneous materials.


Using the identifications of the tangent and cotangent spaces $T_{\gC}\spMetB$ and $T^*_{\gC}\spMetB$ as bundle-valued forms described in Sec. \ref{sec:ext_calc}, the rate of change of $\Eint$ is given by
\begin{equation}\label{eq:dt_Eint}
	\frac{\extd}{\extd t} \Eint = \intB \parD{\EintC}{\gC} \wedgedot \partial_t \gC = \intB (\partial_t \gC)^\sharp \wedgedot \gradEintC,
\end{equation}
where $\parD{\EintC}{\gC} \in \spvecFrmB{n-1}$ denotes the gradient of $\EintC$ with respect to $\gC$, while the sharp and flat operations are with respect to $\gC$ as discussed in Remark \ref{remark:tangent_MB}.
Furthermore, using (\ref{eq:cov_vel_grad_relation},\ref{eq:vel_grad_identity_C_}), one has that
\begin{equation}\label{eq:nabV_stressC}
	\intB \nabC\vC\wedgedot\stC = \intB (\nabC\vfC)^\sharp\wedgedot\stC = \intB (\half \Lie{\vC}{\gC} - \half \extd \vfC)^\sharp\wedgedot\stC.
\end{equation}
Thus, from (\ref{eq:rate_of_strain_Lie},\ref{eq:dt_Eint}) and (\ref{eq:nabV_stressC}), in order for (\ref{eq:Eint_balance}) to hold, under the condition that $(\extd \vfC)^\sharp \wedgedot \stC = 0$, the convective stress $\stC$ should satisfy
\begin{equation}\label{eq:DE_formula_C}
	\stC = 2 \gradEintC.
\end{equation} 
Furthermore, the condition $(\extd \vfC)^\sharp \wedgedot \stC = 0$ is satisfied if
\begin{equation}\label{eq:stC_symmetry}
	(\hat{\alpha}^\sharp \otimes \hat{\beta})\wedgedot \stC = (\hat{\beta}^\sharp \otimes \hat{\alpha})\wedgedot \stC, \qquad \forall \hat{\alpha},\hat{\beta} \in \spFrmB{1}.
\end{equation} 
Equation (\ref{eq:DE_formula_C}) corresponds to the convective counterpart of the well-known Doyle-Erickson formula \cite{Kanso2007OnMechanics} and (\ref{eq:stC_symmetry}) corresponds to the usual symmetry condition on stress. 
Consequently, we have that 
\begin{equation}\label{eq:Eint_balance_C}
	\frac{\extd}{\extd t} \Eint = \intB \nabC\vC\wedgedot\stC = \intB \half (\partial_t \gC)^\sharp\wedgedot\stC= \intB \epC\wedgedot\stC,
\end{equation}
with $\epC$ denoting the convective rate of strain tensor field.

Combined with the equations of motion (\ref{eq:EoM_C_metric},\ref{eq:EoM_C_mom}), the constitutive law (\ref{eq:DE_formula_C}) provides a complete set of equations that describe the motion of the elastic body in an intrinsic manner.
Due to his work highlighting the importance of $\spMetB$ in this intrinsic formulation, we refer to the $\stC$ as the Rougee stress tensor similar\andrea{ly} to \cite{Kolev2021ObjectiveMetrics}.

\begin{remark}
	
	Note that in (\ref{eq:dt_Eint}) the wedge dot should have an alternating property analogously to the standard wedge, so care must be taken when swapping the form legs. 
	However, since we are assuming $n=3$, swapping a 1-form and a 2-form does not change the sign of the product.
\end{remark}

\subsubsection{Material}

Alternatively to (\ref{eq:Eint_balance}), one can attempt to repeat the same line of thought above for the material case.
From (\ref{eq:dt_F},\ref{eq:EnergyBal_M},\ref{eq:integ_by_parts}) and (\ref{eq:Ekin_M_balance}), one has that 
\begin{equation}
	\frac{\extd}{\extd t} \Eint = \intB \nabM\vM\wedgedot\stM = \intB D_t F\wedgedot\stM.
\end{equation}
This equation could immediately tempt one to think that the deformation gradient is a suitable state of deformation and thus one could say that
\begin{equation}
	\Eint[F] := \intB \EintM(X,F),
\end{equation}
where $\map{\EintM}{\cl{B}\times\spvecFrmPhi{1}}{\spFrmB{n}}$ is the material counterpart of $\EintC$.
Consequently, the material version of the Doyle Erickson formula would be \cite{Kanso2007OnMechanics}
\begin{equation}\label{eq:DE_formula_M}
	\stM = \gradEintM,
\end{equation} 
where $\gradEintM \in \spcovFrmPhi{n-1}$ denotes the gradient of $\EintM$ with respect to $F$.

However, the fundamental issue with this description is that unlike the convective metric $\gC$ which characterizes the deformation component of the motion, $F$ characterizes the full motion.
One way to see this is by comparing (\ref{eq:dt_F}) and (\ref{eq:gC_strain_rate}) which indicates that $\gC$ is calculated by the time integration of the symmetric part of the velocity gradient (and thus rate of strain) whereas $F$ is calculated from the full velocity gradient.
Another way to see this is from the axiom of material frame independence or objectivity \cite{Marsden1994MathematicalElasticity}.
This well known result indicates that for the internal energy $\Eint$ to be invariant under arbitrary spatial diffeomorphisms, then $\EintM$ should only depend on $\gC$ and not $F$ \cite[Th. 2.10]{Marsden1994MathematicalElasticity}.
This axiom of invariance is in fact equivalent to the factorization of rigid body motions in the principle bundle structure relating $\spC$ and $\spMetB$ which will be detailed later in Sec. \ref{sec:structures}.
Therefore, this asserts the importance of the space $\spMetB$ as the space of deformations.

\subsubsection{Spatial}
\renewcommand{\cfgPFwd}{\varphi_{*}}
Following the standard construction of \cite[Ch.3]{Marsden1994MathematicalElasticity}, 
the transition to the spatial description is achieved by considering that $\gC = \cfgPleg{}(\gS)$ can be interpreted as a function of $F$ and $\gS$ .
Consequently, one can consider $\Eint$ as the functional
\begin{equation}
	\Eint[F,\gS] := \intS \EintS(x,F,\gS),
\end{equation}
where $\map{\EintS}{\cl{S}\times\spvecFrmPhi{1}\times\cl{M}(\cl{S})}{\spFrmS{n}}$ is the spatial counterpart of $\EintC$, defined such that
$\EintS(x,F,\gS) := \cfgPFwd(\EintC(\varphi^{-1}(x),\cfgPleg{}(\gS))).$
Moreover using the chain rule, one can show that the gradient of $\EintS$ with respect to $\gS$ is given by
$$\parD{\EintS}{\gS} = \cfgPFwd\left(\parD{\EintC}{\gC}\right) \in \spvecFrmS{n-1}.$$
Using the commutative property of the pushforward operation with contractions, the spatial counterpart of (\ref{eq:DE_formula_C}) is given by
\begin{equation}\label{eq:DE_formula_S}
	\stS = \cfgPFwd \left(\stC\right) = \cfgPFwd \left(2 \gC \cdot \parD{\EintC}{\gC}\right) = 2 \cfgPFwd(\gC) \cdot \cfgPFwd\left(\parD{\EintC}{\gC}\right) = 2 \gS \cdot \parD{\EintS}{\gS} = 2 \gradEintS.
\end{equation}
Furthermore, the symmetry condition (\ref{eq:stC_symmetry}) is inherited by $\stSigC$ such that
\begin{equation}\label{eq:stS_symmetry}
	({\alpha}^\sharp \otimes{\beta})\wedgedot \stS = ({\beta}^\sharp \otimes {\alpha})\wedgedot \stS, \qquad \forall {\alpha},{\beta} \in \spFrmS{1}.
\end{equation}
Consequently, we have that 
\begin{equation}\label{eq:Eint_balance_S}
	\frac{\extd}{\extd t} \Eint = \intS \nabS\vS\wedgedot\stS = \intB \epS\wedgedot\stS,
\end{equation}
with $\epS$ denoting the spatial rate of strain tensor field.

By comparing (\ref{eq:Eint_balance_S}) to (\ref{eq:Eint_balance_C}) one can see a high resemblance which can cause a significant confusion between the two descriptions of the constitutive equations.
This deceptive resemblance could lead one to incorrectly think that the metric $\gS$ acts as the state of deformation in the spatial description.
However, $\gS$ is a fixed physical property of the ambient space and thus is neither a time-dependent state not it is the integral of the rate-of-strain tensor $\epfS = \half \Lie{\vS}{\gS}$ !
This misunderstanding can be clarified by explicating the arguments of the Doyle Erickson formula (\ref{eq:DE_formula_S}), which could be written instead as
$$\stS = 2 \gradEintS[F,\gS],$$
to emphasize that the constitutive equations depend on both $\gS$ and $F$, which is indeed the combination that corresponds to $\gC$, the true state of deformation.
Furthermore, with this explicit dependence, one can also see that the spatial constitutive equation (\ref{eq:DE_formula_S}) combined with the equations of motion (\ref{eq:EoM_S_mass}-\ref{eq:EoM_S_mom}) are not well-posed since one requires an extra evolution equation for the deformation gradient $F$.
However, this leads to a combination of the material and spatial descriptions.
This again highlights how only in the convective description can one have a complete intrinsic formulation of the constitutive equations.

\section{Relation with common formulations}
In this final section we show the relation between our proposed exterior calculus formulation and the standard ones in the literature, e.g. in \cite{Marsden1994MathematicalElasticity,Simo1988ThePlates}.
The main difference lies in the choice of using the intensive mass density and velocity variables instead of the extensive mass forms and momentum variables.
Furthermore, common formulations represent the stress as an intensive 2-rank tensor instead of the extensive stress variables we utilized.

In what follows we show how one can recover the usual intensive stress variables from $(\stC,\stM,\stS)$ and how one changes between different representations. Then, we derive the standard governing dynamical equations from the ones we presented in Sec. \ref{subsec:EoM}.

\subsection{From extensive to intensive stress}
Analogous to the relation between the momentum and velocity variables in (\ref{eq:vel_mom_relation}), we can transform the covector-valued $(n-1)$ forms $\stC,\stM,$ and $\stS$ into the following intensive stress variables:
\begin{equation}\label{eq:tau_stresses}
	\begin{split}
		\stTauC :=& \invhodgeC\ \stC \in \spvecFrmB{1} \cong \spVec{^1_1\cl{B}},\\
		\stTauM :=& \invhodgeM\ \stM \in \spvecFrmPhi{1} \cong \spVec{^*\cl{B}\otimes\varphi^*T\cl{S}},\\
		\stTauS :=& \invhodgeS\ \stS \in \spvecFrmS{1} \cong \spVec{^1_1\cl{S}},
	\end{split}
\end{equation}
which can be interpreted as vector-valued 1-forms or equivalently \textit{mixed 2-rank} tensor fields. 
The standard \textit{2-contravariant} stress tensor fields are defined using the mass densities as
\begin{equation}\label{eq:sigma_stresses}
	\begin{split}
		\stSigC :=& \mDC\gC^{-1}\cdot\stTauC \in \spVec{_0^2\cl{B}}, \\  	
		\stSigM :=& \mDM\gM^{-1}\cdot\stTauM \in \spVec{\cl{B}\otimes\varphi^*T\cl{S}}, \\
		\stSigS :=& \mDS\gS^{-1}\cdot\stTauS\in \spVec{_0^2\cl{S}},
\end{split}
\end{equation}
which are known as the convected, 1\textsuperscript{st} Piola-Kirchhoff, and Cauchy stress tensor fields, respectively.
Other common stresses are the 2\textsuperscript{nd} Piola-Kirchhoff stress $J_\varphi \stSigC\in \spVec{_0^2\cl{B}}$ and the Kirchhoff stress $(J_\varphi\circ \varphi^{-1})\stSigS\in \spVec{_0^2\cl{S}}$, where $J_\varphi\in\spFn{\cl{B}}$ is the Jacobian of $\varphi$.

In a local chart, the intensive stresses (\ref{eq:tau_stresses}) and (\ref{eq:sigma_stresses}) are expressed as
\begin{equation*}
	\stTauC = \stTauC^J_I\ E^I\otimes E_J, \qquad \qquad 
	\stTauM = \stTauM^j_I\ E^I\otimes e_j|_\varphi, \qquad \qquad 
	\stTauS = \stTauS^j_i\ e^i\otimes e_j,
\end{equation*}
\begin{equation*}
	\stSigC = \stSigC^{IJ}\ E_I\otimes E_J, \qquad \qquad 
	\stSigM = \stSigM^{Ij}\ E_I\otimes e_j|_\varphi, \qquad \qquad 
	\stSigS = \stSigS^{ij}\ e_i\otimes e_j,
\end{equation*}
with the relation of their components to those of the extensive stress variables in (\ref{eq:ext_stress_expr}) given by
\begin{align}
	\stC_{KAB} =& \stTauC^J_K \mFC_{JAB} = \gC_{KM} \stSigC^{JM}\hat{\omega}_{JAB},\nonumber\\
	\stM_{kAB} =& \gM_{kj}G^{IM}\stTauM^j_I \mFM_{MAB} = \gM_{kj} \stSigM^{Mj} \tilde{\omega}_{MAB},\nonumber\\
	\stS_{kab} =& \stTauS^j_k \mFS_{jab} = \gS_{km}\stSigS^{jm} \omega_{jab},\label{eq:int_stress_relations_compon}
\end{align}
where the indexed $\mu$-\textit{symbols} and $\omega$-\textit{symbols} denote the components of the mass and volume forms, respectively (\cf Table \ref{table:mass_vol_quantities}).

By comparing the different component functions, one can observe that the extensive stress variables $(\stC,\stM,\stS)$ possess mass and volume dependency, whereas the {mixed 2-rank} tensor fields $(\stTauC,\stTauM,\stTauS)$ do not have such dependency and thus represent pure stress information that is determined by the constitutive equations.
On the other hand, the stresses $(\stSigC,\stSigM,\stSigS)$ are also intensive similar to $(\stTauC,\stTauM,\stTauS)$, however they differ by having mass dependency through the mass densities.
A summary of the different stress representations can be found in Table \ref{table:web_of_stress}.

\begin{table}
	\centering
	{\renewcommand{\arraystretch}{1.3}
		\begin{tabular}{|m{2.5cm}|c|c|c|}
			\hline
			&	Convective & Material & Spatial\\\hline
			\rule{0pt}{4ex} Mass and volume dependency &  $\TwoVecText{\stC}{Rougee}$ & $\TwoVecText{\stM}{}$& $\TwoVecText{\stS}{}$\\\hline
			No dependency  & $\TwoVecText{\stTauC}{}$ & $\TwoVecText{\stTauM}{}$& $\TwoVecText{\stTauS}{}$\\\hline
			Mass dependency  & $\TwoVecText{\stSigC}{Convected}$ & $\TwoVecText{\stSigM}{1\textsuperscript{st} Piola-Kirchhoff}$& $\TwoVecText{\stSigS}{Cauchy}$\\\hline
		\end{tabular}
	}
	\caption{Extensive and intensive stress representations}
	\label{table:web_of_stress}
\end{table}

Another key distinction between the extensive and intensive stress variables is the way in which the spatial, convective and material representation are related to each other.
While the extensive stresses are subject to clear intrinsic pullback relations, characterized by (\ref{eq:stress_relations}), the intensive stress variables on the other hand do not have this advantage.
In particular, the transformation from the spatial to the convective representation is expressed locally as
\begin{equation}\label{eq:spatial-to-convective}
	\stTauC_I^J = F^i_I (F^{-1})_j^J \stTauS_i^j, \qquad \qquad \stSigC^{IJ} = (F^{-1})_i^I (F^{-1})_j^J \stSigS^{ij},
\end{equation}
and from the spatial to the material representation as
\begin{equation}\label{eq:spatial-to-material}
	\stTauM_I^j = G_{IM} \gM^{ij}(F^{-1})_m^M \stTauS_i^m, \qquad \qquad \stSigM^{Ij} = J_\varphi (F^{-1})_i^I \stSigS^{ij},
\end{equation}
and from the material to the convective representation as
\begin{equation}\label{eq:material-to-convective}
	\stTauC_I^J =\gC_{IK} G^{JM} (F^{-1})_j^K \stTauM_M^j, \qquad \qquad \stSigC^{IJ} = \frac{1}{J_\varphi} (F^{-1})_j^I \stSigM^{Ji} .
\end{equation}
From the above relations, one can observe that only the spatial-to-convective transformation (\ref{eq:spatial-to-convective}) is a pullback relation, i.e. one has that $\stTauC = \cfgPleg{}(\stTauS)$ and $\stSigC = \cfgPleg{}(\stSigS)$.
This is clearly not the case for (\ref{eq:spatial-to-material}) and (\ref{eq:material-to-convective}) which require, in addition to pulling-back one index, the use of the non-intrinsic quantities $G$ or $J_\varphi$.
In the literature, the relation between $\stSigM$ and $\stSigS$ in (\ref{eq:spatial-to-material}) has been known as the \textit{Piola transformation}, which can be seen to be in fact a non-intrinsic operation.
Again this is an important technical advantage of representing stress as a covector-valued pseudo-form.

\subsection{Governing dynamical equations}

\begin{theorem}[Spatial]
	The counterparts of equations (\ref{eq:EoM_S_mass},\ref{eq:EoM_S_mom}) and (\ref{eq:DE_formula_S}) in terms of the intensive variables $(\mDS,\vS,\stSigS)\in \spFn{\cl{S}}\times\spVecS\times\spVec{_0^2\cl{S}}$ are given by
	\begin{align}
		\partial_t \mDS =& - \Lie{\vS}{\mDS} -\mDS \divrS(\vS) \label{eq:EoM_S_mDS}\\
		\partial_t \vS =& - \nabS_{\vS}\vS + \frac{1}{\mDS} \divrS(\stSigS)\label{eq:EoM_S_vS}\\
		\stSigS =&  2 \mDS \parD{e}{\gS}.
	\end{align}
	Furthermore, the balance of the total energy is expressed as
	\begin{equation}
		\frac{\extd}{\extd t} \intS  \left[\half \gS(\vS,\vS) + e(x,F,\gS)\right] \mDS\vFS = \int_{\Sbound} \stSigS^\flat(\vS,n) {\varsigma}_{\gS},
	\end{equation}
	where $e(x,F,\gS) \in \spFn{\cl{S}}$ denotes the internal energy function, $\stSigS^\flat := \gS\cdot(\gS\cdot\stSigS)  \in \spVec{_2^0\cl{S}}$ denotes the 2-covariant version of the stress, while $\map{n}{\Sbound}{T\cl{S}}$ and $ {\varsigma}_{\gS}:= \mathrm{tr}(\iota_{n}\vFS) \in \spFrm{\Sbound}{2}$ denote, respectively, the unit normal vector field and the area form on the boundary $\Sbound$ induced by the spatial metric $\gS$.
\end{theorem}
\begin{proof}
	
	i) The equivalence of (\ref{eq:EoM_S_mDS}) and (\ref{eq:EoM_S_mass}) is standard and mentioned earlier in the conservation of mass Proposition \ref{prop:cons_mass}.
	
	ii)
	The balance of momentum (\ref{eq:EoM_S_vS}) is derived from (\ref{eq:EoM_S_mom}) such that 
	$$\invhodgeS(\partial_t \momS) = \invhodgeS(- \Lie{\vS}{\momS} + \mFS\otimes\half\extd \iota_{\vS}\vfS) + \invhodgeS\extcdS\hodgeS (\stTauS).$$
	First, using the Leibniz property of the time derivative, we can write
	$$\partial_t \momS = \partial_t(\mFS\otimes\vfS) = \mFS\otimes\gS\cdot (\partial_t \vS) + \partial_t\mFS\otimes\vfS. $$
	Second, using the Leibniz property of the Lie derivative and identity (\ref{eq:vel_grad_identity_S}) we have that
	$$\Lie{\vS}{\momS} = \Lie{\vS}{(\mFS\otimes\vfS)} = \mFS\otimes\Lie{\vS}{\vfS} + \Lie{\vS}{\mFS}\otimes\vfS = \mFS\otimes(\nabS_\vS \vfS + \half \extd \iota_{\vS}\vfS) + \Lie{\vS}{\mFS}\otimes\vfS.$$
	By combining the above two equations with the conservation of mass (\ref{eq:EoM_S_mass}), one has that
	$$\invhodgeS\left(\partial_t \momS + \Lie{\vS}{\momS} - \mFS\otimes\half\extd \iota_{\vS}\vfS\right) = \partial_t \vS + \nabS_\vS\vS$$
	
	The exterior derivative of $\stS$ is the covector-valued top-form expressed locally as
	$$\extcdS \stS = \left(\frac{3}{3!} \stS_{k[ab;c]} e^a\wedge e^b \wedge e^c\right)\otimes e^k,$$
	similar to (\ref{eq:ext_cov_stress_express}), where the semicolon is used as the (standard) shorthand notation for covariant differentiation with respect to $\nabS$.
	Using the relations (\ref{eq:int_stress_relations_compon}), we have that
	$$3 \stS_{k[ab;c]} = 3 (\stTauS^m_k\mFS_{m[ab})_{;c]} = 3 (\gS_{kj}\stSigS^{jm}\omega_{m[ab})_{;c]}.$$
	Using the covariant derivative properties $\nabS\gS = 0$ and $\nabS \vFS = 0$ and the total anti-symmetry of the 3-form $\vFS$, we have that
	$$3 \stS_{k[ab;c]} =3 \gS_{kj}(\stSigS^{jm})_{;[c}\omega_{ab]m} = \gS_{kj}(\stSigS^{jm})_{;m}\omega_{abc} = \gS_{kj}\frac{1}{\mDC}(\stSigS^{jm})_{;m}\mFS_{abc}$$
	Thus, we have that the exterior derivative of $\stS$ to be equivalent to
	$$\extcdS \stS = \extcdS\hodgeS (\stTauS) 
	=\gS_{kj} \frac{1}{\mDS} (\stSigS^{jm})_{;m} \mFS \otimes e^k 
	= \mFS \otimes \frac{1}{\mDS} \left(\gS\cdot \divrS(\stSigS)\right),$$
	and consequently $\invhodgeS\extcdS\hodgeS (\stTauS) = \frac{1}{\mDS} \divrS(\stSigS).$
	This concludes the derivation of (\ref{eq:EoM_S_vS}) from (\ref{eq:EoM_S_mom}).
	
	iii) In standard formulation, the internal energy function is usually expressed as $\EintS(X,F,\gS) = e(X,F,\gS) \mFS,$ with $\map{e}{\cl{S}\times\spvecFrmPhi{1}\times\cl{M}(\cl{S})}{\spFn{\cl{S}}}$ being a scalar function. The gradient of $e$ with respect to $\gS$ is the 2-contravariant tensor field $\parD{e}{\gS} \in \spVec{^2_0\cl{S}}$, and one can interpret its variant $\left(\parD{e}{\gS}\right)^\flat \in \spVec{^1_1\cl{S}} \cong \spvecFrmS{1}$ as a vector-valued 1-form. Consequently, it is related to the gradient of $\EintS$ with respect to $\gS$ by
	$$\gradEintS = \hodgeS \left(\parD{e}{\gS}\right)^\flat \in \spcovFrmS{2}.$$
	In local coordinates, this is represented as
	$$\left(\parD{\EintS}{\gS}\right)^j_{ab} = \left(\parD{e}{\gS}\right)^{jm} \mu_{mab}.$$
	Using (\ref{eq:DE_formula_S}) and $\mFS = \mDS \vFS$, then one has that
	$$\stS_{kab} = 2 \gS_{kj} \left(\parD{\EintS}{\gS}\right)^j_{ab} = 2 \gS_{kj}\left(\parD{e}{\gS}\right)^{jm} \mDS\omega_{mab}.$$
	By comparison to (\ref{eq:int_stress_relations_compon}), one gets $\stSigS^{jm} =  2 \mDS \left(\parD{e}{\gS}\right)^{jm}$.
	
	iv) From the definition of the Hodge star operator (\ref{eq:def_Hodge_star}), one can see that
	$$\invhodgeS\momS \wedgedot \momS = \vS \wedgedot \momS = \vS \wedgedot \hodgeS \vS = \gS(\vS,\vS) \mFS.$$
	Using the definition of the wedge-dot operation and the fact that the boundary normal is a unit vector, i.e. $\gS_{mj} n^j n^l = \delta_m^l$,
	the stress power $\Pst$ can be expressed as
	\begin{align*}
		\tr(\vS\wedgedot\stS) &= \half\tr(\vS^k\stTauS^m_k\mFS_{mab} {e^a\wedge e^b})= \half\tr(\vS^k\stTauS^m_k\delta_m^l\mFS_{lab}{e^a\wedge e^b})\\
		&= \half\tr(\vS^k\stTauS^m_k \gS_{mj} n^j n^l\mFS_{lab}{e^a\wedge e^b})= \tr(\mDS\vS^k\stTauS^m_k\gS_{mj} n^j) \tr(\half n^l {\omega}_{lab}{e^a\wedge e^b})\\
		&= \tr(\gS_{mj}\gS_{kl}\stSigS^{ml}\vS^k n^j) \tr(\iota_{n}\omega_{\gS})	= \tr(\stSigS^\flat(\vS,n)) \varsigma_{\gS},
	\end{align*}
which concludes the proof.
\qed
\end{proof}

\begin{theorem}[Convective]
	The counterparts of equations (\ref{eq:EoM_C_metric},\ref{eq:EoM_C_mom}) and (\ref{eq:DE_formula_C}) in terms of the intensive variables $(\mDC,\vC,\stSigC)\in \spFn{\cl{B}}\times\spVecB\times\spVec{_0^2\cl{B}}$ are given by
	\begin{align}
		\partial_t \gC =& \Lie{\vC}{\gC} \label{eq:gC_dot}\\
		\partial_t \mDC =& - \mDC \divrC(\vC) \label{eq:mDC_dot}\\
		\partial_t \vC =& - \nabC_{\vC}\vC + \frac{1}{\mDC} \divrC(\stSigC)\label{eq:EoM_C_vC}\\
		\stSigC =& 2 \mDC \parD{\hat{e}}{\gC} \label{eq:DE_C_std}
	\end{align}
	Furthermore, the balance of the total energy is expressed as
	\begin{equation}
		\frac{\extd}{\extd t} \intB  \left[\half \gC(\vC,\vC) + \hat{e}(X,\gC)\right] \mDC\vFC = \int_{\Bbound} \stSigC^\flat(\vC,N) \hat{\varsigma}_{\gC},
	\end{equation}
	where $\hat{e}(X,\gC)\in\spFn{\cl{B}}$ denotes the internal energy function,
	$\stSigC^\flat := \gC\cdot(\gC\cdot\stSigC)  \in \spVec{_2^0\cl{B}}$ denotes the 2-covariant version of the stress, while $\map{N}{\Bbound}{T\cl{B}}$ and $\hat{\varsigma}_{\gC} := \tr(\iota_{N}\vFC) \in \spFrm{\Bbound}{2}$ denote, respectively, the unit normal vector field and the area form on the boundary $\Bbound$ induced by the convective metric $\gC$.
\end{theorem}
\begin{proof}
	
	i) The counterpart of (\ref{eq:mDC_dot}) in the extensive representation is $\partial_t \mFC =0$ as presented in Proposition \ref{prop:cons_mass}.
	
	
	ii)	The balance of momentum (\ref{eq:EoM_C_vC}) is derived from (\ref{eq:EoM_C_mom}) such that  
	$$\invhodgeC(\partial_t \momC) = \invhodgeC(\half  \mFC\otimes\extd \iota_{\vC} \vfC) + \invhodgeC\extcdC\hodgeC (\stTauC).$$
	Using (\ref{eq:gC_dot}) and the identities (\ref{eq:vel_grad_identity_C}) and $\Lie{\vC}{\gC} \cdot \vC = \Lie{\vC}{\vfC}$, one can show that
	$$\partial_t (\gC) \cdot \vC = \nabC_{\vC} \vfC + \half \extd \iota_{\vC} \vfC,$$
	and consequently
	$$\partial_t(\gC\cdot \vC) = \gC\cdot\partial_t\vC + \partial_t (\gC) \cdot \vC  =  \gC\cdot\partial_t\vC + \nabC_{\vC} \vfC + \half \extd \iota_{\vC} \vfC.$$
	Using the time independence property of $\mFC$ and (\ref{eq:cov_vel_grad_relation}), we have that
	\begin{align*}
		\invhodgeC(\partial_t \momC) = \invhodgeC (\mFC\otimes\partial_t(\gC\cdot \vC )) 
		=& \partial_t\vC + \gC^{-1}\cdot(\nabC_{\vC} \vfC + \half \extd \iota_{\vC} \vfC) \\
		=& \partial_t\vC + \nabC_{\vC} \vC +  \gC^{-1}\cdot(\half \extd \iota_{\vC} \vfC).
	\end{align*}
	Using the fact that $\invhodgeC(\half  \mFC\otimes\extd \iota_{\vC} \vfC) = \gC^{-1}\cdot(\half \extd \iota_{\vC} \vfC)$, we thus have that 
	$$\invhodgeC(\partial_t \momC- \half  \mFC\otimes\extd \iota_{\vC} \vfC) = \partial_t\vC + \nabC_{\vC} \vC.$$
	
	Since $\nabC\gC = 0$ and $\nabC \vFC = 0$, one can show, analogously to the spatial case in the previous theorem that the exterior derivative of the convective stress $\stC$ is equal to
	 $$\extcdC \stC = \extcdC\hodgeC (\stTauC) 
	 = \gC_{KJ} \frac{1}{\mDC} (\stSigC^{JM})_{;M} \mFC \otimes E^K 
	 = \mFC \otimes \frac{1}{\mDC} \left(\gC\cdot \divrC(\stSigC)\right),$$
	 where the semicolon is used as the shorthand notation for covariant differentiation with respect to $\nabC$.
	 Thus, we have that $\invhodgeC\extcdC\hodgeC (\stTauC) = \frac{1}{\mDC} \divrC(\stSigC).$
	 This concludes the derivation of (\ref{eq:EoM_C_vC}) from (\ref{eq:EoM_C_mom}).


	iii) The derivation of (\ref{eq:DE_C_std}) from (\ref{eq:DE_formula_C}) is identical to the spatial case in Theorem~1 starting from $\EintC(X,\gC) = \hat{e}(X,\gC) \mFC$.
	
	iv) Finally analogously to the spatial case, using the fact that the normal is a unit vector, i.e. $\gC_{MJ} N^J N^L = \delta_M^L$, one has that 
	$$\tr(\vC\wedgedot\stC) = tr(\gC_{MJ}\gC_{KL}\stSigC^{ML}\vC^K N^J) \tr(\iota_{N}\hat{\omega}_{\gC})	= \tr(\stSigC^\flat(\vC,N)) \varsigma_{\gC},$$
	which concludes the proof.
	\qed
\end{proof}

\begin{theorem}[Material]
	The counterparts of equations (\ref{eq:EoM_M_cfg},\ref{eq:EoM_M_mom}) and (\ref{eq:DE_formula_M}) in terms of the intensive variables $(\vM,\stSigM)\in \spVecPhi\times\spVec{\cl{B}\otimes\varphi^*T\cl{S}}$ are given by
	\begin{align}
		\partial_t \varphi =& \vM\\
		D_t F =& \nabM \vM\\
		D_t \vM =& \frac{1}{\mDM} \divrM(\stSigM)\label{eq:EoM_M_vM}\\
		\stSigM =& \mDM\ {\gM}^{-1}\cdot\parD{\tilde{e}}{F} .
	\end{align}
	Furthermore, the balance of the total energy is expressed as
	\begin{equation}
		\frac{\extd}{\extd t} \intB  \left[\half \gM(\vM,\vM) + \tilde{e}(X,F)\right] \mDM \omega_G = \int_{\Bbound} \stSigM^\flat(\vM,\tilde{n}) \tilde{\varsigma}_{G},
	\end{equation}
	where $\tilde{e}(X,F) \in \spFn{\cl{B}}$ denotes the internal energy function,
	$\stSigM^\flat := \gM\cdot(G\cdot\stSigM)  \in \spVec{^*\cl{B}\otimes\varphi^*T^*\cl{S}}$ denotes the 2-covariant version of the stress, while $\map{N}{\Bbound}{T\cl{B}}$ and $\tilde{\varsigma}_{G} := \mathrm{tr}(\iota_{N} \omega_G) \in \spFrm{\Bbound}{2}$ denote, respectively, the unit normal vector field and the area form on the boundary $\Bbound$ induced  by the reference metric $G$.
\end{theorem}
\begin{proof}
	i) 
	Since we can identify $\momM$ by $\mFM\otimes\vfM$, then one can use the time-independence of $\mFM$ to write the momentum balance (\ref{eq:EoM_M_mom}) as
	$$D_t \momM = \mFM\otimes\gM \cdot(D_t \vM).$$
	Thus, it follows immediately that $\invhodgeM(D_t \momM) = D_t \vM$.

	The exterior derivative of $\stM$ is the covector-valued top-form expressed locally as
	$$\extcdM \stM = \left(\frac{3}{3!} \stM_{k[AB;C]} E^A\wedge E^B \wedge E^C\right)\otimes e^k|_\varphi,$$
	where the semicolon is used as the shorthand notation for covariant differentiation with respect to $\nabM$.
	Using the relations (\ref{eq:int_stress_relations_compon}), we have that
	$$3 \stM_{k[AB;C]}= 3 (\gM_{km} G^{JM} \stTauM^m_J\mFM_{M[AB})_{;C]} = 3 (\gM_{km} \stSigM^{mM}\tilde{\omega}_{M[AB})_{;C]}.$$
	Recall from Sec. \ref{sec:coord_kinematics} that $\nabM\gM =0$. Furthermore, one has that $\nabM \vFM =0$ since $\nabM$ was constructed using the connection coefficients of $G$ and thus it is compatible with the volume form induced by $G$ (\cf Remark \ref{remark:material_metric}).
	Consequently, similar to the spatial and convective cases, one has that
	$$3 \stM_{k[AB;C]}= \gM_{km} (\stSigM^{mM})_{;M} \tilde{\omega}_{ABC}.$$
	Thus, we have that the exterior derivative of $\stM$ to be equivalent to
	$$\extcdM \stM = \extcdM\hodgeM (\stTauM) = \mFM \otimes \frac{1}{\mDM} \left(\gM\cdot \divrM(\stSigM)\right),$$
	and consequently $\invhodgeM\extcdM\hodgeM (\stTauM) = \frac{1}{\mDM} \divrM(\stSigM).$
	This concludes the derivation of (\ref{eq:EoM_M_vM}) from (\ref{eq:EoM_M_mom}).

	iii) Following the same line of thought as in the proof of the spatial case, we have that $\EintM(X,F) = \tilde{e}(X,F) \mFM$. The gradient of $\tilde{e}$ with respect to $F$, denoted by $\parD{\tilde{e}}{F} \in \spVec{\cl{B}\otimes \varphi^*T^*\cl{S}}$, is related to the gradient of $\EintM$ with respect to $F$ by
	$$ \gradEintM_{kAB}= \left(\parD{\tilde{e}}{F}\right)^M_k \tilde{\mu}_{MAB}.$$
	Consequently, by comparison to (\ref{eq:int_stress_relations_compon}) one gets $\stSigM^{Mj} =  \gM^{jk}\mDM \left(\parD{\tilde{e}}{F}\right)^M_k$.
	
	iv) Finally analogously to the spatial and convective cases, using the fact that the normal is a unit vector, i.e. $G_{MJ} N^J N^L = \delta_M^L$, one has that 
	\begin{align*}
		\tr(\vM\wedgedot\stM) &= \half\tr(\vM^k\gM_{ki}G^{LM}\stTauM^i_L\mFM_{MAB} {E^A\wedge E^B})=
		\half\tr(\stSigM^{iM}\vM^k \gM_{ki} \tilde{\omega}_{MAB} {E^A\wedge E^B})\\
		&= \half\tr(\gM_{ki}\stSigM^{iM}\vM^k G_{MJ} N^J N^K\tilde{\omega}_{KAB} {E^A\wedge E^B})\\
		&= \tr(\gM_{ki}G_{MJ}\stSigM^{iM}\vM^k N^J) \tr(\half N^K\tilde{\omega}_{KAB} {E^A\wedge E^B})\\
		&= \tr(\stSigM^\flat(\vM,N)) \tr(\iota_{N}\vFM)	= \tr(\stSigM^\flat(\vM,N)) \tilde{\varsigma}_{G},
	\end{align*}
	which concludes the proof.
\qed
\end{proof}

\begin{remark}[\textbf{Intrinsicality of the material representation}]\label{remark:intrinsic_material}
	
	A very important distinction should be made between the extensive momentum balance (\ref{eq:EoM_M_mom}) and the intensive momentum balance (\ref{eq:EoM_M_vM}). Both equations are in fact a statement of Newton's second law, i.e. the rate of change of momentum are equal to forces due to stress. These forces are computed by spatial differentiation of the stress tensor, represented by the exterior covariant derivative $\extcdM$ in (\ref{eq:EoM_M_mom}) and the divergence operator $\divrM$ in (\ref{eq:EoM_M_vM}). However, a key difference is that the extensive representation (\ref{eq:EoM_M_mom}) is intrinsic and independent of any reference configuration while (\ref{eq:EoM_M_vM}) is not.
	
	The non-intrinsicality of (\ref{eq:EoM_M_vM}) can be seen from the expression of $\divrM$ constructed from the covariant differential $\nabM\stSigM$ in (\ref{eq:cov_der_def_mat_tensor}) which uses the Levi-Civita connection of the reference metric $G$ as explained in Remark \ref{remark:material_metric}.
	On the other hand, the exterior covariant derivative $\extcdM$ is constructed by anti-symmetrization of the covariant derivative $\nabM$ applied to a collection of covector fields, as explained in Remark \ref{remark:extd_coord}, which does not require a reference metric (\cf the coordinate expression of $\nabM$ applied to a covector field in Table \ref{table:motion_kinematics} compared to $\nabM$ applied to a second rank tensor field in (\ref{eq:coord_exp_cov_M_P})).
	
	This non-intrinsicality is a consequence of the fundamental difference between the extensive stress tensor $\stM$ and its intensive counterpart $\stSigM$ as depicted in (\ref{eq:int_stress_relations_compon}).
	The extensive mass form $\mFM$ incorporated in the definition of $\stSigM$ is intrinsically defined for the body manifold. On the other hand, $\stSigM$ is constructed from the intensive mass density $\mDM$, which is depends on the reference configuration chosen.
\end{remark}

\section{Underlying structures}\label{sec:structures}
In this section, we conclude by emphasizing two types of structures underlying the theory of nonlinear elasticity. Namely, 1) the principle bundle structure relating the configuration space $\mathscr{C}$ to the deformation space $\spMetB$, and 2) the de Rham complex structure relating the spaces of bundle-valued forms to each other.

\subsection{Principal fiber bundle structure of $\spC$ and $\spMetB$}
Recall from Sec. \ref{sec:deformation} the map $\map{\pi_g}{\mathscr{C}}{\spMetB}$ that associates to any configuration $\cfgT$ a Riemannian metric on $\cl{B}$.
The curve $c_{\varphi}:t\mapsto \cfgT$ in $\mathscr{C}$ characterizing the motion of the body induces a curve $c_{\gC}:t\mapsto \gC_t = \pi_g(\cfgT)$ in $\spMetB$.

A key distinction should be made between the curve $c_{\varphi}$ on the configuration space $\mathscr{C}$ and its image $c_{\gC} =\pi_g(c_{\varphi})$ on the deformation space $\spMetB$.
While $c_{\varphi}$ represents a motion of the body in the ambient space that consists of both rigid body motion (i.e. simultaneous translation and rotation) and deformation, the curve $c_{\gC}$ represents only the deformation component of the motion.
This can be seen from the fact that $\pi_g$ is a projection map that is not injective.
Let $\eta \in \spIsomA$ be an isometry of the ambient space, i.e.
$\map{\eta}{\mathscr{A}}{\mathscr{A}} , \mathrm{such\ that}, \eta^*\gS = \gS,$
which represents physically a rigid body motion.
If any two configurations $\varphi_1,\varphi_2 \in \mathscr{C}$ are related by $\varphi_2 = \eta \circ \varphi_1$ (equivalently $\varphi_1 = \eta^{-1} \circ \varphi_2$), then we have that
$$\pi_g(\varphi_2) = \varphi_2^*(\gS) = (\eta \circ \varphi_1)^*\gS = \varphi_1^*(\eta^*\gS) = \varphi_1^*(\gS) = \pi_g(\varphi_1).$$
In this case, $\varphi_2$ represents a superposed rigid body motion of the configuration $\varphi_1$.

As shown in \cite{Stramigioli2022TheMechanics}, this observation is a consequence of a geometric structure that underlies the triplet $(\mathscr{C},\pi_g,\spMetB)$, namely a \textit{principal fiber bundle} structure which can be formulated as follows.
Let $G:=\spIsomA$ denote the group of isometries on $\mathscr{A}$ with the group operator being composition, which can be shown to be a Lie group.
Consider the right action of $G$ on the configuration space $\mathscr{C}$ defined by:
\begin{equation}
	\fullmap{\triangleleft}{\mathscr{C}\times G}{\mathscr{C}}{(\varphi,\eta)}{\varphi\triangleleft\eta:= \eta^{-1}\circ \varphi.}
\end{equation}
One can check that $\triangleleft$ is a free $G$-action since $\forall \varphi\in\mathscr{C}$ the only rigid body motion that does not change $\varphi$ is the identity of $G$.

For any $\varphi\in\mathscr{C}$, we define its orbit under the action $\triangleleft$ as the set
$$\mathscr{O}_{\varphi} := \{ \bar{\varphi} \in \mathscr{C}\ |\ \exists \eta \in G : \varphi \triangleleft \eta = \bar{\varphi}\}.$$
Let $\varphi \sim \bar{\varphi}$ be equivalent if $\bar{\varphi} \in \mathscr{O}_{\varphi}$. It is straightforward to show that this is an equivalence relation (i.e. reflexive, symmetric, and transitive).
The orbit space of $\mathscr{C}$ is defined as the quotient space $\mathscr{C}/G :\{\mathscr{O}_{\varphi} | \varphi \in \mathscr{C}\}$ where each element $\mathscr{O}_{\varphi} = [\varphi]$ is an equivalence class of $\varphi$.
Intuitively, the set $\mathscr{O}_{\varphi}$ consists of all configurations $\bar{\varphi}$ that are rigid body motions of $\varphi$ and thus they all have the same \enquote{shape}.
Consequently, each orbit $\mathscr{O}_{\varphi}$ is associated to one deformation state $\gC_{\varphi} := \pi_g(\varphi) \in \spMetB$ in a bijective manner.
Thus, we have that the map
\begin{equation*}
	\fullmap{\zeta}{\spMetB}{\mathscr{C}/G}{\gC_{\varphi}}{[\varphi],}
\end{equation*}
is a diffeomorphism, which in addition satisfies $\zeta \circ \pi_g = \pi$, where $\map{\pi}{\mathscr{C}}{\mathscr{C}/G }$ denotes the canonical projection map that maps a member of the set $\mathscr{C}$ to its equivalence class. 
Therefore, this proves that the bundle $\map{\pi_g}{\mathscr{C}}{\spMetB}$ is a principal fiber $G$-bundle.

\begin{figure}
	\centering
	\includegraphics[width=0.7 \textwidth]{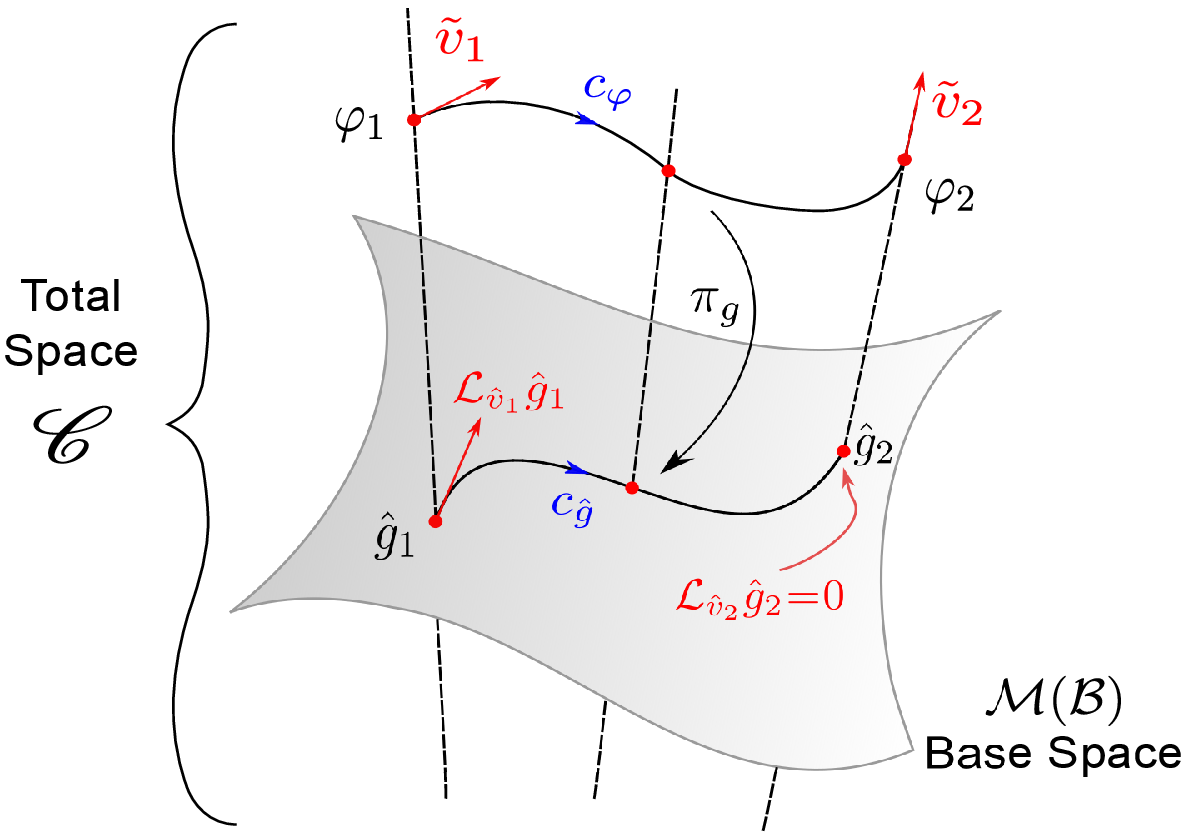}
	\caption{Principal fiber bundle structure of $\map{\pi_g}{\mathscr{C}}{\spMetB}$.}
	\label{fig:princ_fiber_bundle}
\end{figure}

A consequence of this extra bundle structure that relates the right $G$-space $\mathscr{C}$ and the space of deformations $\spMetB$ is that (locally) at any point $\gC\in\spMetB$ is attached a fiber that is isomorphic to $G$. This follows from the fact that $\pi ^{-1}(\mathscr{O}_{\varphi}) = \mathscr{O}_{\varphi} \cong G$ since $\triangleleft$ is a free $G$-action.
This principal bundle structure, depicted in Fig.~\ref{fig:princ_fiber_bundle}, is the formalization of how the full motion of the elastic body is represented by the curve $c_\varphi$ in the total space $\mathscr{C}$, whereas its projection under $\pi_g$ to the base space $\spMetB$ corresponds to the deformation component and motion along the fibers corresponds to the rigid body component.

This motion along the fibers can be formulated geometrically for the curve $\map{c_\varphi}{t}{\cfgT}$ as follows.
Consider the tangent vector $\vM_t \in T_{\cfgT}\mathscr{C}$ to the curve and its induced spatial vector field $\vS_t \in \spVecS$ defined in (\ref{eq:vel_relations}). 
Let the time dependent flow generated by $\vS_t$ be defined as $\map{\psi_{t,s}:= \cfgT\circ \varphi_s{\scriptstyle^{-1}}}{\varphi_s(\cl{B})}{\varphi_t(\cl{B})}$.
If each map $\psi_{t,s}$ is an isometry (i.e. an element of $G$) then $\vS_t$ is generated by an element of the Lie algebra of $G$ and is called a Killing vector field satisfying $\Lie{\vS_t}{\gS} = 0$.
In this case, we have from (\ref{eq:rate_of_strain_Lie})  that 
$$\Lie{\vS_t}{\gS} = 0 \implies \Lie{\vC_t}{\gC_t} = \partial_t \gC_t  =0.$$
In other words, if the tangent vector $\vM_t \in T_{\cfgT}\mathscr{C}$ in the total space is purely along a fiber, then its pushforward under the map $\pi_g$ would be zero and thus would correspond to a pure rigid body motion (\cf point $\varphi_2$ in Fig. \ref{fig:princ_fiber_bundle}).

If a connection is given on the principal $G$-bundle $\map{\pi_g}{\mathscr{C}}{\spMetB}$, one could uniquely decompose the tangent space $T_{\varphi}\mathscr{C}$ as
$T_{\varphi}\mathscr{C} = V_{\varphi}\mathscr{C} \oplus H_{\varphi}\mathscr{C},$
where the vertical subspace $V_{\varphi}\mathscr{C}$ is defined as the kernel of the tangent map of $\pi_g$ and represents the rigid body motion, while the horizontal subspace $H_{\varphi}\mathscr{C}$ would correspond to the pure deformation.
Based on this construction, one can define parallel transport on $\mathscr{C}$ and covariant differentiation on $\spMetB$.
Using the Riemannian structure of $\spMetB$, one has a unique (Levi-Civita) connection that allows such operations.
Consequently, one can define geodesics between any two states of deformation on $\spMetB$ and by doing so obtain a consistent definition of strain \cite{Fiala2011GeometricalMechanics,Fiala2016GeometryAnalysis}.

\begin{remark}
	We emphasize that the principal bundle structure $\map{\pi_g}{\mathscr{C}}{\spMetB}$ has been overlooked in the literature. Instead, the starting point in \cite{Rougee2006AnStrain,Fiala2011GeometricalMechanics,Fiala2016GeometryAnalysis} was the Levi-Civita connection on $\spMetB$. However, it would be interesting to investigate in the future starting from a connection one-form on the principal $G$-bundle and explore the insight it could provide for developing geometric time integration schemes, similar to \cite{Fiala2016GeometryAnalysis}. Furthermore, it would be interesting to explore the relation to the screw-theory formulation presented in \cite{Stramigioli2022TheMechanics}.
\end{remark}

\subsection{De Rham complex structure}

One key advantage of formulating nonlinear elasticity using bundle-valued forms is that it highlights its underlying complex structure which is fundamental to many analytical and computational tools. On {the} one hand, such structure provides valuable information for solving and analyzing PDEs by linking its topological and geometric properties.
On the other hand, by extending this structure to general Hilbert complexes using Sobolev spaces, one can identify suitable solution spaces for mixed finite-element formulations of nonlinear elasticity leading to stable structure-preserving numerical schemes \cite{FaghihShojaei2019Compatible-strainElasticity}.
In what follows, we present the vector-valued de Rham complexes for the spatial, material and convective representations. The same constructions can be trivially also extended to covector-valued forms.

\begin{figure}
	\centering
	\begin{adjustbox}{width=\textwidth}
		\begin{tikzcd}[row sep=normal, column sep=tiny,labels={font=\everymath\expandafter{\the\everymath\textstyle}}]
			& \spcovFrmB{3-k} \arrow[from=dl,"\hodgeC"red] \arrow[from=rr, "\cfgPleg{v}"{red}] \arrow[from=dd,"\extcdC^k"{near start , red}] & & \spcovFrmPhi{3-k} \arrow[from=dl,"\hodgeM"red] \arrow[from=rr, "\cfgPleg{f}"{red}] \arrow[from=dd,"\extcdM^k"{near start , red}] & & \spcovFrmS{3-k} \arrow[from=dd,"\extcdS^k"{red}]\arrow[from=dl,"\hodgeS"red]\\
			\spvecFrmB{k} \arrow[from=rr, crossing over, "\cfgPleg{v}"{near start , red}] \arrow[dd,"\extcdC^k"{red}] & & \spvecFrmPhi{k} \arrow[from=rr, crossing over, "\cfgPleg{f}"{near start , red}] & & \spvecFrmS{k} \\
			& \spcovFrmB{2-k} \arrow[from=dl,"\hodgeC" red] \arrow[from=rr, "\cfgPleg{v}"{near end , red}] & & \spcovFrmPhi{2-k} \arrow[from=dl,"\hodgeM" red] \arrow[from=rr, "\cfgPleg{f}"{near end , red}] & & \spcovFrmS{2-k} \arrow[from=dl,"\hodgeS" red]\\
			\spvecFrmB{k+1} \arrow[from=rr, "\cfgPleg{v}"red] & & \spvecFrmPhi{k+1} \arrow[from=uu,"\extcdM^k"{near start , red}, crossing over] \arrow[from=rr, "\cfgPleg{f}"{red}] & & \spvecFrmS{k+1}\arrow[from=uu,"\extcdS^k"{near start , red}, crossing over]\\
		\end{tikzcd}
	\end{adjustbox}
	\caption{Double de Rham complexes in three-dimensional space}
	\label{fig:double_deRham}
\end{figure}
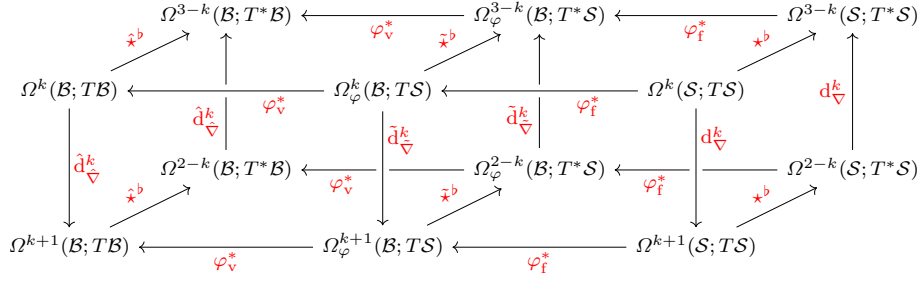

The sequence of vector spaces $\spvecFrmS{k}$ along with the differential operators $\extcdS^k$ comprise what is known in algebraic topology as a \textit{co-chain complex}, provided that $\extcdS^k \circ \extcdS^{k-1} = 0$.
This condition is satisfied if and only if the ambient space $\mathscr{A}$ is flat i.e. with no intrinsic curvature. This can be seen for the spatial exterior covariant derivative since for any $\alpha\in \spvecFrmS{0}$ and $\beta\in \spvecFrmS{1}$ we have that \cite{Angoshtari2013GeometricElasticity}
\begin{align*}
	(\extcdS^1\circ\extcdS^0(\alpha))(u_0,u_1) &= \cl{R}(u_0,u_1) \alpha,\\
	(\extcdS^2\circ\extcdS^1(\beta))(u_0,u_1,u_2) &= \cl{R}(u_0,u_1) \beta(u_2) - \cl{R}(u_0,u_2) \beta(u_1) + \cl{R}(u_1,u_2) \beta(u_0),
\end{align*} 
where $\cl{R}$ denotes the curvature tensor of the connection $\nabS$ on $\cl{S}\subset \mathscr{A}$.
Hence, $\extcdS$ is a differential operator that satisfies $\extcdS^k \circ \extcdS^{k-1} =0$ if and only if $\mathscr{A}$ is a flat space with $\cl{R}=0$.
In addition, since the convective connection $\nabC$ was induced by $\nabS$ through the map $\cfgT$, one can show that the curvature of $\nabC$ is given by $\hat{\cl{R}} := \cfgP\cl{R}$ \cite{Angoshtari2013GeometricElasticity}.
Thus, the flatness of $\mathscr{A}$ implies also that the convective exterior covariant derivative satisfies $\extcdC^k \circ \extcdC^{k-1} =0$, where the same conclusion holds as well for the material exterior covariant derivative \cite{Angoshtari2013GeometricElasticity}.
Consequently, each of the pairs $(\spvecFrmS{k}, \extcdS^k)$, $(\spvecFrmB{k}, \extcdC^k)$ and  $(\spvecFrmPhi{k}, \extcdM^k)$ gives rise to a bundle-valued de Rham complex.
We shall refer to these three complexes as the \textit{spatial, convective,} and \textit{material de Rham complex}, respectively.
All of the aforementioned complex structure for vector-valued forms is also applicable to vector-valued pseudo-forms using the Hodge star operators (\ref{eq:hodge_stars}), as depicted in Fig. \ref{fig:double_deRham}.

Recall the commutative property of the exterior covariant derivative with pullbacks (\ref{eq:comm_prop_ext_cov_derv})
which is depicted in Fig. \ref{fig:deRham_complexes_comm_diag} showing the spatial, material and convective de Rham complexes.
This key commutative property indicates that $\cfgPleg{v}$ and $\cfgPleg{f}$ are in fact complex isomorphisms. Thus, this indicates that any analytical result holding for one complex, e.g. its de Rham cohomology groups, should have a counterpart in the other ones. Additionally, as will be seen later, one can unify the three representations of the governing equations of nonlinear elasticity and change from one to the other in an elegant manner.
This shows again the technical advantage of using bundle-valued forms to mathematically represent nonlinear elasticity and continuum mechanics in general.

\begin{figure}
	\centering
	\begin{tikzcd}[column sep=scriptsize]
		0  \arrow[r,sep=small] 	& \spvecFrmS{0}  \arrow[r,"\extcdS^0"] \arrow[d, "\cfgPleg{f}"] & \spvecFrmS{1}  \arrow[r,"\extcdS^1"] \arrow[d, "\cfgPleg{f}"] & \spvecFrmS{2}  \arrow[r,"\extcdS^2"] \arrow[d, "\cfgPleg{f}"] & \spvecFrmS{3}  \arrow[r] \arrow[d, "\cfgPleg{f}"]	& 0 & {\color{red}\mathrm{(spatial)} }\\
		0  \arrow[r] 	& \spvecFrmPhi{0}   \arrow[r,"\extcdM^0"] \arrow[d, "\cfgPleg{v}"] & \spvecFrmPhi{1}   \arrow[r,"\extcdM^1"] \arrow[d, "\cfgPleg{v}"]  & \spvecFrmPhi{2}   \arrow[r,"\extcdM^2"] \arrow[d, "\cfgPleg{v}"] 	& \spvecFrmPhi{3}   \arrow[r] \arrow[d, "\cfgPleg{v}"]  & 0 & {\color{red}\mathrm{(material)}}\\
		0  \arrow[r] 	& \spvecFrmB{0}  \arrow[r,"\extcdC^0"]  & \spvecFrmB{1}  \arrow[r,"\extcdC^1"] & \spvecFrmB{2}  \arrow[r,"\extcdC^2"]&  \spvecFrmB{3}  \arrow[r] 	& 0 & {\color{red}\mathrm{(convective)}}
	\end{tikzcd}
	\caption{Commutative diagram of the spatial, material and convective de Rham complexes shown from top to bottom, respectively.}
	\label{fig:deRham_complexes_comm_diag}
\end{figure}


\section{Conclusion}
We presented in this paper a formulation of nonlinear elasticity using vector-valued and covector-valued differential forms.
All three representations of the motion have been considered and the transformation of all physical variables from one representation to the other has been geometrically identified.
It has been emphasized throughout the paper how an identification of the body with a reference configuration in the ambient space is not needed for describing the equations of motion, unless one represents the momentum and stress as thermodynamically intensive variables in the material representation.
The underlying de Rham complex and principle bundle structure relating the configuration space $\spC$ to the space of Riemannian metrics on the body manifold $\spMetB$ have been highlighted, which emphasize the significance of $\spMetB$ as an intrinsic space of deformations.

In a sequel of this paper, we shall reformulate the theory of nonlinear elasticity in the port-Hamiltonian framework which will highlight the energetic structure underlying the equations of motion in addition to deriving these equations from first principles using Hamiltonian reduction techniques.

\section{Appendix}
\subsection{Fibre bundles}\label{appendix:bundles}
The theory of fibre bundles is essential for precisely defining the mathematical objects used to describe the physical quantities of nonlinear elasticity. These include vector fields, tensor fields, differential forms, and two point tensor fields. In what follows we provide a brief introduction of the topic. For further exposition see \cite{Abraham1988ManifoldsApplications}.

\newcommand{\spE}{\bb{E}}
\newcommand{\spF}{\bb{F}}
\newcommand{\spB}{\cl{B}}
\newcommand{\spS}{\cl{S}}

Let $\spE,\spB$ be smooth manifolds and let $\map{\pi}{\spE}{\spB}$ be a smooth surjection.
Then the triple $(\spE,\pi,\spB)$ is called a \textit{fibre bundle over} $\spB$, $\spE$ is referred to as the \textit{total space}, $\spB$ as the \textit{base space}, and $\pi$ as the \textit{projection map}.
The preimage of a point $X\in \spB$ under $\pi$ is called the the \textit{fibre at} $X$ and denoted by $\spE_X := \pi^{-1}(X)$.
A frequent notation for bundles is either $(\spE,\pi,\spB)$, $\map{\pi}{\spE}{\spB}$, or simply $\spE$ if it is clear from the context.
A bundle $(\spE,\pi,\spB)$ is said to be \textit{trivial} if it is isomorphic (as bundles) to a product bundle (i.e. $\spE = \spB\times \cl{F}$ for some manifold $\cl{F}$).

The tangent and cotangent bundles are defined as the disjoint unions of all tangent and cotangent space to $\spB$, respectively:
$$T\spB := \bigsqcup_{X\in \spB} T_X\spB, \qquad \qquad T^*\spB := \bigsqcup_{X\in \spB} T^*_X\spB.$$
A type $(p,q)$ tensor at $X\in \spB$ is a multilinear map
$$\map{\zeta}{\underbrace{T_X^*\spB\cdots T_X^*\spB}_{p\mathrm{-copies}}\times\underbrace{T_X\spB\cdots T_X\spB}_{q\mathrm{-copies}}}{\bb{R}}.$$
The bundle of all $(p,q)$ tensors is denoted by 
$$T^p_q\spB := \bigsqcup_{X\in \spB} T^p_{q,X}\spB,$$
where $T^p_{q,X}\spB$ denotes the set of all $(p,q)$ tensors at $X\in \spB$.
The set of all totally anti-symmetric $(0,k)$ tensors at $X\in \spB$ is denoted by $\Lambda^k T_X^*\spB$.
The disjoint union of all these spaces defines the bundle
$$\Lambda^k T^*\spB := \bigsqcup_{X\in \spB} \Lambda^k T_X^*\spB \subset T^0_k\spB.$$
Similarly, one can construct the bundle of all symmetric (0,2) tensors denoted by $ST^0_2\spB\subset T^0_2\spB$.

Now consider the two bundles $(\spE,\pi_\spE,\spB)$ and $(\spF,\pi_\spF,\spS)$ and the smooth map $\map{f}{\spB}{\spS}$.
We can combine the two bundles and define a new one over $\spB$ as the disjoint union 
$$\spF \otimes_f \spE:=\bigsqcup_{X \in \spB } \pi_\spE^{-1}(X) \otimes \pi_\spF^{-1}(f(X)).$$
We emphasize in the notation of the new bundle its dependence on the map $f$.
An alternative notation for $\spF \otimes_f \spE$ is also $\spF \otimes f^*\spE$.
The same procedure can be used to combine bundles over the same manifold (i.e. $\spS=\spB$) and the same point ($f = \textrm{id}_\spB$). In this case, the notational dependency on the map $f=\textrm{id}_\spB$ is usually suppressed.

Consider the tensor bundles $T^p_q\spB$ and $T^r_s\spS$ over $\spB$ and $\spS$, respectively.
A type ${\small\TwoTwoMat{p}{r}{q}{s}}$ \textit{two-point tensor} at $X\in \cl{B}$ over the map $\map{\varphi}{\cl{B}}{\cl{S}}$ is a multilinear map\cite[Pg. 70]{Marsden1994MathematicalElasticity}
$$\map{\xi}{\underbrace{T_X^*\cl{B}\cdots T_X^*\cl{B}}_{p\mathrm{-copies}}\times
	\underbrace{T_X\cl{B}\cdots T_X\cl{B}}_{q\mathrm{-copies}}\times
	\underbrace{T_{\varphi(X)}^*\cl{S}\cdots T_{\varphi(X)}^*\cl{S}}_{r\mathrm{-copies}}\times
	\underbrace{T_{\varphi(X)}\cl{S}\cdots T_{\varphi(X)}\cl{S}}_{s\mathrm{-copies}}}{\bb{R}}.$$
The bundle of ${\small\TwoTwoMat{p}{r}{q}{s}}$ two point tensors over $\spB$ will be denoted by
$T^p_q\spB\otimes \varphi^* T^r_s\spS \equiv T^p_q\spB\otimes_\varphi T^r_s\spS$.
For the case $p=q=0$, the bundle $\varphi^* T^r_s\spS$ is called the \textit{pullback or induced bundle} of $T^r_s\spS$ by $\varphi$, i.e.
$$\varphi^* T^r_s\spS := \bigsqcup_{X \in \spB }  T^r_{s,\varphi(X)} \spS.$$

Let $(\spE,\pi,\spB)$ be a fibre bundle, the map $\map{\sigma}{\spB}{\spE}$, such that $\pi \circ \sigma = \textrm{id}_\spB$, is called a \textit{section of the bundle}.
The set of all sections of $\spE$ is denoted by $\Gamma(\spE)$.
For example, $\spVec{\spB}$ is the set of vector fields, $\spVec{^p_q\spB}$ is the set of $(p,q)$ tensor fields,  and $\spVec{^p_q\spB\otimes \varphi^* T^r_s\spS}$ is the set of two-point tensor fields on $\spB$.
An element $\vM \in \spVecPhi$ is called a vector field over the map $\varphi$ while an element $\omega \in \Gamma(\Lambda^k T^*\spB)$ is called a (scalar-valued) differential $k$-form.
It is standard to denote $\Gamma(\Lambda^k T^*\spB)$ by $\spFrmB{k}$.

%
%
%
%
%
%
%

\subsection{Material time derivative}\label{append:mat_time_deriv}
\newcommand{\zT}{\tilde{\zeta}}
\newcommand{\DzT}{\frac{\textrm{D} \zT}{\extd t}}
Let $(M,g)$ be a Riemannian manifold with $\nabS$ its Levi-Civita connection.
Given a curve $\map{c}{I\subset \bb{R}}{M}$, a vector field along $c$ is a map
\begin{equation}
	\fullmap{\zT}{I}{TM}{t}{\zT(t) \in T_{c(t)}M.}
\end{equation}
One has that $\zT\in \Gamma(c^*TM)$ to be a section of the pullback bundle $c^*TM$.
Let $\xi\in \spVec{M}$ be any vector field such that $\zT(t) = \xi(c(t)) \in T_{c(t)}M$ (which in principle doesn't need to be defined on all of $M$ but only along $c$).
The covariant derivative of $\zT\in\Gamma(c^*TM)$ along $c$ is the vector field along $c$, denoted as $\DzT\in\Gamma(c^*TM)$, and defined by
\begin{equation}\label{eq:mat_deriv_def}
	\fullmap{\DzT}{I}{TM}{t}{\DzT(t):= (\nabS_{c'(t)}\xi)(c(t)),}
\end{equation}
where $\map{c'}{I}{TM}$ is the tangent vector field of $c$ with $c'(t)\in T_{c(t)}M$.

In continuum mechanics, the vector field $\DzT\in\Gamma(c^*TM)$ over $c$ is referred to as the material time derivative of $\zT\in\Gamma(c^*TM)$.
Note that it is common to use a different notation for covariant differentiation of vector fields along a curve than that used for covariant differentiation of true vector fields. Other popular notations are $\frac{\nabS}{\extd t}$ in \cite{Frankel2019ThePhysics} and $D_t $ in \cite{Kolev2021ObjectiveMetrics}.
In this work we shall opt for $D_t$ to denote the material time derivative.

Consider a chart $(U,x^i)$, such that $U\subset M$ and $\map{x^i}{M}{\bb{R}}$.
Let the curve $c$ have components $c^i(t) := x^i\circ c(t) \in \bb{R}$ and $c'(t)$ to be locally represented by ($c^i(t),\dot{c}^i(t)$).
If the components of $\zT$ are denoted by $\zT^i$, then the components of $D_t \zT$ are given by
$$\left(D_t \zT\right)^i = \partial_t \zT^i + (\Gamma^i_{jk} \circ c) \dot{c}^j \zT^k.$$

\subsection{Proof of Prop. \ref{prop:covDiff_decomp}}\label{propProof:covDiff_decomp}
Let $u,v,w \in \spVecS$ be any vector fields on the manifold $\cl{S}$ and consider the identity 
\begin{equation}\label{eq:Lie_deriv_identity_cov}
	\Lie{u}{v} = \nabS_u v - \nabS_v u, \qquad\forall u,v\in \spVecS.
\end{equation}
Using the fact that $\Lie{u}{f} = \nabS_u f = u(f), \forall f\in \spFn{\cl{S}}$, the Leibniz rule for the Lie derivative (over contraction and tensor product), the Leibniz rule for the covariant derivative (over tensor product), and (\ref{eq:Lie_deriv_identity_cov}), one can show that
\begin{align}
	\Lie{u}{g}(v,w) &= \Lie{u}{(g(v,w))} - g(\Lie{u}{v},w) - g(v,\Lie{u}{w}), \nonumber \\
					&= \nabS_u{(g(v,w))} - g(\Lie{u}{v},w) - g(v,\Lie{u}{w}), \nonumber \\
					&= g(\nabS_u v,w) + g(v,\nabS_uw) - g(\Lie{u}{v},w) - g(v,\Lie{u}{w}), \nonumber \\
					&= g(\nabS_u v - \Lie{u}{v},w) + g(v,\nabS_uw - \Lie{u}{w}), \nonumber \\
					&= g(\nabS_v u,w) + g(v,\nabS_w u) = \nabS_v u^\flat (w) + \nabS_w u^\flat (v) .\label{eq:proofApp_1}
\end{align}
Now consider the coordinate free definition of the exterior derivative which can be expressed using the Leibniz rule for the covariant derivative and (\ref{eq:Lie_deriv_identity_cov}) as
\begin{align}
	\extd u^\flat(v,w) &= v(u^\flat(w)) -  w(u^\flat(v)) - u^\flat(\Lie{v}{w})
						= \nabS_v(u^\flat(w)) -  \nabS_w(u^\flat(v)) - u^\flat(\nabS_v w - \nabS_w v) \nonumber\\
						&= \nabS_v u^\flat(w) + u^\flat(\nabS_v w) - \nabS_w u^\flat(v) - u^\flat(\nabS_w v) - u^\flat(\nabS_v w) + u^\flat(\nabS_w v)\nonumber\\
						&= \nabS_v u^\flat(w) - \nabS_w u^\flat(v) .\label{eq:proofApp_2}
\end{align}
Summing (\ref{eq:proofApp_1}) and  (\ref{eq:proofApp_2}) yields
$$2 \nabS_v u^\flat(w) =2 \nabS u^\flat(v,w)=\Lie{u}{g}(v,w)+ \extd u^\flat(v,w),$$
which proves (\ref{eq:vel_grad_identity_S_}) and (\ref{eq:vel_grad_identity_C_}) due to arbitrariness of $(v,w)$.

Now consider 
$$\nabS v^\flat(v) = \half \Lie{v}{g}(v) + \half \extd v^\flat(v).$$
Using $\Lie{v}{g}(v) = \Lie{v}{v^\flat}$ and Cartan's identity (\ref{eq:Cartan_Lie_deriv}), one can show that
\begin{align*}
	\nabS v^\flat(v) = \half \Lie{v}{v^\flat} + \half \iota_v \extd v^\flat 
					= \half \Lie{v}{v^\flat} + \half \Lie{v}{v^\flat} - \half  \extd \iota_v v^\flat = 
					\Lie{v}{v^\flat} - \half  \extd \iota_v v^\flat,
\end{align*}
which proves (\ref{eq:vel_grad_identity_S}) and (\ref{eq:vel_grad_identity_C}).



\begin{acknowledgements}
This work was supported by the PortWings project funded by the European Research Council
[Grant Agreement No. 787675]
\end{acknowledgements}

\bibliographystyle{ieeetr}
\bibliography{references}

\end{document}